\documentclass[aps,prd,showpacs,eqsecnum,nofootinbib,superscriptaddress]{revtex4-1}

\usepackage{amsmath,amssymb,graphicx,color,ulem,multirow}
\usepackage{array,tikz-cd}
\usepackage{comment}
\usepackage{graphicx}
\usepackage{natbib} 
\usepackage[caption=false]{subfig}

\newcommand{\col}{\mathfrak{C}}
\newcommand{\R}{r}
\newcommand{\G}{g}
\newcommand{\B}{b}

\definecolor{labelkey}{rgb}{0.4,0.4,0.4}

\newcommand{\hf}{\frac{1}{2}}

\newcommand{\Tr}{{\rm Tr~}}

\newcommand{\trans}{\hat{T}}
\newcommand{\om}{\hat{\Omega}}

\newcommand{\reflection}{\hat{\Upsilon}}

\newcommand{\projz}{\mathcal{P}^{\Omega}}
\newcommand{\projt}{\mathcal{P}^{T} }
\newcommand{\projec}{\mathcal{P} }

\newcommand{\as}{a}
\newcommand{\ad}{a^2}
\newcommand{\at}{a^3}
\newcommand{\bs}{b}
\newcommand{\bd}{b^2}
\newcommand{\bt}{b^3}
\newcommand{\cs}{c}
\newcommand{\cd}{c^2}
\newcommand{\ct}{c^3}
\newcommand{\ds}{d}
\newcommand{\dd}{d^2}
\newcommand{\dt}{d^3}

\def\nn{\nonumber}
\def\beq{\begin{equation}}
\def\eeq{\end{equation}}
\def\bea{\begin{eqnarray}}
\def\eea{\end{eqnarray}}

\def\bec{\begin{center}}
\def\eec{\end{center}}

\begin{document}

\title{Abelian Tensor Models on the Lattice}

\author{Soumyadeep Chaudhuri}
\author{Victor I. Giraldo-Rivera}
\author{Anosh Joseph}
\author{R. Loganayagam}
\author{Junggi Yoon}

\affiliation{International Centre for Theoretical Sciences (ICTS-TIFR), \\
				Tata Institute of Fundamental Research, \\
				Shivakote, Bangalore 560089 INDIA} 

\date{\today}

\begin{abstract}

We consider a chain of Abelian Klebanov-Tarnopolsky fermionic tensor models coupled through quartic nearest-neighbor interactions. We characterize the gauge-singlet spectrum for small chains ($L=2,3,4,5$) and observe that the spectral statistics exhibits strong evidence in favor of quasi-many-body localization. 

\end{abstract}


\maketitle

\section{Introduction}
\label{sec:intro}

Tensor models~\cite{Gurau:2011xq,Bonzom:2012hw,Carrozza:2015adg, Gurau:2016cjo, Gurau:2016lzk, Klebanov:2016xxf} of fermions provide a novel class of quantum mechanical models where a qualitatively new kind of large-$N$ limit could be studied. In their large $N$ limits, they are intermediate between the familiar class of vector models and matrix models. On one hand, they share with vector models the ease of solvability: the dominating diagrams at  large-$N$ are of a simple type and can easily be resummed to give tractable Schwinger-Dyson equations. On the other hand, they share with matrix models the feature of nontrivial large-$N$ dynamics.

The recent interest in tensor models is triggered by Witten's observation~\cite{Witten:2016iux} that they are large-$N$ equivalent to the maximally chaotic, disordered, vectorlike fermion models of Sachdev-Ye-Kitaev (SYK)~\cite{Sachdev:1992fk, kitaevfirsttalk, KitaevTalks}. The interest on SYK from a quantum gravity viewpoint stems in turn from their similarity to black holes~\cite{Shenker:2013pqa} in being maximally chaotic~\cite{kitaevfirsttalk, KitaevTalks, Maldacena:2015waa, Maldacena:2016upp}.

At finite $N$, unlike SYK models, tensor models have the merit of being completely unitary quantum mechanical models using which, for example, one might hope to understand more about finite-$N$ restoration of unitarity in black holes. The disadvantage of tensor models lies in their relative unfamiliarity compared to matrix models and the rapid growth of the degrees of freedom with $N$, making numerical computations at finite $N$ using existing techniques very expensive. There are also many fundamental questions about tensor models including the general structure of gauge invariants, in particular the set of them that dominate large-$N$ limit (viz. the analogues of `single-trace' operators) which remain unsolved. There is a clear necessity for coming up with effective and efficient ways of tackling tensor models, which would allow us to work out the finite-$N$ physics of these class of models.

In this work, we will embark on the analysis of what is perhaps the simplest of tensor models: those whose gauge groups are Abelian. Our main focus would be lattice versions of the uncolored tensor model \emph{a la} Klebanov-Tarnopolsky~\cite{Klebanov:2016xxf}\footnote{The spectrum of Abelian Gurau-Witten models and $N = 3$ Klebanov-Tarnopolsky models have been studied recently in~\cite{Krishnan:2016bvg, Krishnan:2017ztz}.} (henceforth, we call it KT model.)  Such an uncolored tensor model was introduced by Carrozza and Tanasa~\cite{Carrozza:2015adg} with 0-dimensional bosonic $O(N)^3$ random tensors, and Klebanov and Tarnopolsky generalized it into the 0+1-dimensional fermionic $O(N)^3$ tensors~\cite{Klebanov:2016xxf}.

These lattice tensor models (which we will refer to as KT chains here on) are unitary counterparts of lattice SYK model studied by Gu-Qi-Stanford \cite{Gu:2016oyy}\footnote{Other lattice generalizations of SYK model include \cite{Berkooz:2016cvq, Davison:2016ngz, Banerjee:2016ncu, Jian:2017unn, Jian:2017jfl}.}. The large $N$ versions of these KT chains will be studied in detail in an adjoining paper~\cite{Narayan:2017qtw} by a different set of authors. We will refer the reader there for a description of how large $N$ KT chains largely reproduce the phenomenology of Gu-Qi-Stanford model with its maximal chaos and characteristic Schwarzian diffusion. Our aim here would be to study the Abelian counterparts of the models in~\cite{Narayan:2017qtw} with a special focus on the singlet spectrum.

The Abelian KT chains do not exhibit maximal chaos. In fact, as we will argue in the following sections, a variety of spectral diagnostics of singlet states show it to be closer to being integrable -- may be even many-body-localized, though the lattice sizes we study here are unfortunately too small to resolve that question. In this, the Abelian KT tensor chains are qualitatively different from their non-Abelian large $N$ cousins whose large time dynamics and spectral statistics of singlet states are governed by random-matrix-like behavior~\cite{Cotler:2016fpe, Garcia-Garcia:2017pzl}. Given this qualitative difference, the Abelian KT chains  are far from the classical black-hole-like behavior that sparked the recent interest in tensor models. We will thus begin by explaining our motivations for studying these models.

First of all, it is logical to begin a finite $N$ study of tensor models with the study of Abelian models. Given the intricate and unfamiliar structure of singlet observables in tensor models, the Abelian model provides a toy model to train our intuition. These Abelian tensor chains are simple enough for us to exactly solve for their gauge-invariant spectrum (at least for small chain lengths). We hope that the results presented in this work would serve as a stepping stone for a similar analysis in non-Abelian tensor models.

A second broader (if more vague) motivation is to have a toy model to see whether tensor models can be embedded within string theory. This is an outstanding and a crucial question on which hinges the utility of tensor models for quantum gravity and black holes: can we find exact holographic duals of tensor models with black hole solutions?~\footnote{See ~\cite{Maldacena:2016upp, Mandal:2017thl, Das:2017pif} for holographic models reproducing SYK-like spectrum. See~\cite{Peng:2016mxj} for a supersymmetric version of tensor models.} This may well require an embedding into string theory, however it is unclear at present how this might come about. Our hope is that the study of Abelian tensor models can give us some intuition on the analogue of the `coulomb branch' for tensor models. Like the Abelian gauge theories which describe D-branes, Abelian tensor models may give us  intuitions about string theory particles on which the tensor models live.

A third motivation is from the viewpoint of many-body localization~(MBL)~\cite{2006AnPhy.321.1126B, 2016JSP...163..998I, 2015ARCMP...6...15N}\footnote{For works which have studied MBL-like and other insulating phases using large $N$ SYK model, see~\cite{Jian:2017unn, Jian:2017jfl}.}. Many-body localization is a phenomenon by which a quantum system (often with a quenched disorder) fails to thermalize in the sense of Eigenstate thermalization hypothesis~(ETH)~\cite{1991PhRvA..43.2046D, 1994PhRvE..50..888S, 2008Natur.452..854R}.

ETH posits that in an energy eigenstate of an isolated quantum many body system, any smooth local observable will eventually evolve to its corresponding microcanonical ensemble average. The idea of ETH is to hypothesize that, in this sense, every energy eigenstate behaves like a thermal bath for its subsystems and the subsystem is effectively in a thermal state. By now, many low dimensional disordered systems are known where ETH has been known to fail, thus leading to a many body localized (MBL) phase where even interactions fail to thermalize the system. MBL behavior signals a breakdown of ergodicity in the system and is often  associated with integrability or near-integrability. Its name derives from the fact that it is the many-body and Hilbert space analogue of Anderson localization~\cite{1958PhRv..109.1492A} whereby in low dimensions, a single particle moving in a disordered potential gets spatially localized.

MBL phase is a novel nonergodic state of matter where standard statistical mechanical intuitions fail. Thus, the failure of thermalization and the emergence of MBL behavior has drawn a great amount of interest recently. Given the variety of disordered models which have been studied in the context of MBL, it is a natural question to enquire whether a quenched disorder is strictly necessary for MBL-like behavior. An interesting question is to enquire whether one can achieve MBL behavior in a translation invariant unitary model~\cite{kagan1984localization, 2013arXiv1305.5127D, 2014CMaPh.332.1017D, 2014PhRvB..90p5137D, 2014AIPC.1610...11S, 2014JSMTE..10..010G}. This question has been vigorously debated in the recent literature with many authors~\cite{2014arXiv1409.8054D, 2016PhRvL.117x0601Y, 2015AnPhy.362..714P} concluding that MBL-like behavior in translation invariant systems is likely to be not as robust as the localizing behavior observed in disordered systems. When one tries to construct a unitary model which can naively exhibit MBL-like behavior, one ends up instead with a quasi-many-body localized state (qMBL)~\cite{2016PhRvL.117x0601Y} where a many-body localizationlike behavior persists for long but finite times, but thermalization does happen eventually. For example, in the systems studied by~\cite{2016PhRvL.117x0601Y}, the time scales involved for thermalization of modes with a small wave number $k$ are nonperturbatively long (i.e., $\tau \sim \exp[1/(k\xi)]$) but finite even as system size is taken to be infinite. Such slowly thermalizing, almost MBL like systems are interesting on their own right since they show a transition from MBL-like behavior to ergodic behavior as they evolve in time. The Abelian KT chains that we study in this work share many similarities with the model described in~\cite{2015AnPhy.362..714P, 2016PhRvL.117x0601Y} and we expect a similar low temperature phase with anomalous diffusion in the thermodynamic limit.

In this work, we will present preliminary evidence that Abelian KT chains at small $L$ indeed seem to exhibit the necessary features to exhibit a quasi-many-body localized behavior. Since our main concern here would be the singlet spectrum, the main evidence we will present here will be the characteristically large degeneracies in the middle part of the singlet spectrum \footnote{We note that previous studies of Abelian quantum mechanical tensor models without the singlet condition~\cite{Krishnan:2016bvg, Krishnan:2017ztz} (as opposed to tensor models on lattice with singlet condition studied here) had also reported huge degeneracies in the middle part of the spectrum.} and a Poisson-like spectral statistics (showing near integrability and consequently MBL-like behavior). We will leave to future work a more detailed analysis of possible quasi-localization in these sets of models (like transport, entanglement etc.) at thermodynamic limit.

Our main concern in this work would be to study the spectrum of the fermionic tensor chain built out of tensor models by Klebanov-Tarnopolsky~\cite{Klebanov:2016xxf}. KT model is a quantum mechanical theory of a real fermionic field $\psi_{ijk}$ which transforms in the tri-fundamental representation of an $SO(N)^3$ gauge group. These fermions interact via a Hamiltonian
\beq
H = \frac{g}{4}~\psi_{ijk} \psi_{ilm} \psi_{njm} \psi_{nlk}\ ,
\eeq
where we will be interested in the simplest case of $N = 2$, which we refer to as the Abelian KT model. The Hilbert space of this model is exceedingly simple with just $16$ states. Of these $14$ are degenerate and lie in the middle of the spectrum whereas the rest two states are split off from these midspectral states by an energy gap $\pm 2g$. We will choose the zero of our energy to lie in the middle of the spectrum so that whenever the spectrum is symmetric about the middle, the corresponding spectral reflection symmetry is manifest. We will use this simple model as the ``atom'' to build an Abelian KT chain. 

The Abelian KT chain is made of $L$ copies of Abelian KT models arranged on a circle and with a Gu-Qi-Stanford type hopping term connecting the nearest neighbors:
\beq
\begin{split}
H = \frac{g}{4} \sum_{a=1}^L \psi^{(a)}_{ijk} \psi^{(a)}_{ilm}
\psi^{(a)}_{njm} \psi^{(a)}_{nlk} + \frac{\lambda_r}{4}
\sum_{a=1}^L\psi^{(a)}_{ijk} \psi^{(a)}_{ilm}
\psi^{(a+1)}_{njm} \psi^{(a+1)}_{nlk} \\ + \frac{\lambda_g}{4}
\sum_{a=1}^L \psi^{(a)}_{ijk} \psi^{(a+1)}_{ilm} \psi^{(a)}_{njm}
\psi^{(a+1)}_{nlk} +
\frac{\lambda_b}{4} \sum_{a=1}^L \psi^{(a)}_{ijk} \psi^{(a+1)}_{ilm}
\psi^{(a+1)}_{njm} \psi^{(a)}_{nlk},
\end{split}
\eeq
where we impose periodic boundary conditions for fermions: $\psi^{(L+1)} = \psi^{(1)}$. Here, $\lambda_{r,g,b}$ are the three Gu-Qi-Stanford couplings which differ from each other in which of the three $SO(2)^3$ index is contracted across the sites.

The outline of this paper is as follows: after a brief review of SYK and tensor models in Sec~\ref{sec:review-tensor-models} and Abelian KT models in Sec~\ref{ssec:KT1site}, in Sec~\ref{ssec:KT2site} we will present a complete singlet spectrum of the simplest KT chain: the 2 site Abelian KT chain. This is followed by a detailed analysis of the gauge singlet spectrum of the 3 site and 4 site Abelian KT chains in Secs~\ref{sec:KT3site} and \ref{sec:KT4site}, respectively. In these sections, we will focus on the various structural features of the spectrum which are generic to the Abelian tensor chains. After a brief description of how these features generalize to the 5-site case in Sec~\ref{sec:KT5site}, we will analyze the spectra of these models and the associated thermodynamics in Sec~\ref{sec:spectra} and argue for the near-integrability of Abelian KT chains. We will conclude in Sec~\ref{sec:conclusions} with discussions on further directions. Some of the technical details about the spectrum of $4$ site KT chain are relegated to Appendix~\ref{app:4site}.

\section{Review: Tensor Model and SYK Model}
\label{sec:review-tensor-models}

\subsection{Klebanov-Tarnopolsky model (KT tensor model)}
\label{sec:KT-Model}

We begin with a review of the Klebanov-Tarnopolsky~(KT) model~\cite{Klebanov:2016xxf} which is the simplest tensor model exhibiting maximal chaos in large $N$. The KT model is a unitary quantum mechanical model of a real fermion field $\psi_{ijk}$ ($i,j,k = 1,2, \cdots N$) which transforms in the trifundamental representation of $SO(N)^3$ gauge group. We will find it convenient to distinguish three $SO(N)$ gauge groups by RGB color. i.e., $r$, $g$ and $b$ denote the color of the first, second and the third $SO(N)$ gauge groups. We will also correspondingly take the first, second and third gauge indices $i$, $j$ and $k$ of $\psi_{ijk}$ to also be of colors red, green and blue respectively. 
 
\begin{figure}[htp]

\centering
\includegraphics[width=2.1in]{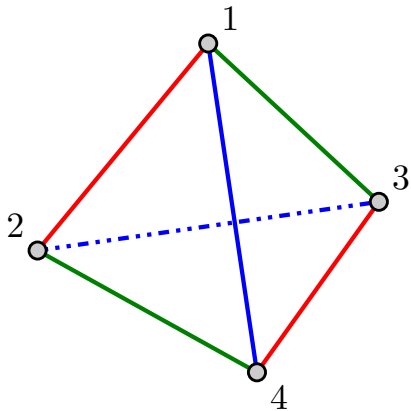}
\caption{The tetrahedron contraction of the gauge indices in the Hamiltonian of the KT model~\eqref{def:KT model Hamiltonian}. Each vertex of the tetrahedron corresponds to the fermion $\psi_{ijk}$, and the each edge represents the gauge contraction of the corresponding color between two fermions.}
\label{fig:tetrahedron-diagram}

\end{figure}

The Hamiltonian of the KT model is given by
\beq
H = \frac{g}{4} \; \psi_{ijk} \psi_{ilm} \psi_{njm} \psi_{nlk}\ .
\label{def:KT model Hamiltonian}
\eeq
Note that the four-fermion gauge index contractions in tensor models with SYK-like behavior have a tetrahedronlike structure. Henceforth, we will call it tetrahedron interaction. Fig.~\ref{fig:tetrahedron-diagram} represents the gauge contraction of four fermions in the Hamiltonian~\eqref{def:KT model Hamiltonian}. Each vertex of the tetrahedron represents a fermion field $\psi_{ijk}$ whereas the edges denote their gauge indices. The tetrahedron is then a geometric representation of how the color indices contract.  
%
%
%

The tetrahedral structure of the gauge contraction is crucial to the dominance of melonic diagrams in large $N$, which in turn enables us to solve the model in the strong coupling limit. In the strong coupling limit, like SYK model, KT model also exhibits an emergent reparametrization symmetry. This reparametrization symmetry is (explicitly and spontaneously) broken, and the associated Goldstone boson leads to the characteristic Schwarzian (and {\it inter alia} maximally chaotic) behavior of the model.

This maximal chaos is not restricted to the KT model, but has been found in a wide class of tensor models with tetrahedron interaction. For example, Gurau-Witten model~\cite{Witten:2016iux, Gurau:2016lzk} is also maximally chaotic. Fermions of the Gurau-Witten model have an additional flavor index, and similar features are observed in large $N$: the emergence and the breaking of reparametrization symmetry and maximal chaos. We refer the reader to~\cite{Narayan:2017qtw} for a more detailed description of these models and a general technique which can be used to show maximal chaos not only in large $N$ KT model but also in large $N$ Gurau-Witten model and lattice generalizations thereof.

\subsection{SYK model and its extension to a $1d$ model}
\label{sec:review of SYK model}

We will now very briefly review the SYK model~\cite{Sachdev:1992fk, kitaevfirsttalk, KitaevTalks, Sachdev:2015efa} and its lattice generalization by Gu-Qi-Stanford \cite{Gu:2017ohj} which inspired the lattice models of this work. SYK model is a vectorlike quantum mechanical model of $N_{\text{\tiny SYK}}$ real fermions $\psi_i$ ($i = 1,2,\cdots, N_{\text{\tiny SYK}})$, but with disorder in form of a random four-fermion interaction. The Hamiltonian is given by
\beq
H_{\text{SYK}} \equiv \sum_{1 \leq k < l < m< n \leq N_{\text{\tiny SYK}}} j_{k l m n} \psi_{k} \psi_{l} \psi_{m} \psi_{n}\ ,
\eeq
where $j_{ k l m n}$ is a Gaussian-random coupling with variance $\overline{ j_{klmn}^2 } = \frac{3! J^2 }{N^3_{\text{\tiny SYK}}}$. After disorder average, the  melonic diagrams dominate two point functions in large $N_{\text{\tiny SYK}}$, and reparametrization symmetry emerges in the strong coupling limit~\cite{KitaevTalks, Polchinski:2016xgd, Jevicki:2016bwu, Maldacena:2016hyu, Jevicki:2016ito}. Like the KT model, the reparametrization symmetry is broken explicitly and spontaneously, which, in turn, leads to maximal chaos due to the corresponding Schwarzian pseudo-Goldstone boson~\cite{Maldacena:2016hyu}.

A lattice generalization of the SYK model was  studied by Gu-Qi-Stanford in~\cite{Gu:2017ohj} where $L$ copies of SYK models on a $L$-site lattice are interacting via the nearest neighbor interaction:
\beq
H =\sum_{1 \leq k < l < m < n \leq N}  \;\;\sum_{a=1}^{L}  \Big[ j_{klmn} \psi^{(a)}_{k} \psi^{(a)}_{l} \psi^{(a)}_{m} \psi^{(a)}_{n} +   j^{\prime}_{klmn} \psi^{(a)}_{k} \psi^{(a)}_{l} \psi^{(a+1)}_{m} \psi^{(a+1)}_{n} \Big]\ .
\eeq
Here $j_{klmn}$ and $j^{\prime}_{klmn}$ are two Gaussian-random couplings with variances $\overline{ j_{klmn}^2 } = \frac{3! J^2_0 }{N^3}$ and $\overline{ j^{\prime \; 2}_{klmn} } = \frac{J_1^2}{N^3}$, respectively. Periodic boundary conditions are imposed on the fermions: $\psi^{(L+1)} = \psi^{(1)}$. Note that this model has a $\mathbb{Z}_2^L$ global symmetry which acts by flipping the sign of all fermions in a given site $\psi^{(a)} \mapsto -\psi^{(a)}$ for each lattice index $a$ ($ = 1, 2, \cdots, L$).
This model also exhibits maximal chaos in large $N$. Furthermore, in this lattice generalization of the SYK model, one can evaluate the speed with which chaos propagates in space (the butterfly velocity).

In this work, we construct a similar lattice model as above where the SYK model is replaced with a KT tensor model instead. Further, our focus will be on the opposite limit to the large $N$ limit studied in the works referenced above. Consequently, the phenomenology of our lattice chains would be very different from their large $N$ cousins described above.

\section{Abelian KT Model}

\subsection{Hamiltonian}
\label{ssec:KT1site}

As mentioned before, the Hamiltonian of the KT model is given by
\beq \label{eq:KTpsiH}
H = \frac{g}{4} \psi_{ijk} \psi_{ilm} \psi_{njm} \psi_{nlk}\ ,
\eeq
where $\{ \psi_{ijk}, \psi_{lmn} \} = \delta_{il} \delta_{jm} \delta_{kn}$.

Here we consider the Abelian case that is, when $N = 2$. Since $SO(2)^3\cong U(1)^3$, we will find it convenient to think of the gauge group as $U(1)^3$ instead. We can then use these charges to define components of $\psi$ with definite $U(1)^3$ charges. In this work, we will mostly be interested in $U(1)^3$ gauge-singlet states and the spectrum in the singlet sector.

We will find it convenient to introduce the following creation and annihilation operators which we will use from here on:
\bea
a_1 &\equiv& \psi_{\frac{(1-i2)}{\sqrt{2}}\frac{(1-i2)}{\sqrt{2}}\frac{(1-i2)}{\sqrt{2}}}\ , \nn \\
a_1^\dag &\equiv& \psi_{\frac{(1+i2)}{\sqrt{2}}\frac{(1+i2)}{\sqrt{2}}\frac{(1+i2)}{\sqrt{2}}}\ , \nn \\
a_2 &\equiv& \psi_{\frac{(1-i2)}{\sqrt{2}}\frac{(1+i2)}{\sqrt{2}}\frac{(1+i2)}{\sqrt{2}}}\ , \nn \\
a_2^\dag &\equiv& \psi_{\frac{(1+i2)}{\sqrt{2}}\frac{(1-i2)}{\sqrt{2}}\frac{(1-i2)}{\sqrt{2}}}\ , \\
a_3 &\equiv& \psi_{\frac{(1+i2)}{\sqrt{2}}\frac{(1-i2)}{\sqrt{2}}\frac{(1+i2)}{\sqrt{2}}}\ , \nn \\
a_3^\dag &\equiv& \psi_{\frac{(1-i2)}{\sqrt{2}}\frac{(1+i2)}{\sqrt{2}}\frac{(1-i2)}{\sqrt{2}}}\ , \nn \\
a_4 &\equiv& \psi_{\frac{(1+i2)}{\sqrt{2}}\frac{(1+i2)}{\sqrt{2}}\frac{(1-i2)}{\sqrt{2}}}\ , \nn \\
a_4^\dag &\equiv& \psi_{\frac{(1-i2)}{\sqrt{2}}\frac{(1-i2)}{\sqrt{2}}\frac{(1+i2)}{\sqrt{2}}}\ , \nn 
\eea
where 
\bea
\psi_{\frac{(1+\sigma_1 i2)}{\sqrt{2}}\frac{(1+\sigma_2 i2)}{\sqrt{2}}\frac{(1+\sigma_3 i2)}{\sqrt{2}}} &=& \frac{1}{2\sqrt{2}} \Big(\psi_{111} - \sigma_2 \sigma_3 \psi_{122} - \sigma_3 \sigma_1 \psi_{212} - \sigma_1 \sigma_2 \psi_{221} \nn \\
&&\qquad+ i(\sigma_1 \psi_{211} + \sigma_2 \psi_{121} + \sigma_3 \psi_{112} - \sigma_1 \sigma_2 \sigma_3 \psi_{222})\Big)\ .
\eea

These operators satisfy the relations
\beq
\{a_i, a_j\} = 0,~~\{a_i^\dag, a_j^\dag\} = 0,~~\{a_i, a_j^\dag\} = \delta_{ij}\ .
\eeq
\begin{table}[h!]
\centering
\begin{tabular}{|c|ccc|}
\hline
 & $U(1)_r$ & $U(1)_g$ & $U(1)_b$ \\   
\hline
\hline
 $a_1$ & $-$ & $-$ & $-$ \\
 $a_1^\dag$ & $+$ & $+$ & $+$ \\
 $a_2$ & $-$ & $+$ & $+$ \\
 $a_2^\dag$ & $+$ & $-$ & $-$ \\
 $a_3$ & $+$ & $-$ & $+$ \\
 $a_3^\dag$ & $-$ & $+$ & $-$ \\
 $a_4$ & $+$ & $+$ & $-$ \\
 $a_4^\dag$ & $-$ & $-$ & $+$ \\
\hline
\end{tabular}
\caption{The $U(1)^3$ charges of creation and annihilation operators of the Abelian KT model.} 
\label{tab:u1-cube-charges}
\end{table}
The creation and annihilation operators that we have formed out of the fermionic fields have definite charges under the $U(1)^3$ gauge symmetry as given in Table~\ref{tab:u1-cube-charges}.

In terms of these creation and annihilation operators, the Hamiltonian has the form
\beq
H = 2g~\Big( a_1^\dag a_2^\dag a_3^\dag a_4^\dag+a_4 a_3 a_2 a_1 \Big)\ .
\eeq
In writing this expression, we have removed an irrelevant constant energy shift by $g$ from the Hamiltonian given in~\eqref{eq:KTpsiH}. Written in this form, this Hamiltonian exhibits a spectral reflection symmetry and the energy eigenstates are symmetrically distributed on either side
of $E = 0$ .

It is interesting to see that this Hamiltonian also has two other symmetries:
\begin{itemize}
\item $\mathbb{Z}_3$ symmetry generated by $\hat{\Omega}$:
\begin{equation}
\hat{\Omega} a_{i}\hat{\Omega}^{-1} =a_{(234)\cdot i} \hspace{1cm} (i = 1, 2, 3, 4)
\end{equation}
where $(234)\cdot i$ is the cyclic permutation of $i$ by $(234)\in S_3$ ($i = 1, 2, 3, 4$), e.g.
%
\beq
\label{eq: example of omega action}
\begin{split}
\hat{\Omega} \ a_1 \ \hat{\Omega}^{-1} = a_1 \ , \qquad
\hat{\Omega} \ a_2 \ \hat{\Omega}^{-1} = a_3\ , \\ 
\hat{\Omega} \ a_3 \ \hat{\Omega}^{-1} = a_4 \ , \qquad
\hat{\Omega} \ a_4 \ \hat{\Omega}^{-1} = a_2\ .
\end{split}
\eeq

\item A spectral reflection under an operation $\hat{\Sigma}$ where
\beq
\hat{\Sigma} \ a_i \ \hat{\Sigma}^{-1} = e^{i\pi \over 4} a_i \hspace{1cm} (i = 1, 2, 3, 4)\ .
\label{def: spectral reflection}
\eeq
%
%
%
Note that $\hat{\Sigma} H \hat{\Sigma}^{-1} = -H$. The operator $\hat{\Sigma}$ generates $\mathbb{Z}_8$ group whereas the Hamiltonian is invariant under a $\mathbb{Z}_4$ subgroup generated by $\hat{\Sigma}^2$. The remaining elements of $\mathbb{Z}_8$ group ($\hat{\Sigma}$, $\hat{\Sigma}^3$, $\hat{\Sigma}^5$, and $\hat{\Sigma}^7$) flip the  sign of the Hamiltonian.
\item We note that the $\mathbb{Z}_2$ action which maps $\psi_{ijk}$ to $-\psi_{ijk}$ is generated by the element $\hat{\Sigma}^4$. 
\end{itemize}
We find it useful to define the following notations for the products of the creation operators:
\beq
\begin{split}
A_\pm^\dag & \equiv \frac{1}{\sqrt{2}} (1 \pm a^{\dag}_1 a^{\dag}_2 a^{\dag}_3 a^{\dag}_4)\ , \\
(a^{\dag^2})_{ij} & \equiv a^{\dag}_i a^{\dag}_j\ , \\
(a^{\dag^2})_{\widehat{ij}} & \equiv \frac{1}{2!} \sum_{k,l} \epsilon_{ijkl} a^{\dag}_k a^{\dag}_l\ , \\
(a^{\dag^3})_{\hat{i}} & \equiv \frac{1}{3!} \sum_{j,k,l} \epsilon_{ijkl} a^{\dag}_j a^{\dag}_k a^{\dag}_l\ . \\
\end{split}
\label{def:notations for oscillators}
\eeq
We define a vacuum $|\  \rangle$ which is annihilated by all the $a_i$'s. (i.e., $a_i | \  \rangle = 0$.) The Hilbert space for this theory is 16 dimensional. We find it useful to work with the following basis:
\beq
\begin{split}
| A_+ \rangle & \equiv A_+^\dag |\ \rangle\ , \\
| A_- \rangle & \equiv A_-^\dag |\ \rangle\ , \\
|\as_i \rangle & \equiv \as_i^\dag|\ \rangle\ , \\
|\ad_{ij} \rangle & \equiv (a^{\dag^2})_{ij}|\ \rangle\ , \\
|\at_{\hat{i}} \rangle & \equiv (a^{\dag^3})_{\hat{i}}|\ \rangle\ . \\
\end{split}
\eeq
Here we have chosen a convenient notation for the basis states, which will be useful for later sections. More explicitly, we have a mapping as given in Table~\ref{tab:ijhat}.
\begin{table}[h!]
\centering

\begin{tabular}{ |c|c| } 
\hline
\multirow{2}{*}{$i$} & \multirow{2}{*}{$\hat{i}$} \\
& \\
\hline
1 & 234 \\ 2 & 143 \\ 3 & 124 \\ 4 & 132 \\
\hline
\end{tabular}
\qquad
\begin{tabular}{ |c|c| } 
\hline
 & \\
$ij$ & $\widehat{ij}$ \\
 & \\
\hline
12 & 34 \\ 13 & 42 \\ 14 & 23 \\
\hline
\end{tabular}

\caption{Definition of $\hat{i}$ and $\widehat{ij}$ for defining $(a^{\dag^3})_{\hat{i}}$ and $(a^{\dag^2})_{\widehat{ij}}$ respectively. If  $\widehat{ij}=kl$, then  $\widehat{kl}=ij$ and $\widehat{ji}=lk$.}
\label{tab:ijhat}
\end{table}

The spectrum of this model is given in Table~\ref{tab:spectrum of 1site KT model}.
\begin{table}[h!]
\centering
\begin{tabular}{ |c|c| } 
\hline
Eigenvalues & Degeneracy \\
\hline
0 & 14 \\
\hline
$\pm 2g$ & 1 \\
\hline
\end{tabular}
\caption{Spectrum for the one-site Abelian KT model. }
\label{tab:spectrum of 1site KT model}
\end{table}
There are $14$ middle states (states with zero energy), and they are given by four one-fermion states of the form $|\as_i\rangle$, six two-fermion states of the form $|\ad_{ij}\rangle$ and four three-fermion states of the form $|\at_{\hat{i}}\rangle$. The other states $|A_+\rangle$ and $|A_-\rangle$ have energies $+2g$ and $-2g$, respectively.

In Table~\ref{tab:energy-actions-etc}, we provide these 16 states, the actions of operators $\hat{\Sigma}$ and $\hat{\Omega}$ on each state, the corresponding energies of the states, and the $U(1)^3$ charges.
\begin{table}[t!]
\centering
\begin{tabular}{|c|c|c|c|c|}
\hline
State & Action of $\hat{\Sigma}$ & Action of $\hat{\Omega}$ & Energy & $U(1)^3$ charges \\
\hline
\hline
$|A_+\rangle$ & $|A_-\rangle$ & $|A_+\rangle$ & $2g$ & $(0,0,0)$ \\
\hline
$|A_-\rangle$ & $|A_+\rangle$ & $|A_-\rangle$ & $-2g$ & $(0,0,0)$ \\
\hline
$|\as_1\rangle$ & $\sqrt{-i} |\as_1\rangle$ & $|\as_1\rangle$ & $0$ & $(1,1,1)$ \\
\hline
$|\as_2\rangle$ & $\sqrt{-i} |\as_2\rangle$ & $|\as_3\rangle$ & $0$ & $(1,-1,-1)$ \\
\hline
$|\as_3\rangle$ & $\sqrt{-i} |\as_3\rangle$ & $|\as_4\rangle$ & $0$ & $(-1,1,-1)$ \\
\hline
$|\as_4\rangle$ & $\sqrt{-i} |\as_4\rangle$ & $|\as_2\rangle$ & $0$ & $(-1,-1,1)$ \\
\hline
$|\ad_{12}\rangle$ & $-i |\ad_{12}\rangle$ & $|\ad_{13}\rangle$ & $0$ & $(2,0,0)$ \\
\hline
$|\ad_{13}\rangle$ & $-i |\ad_{13}\rangle$ & $|\ad_{14}\rangle$ & $0$ & $(0,2,0)$ \\
\hline
$|\ad_{14}\rangle$ & $-i |\ad_{14}\rangle$ & $|\ad_{12}\rangle$ & $0$ & $(0,0,2)$ \\
\hline
$|\ad_{23}\rangle$ & $-i |\ad_{23}\rangle$ & $|\ad_{34}\rangle$ & $0$ & $(0,0,-2)$ \\
\hline
$|\ad_{34}\rangle$ & $-i |\ad_{34}\rangle$ & $|\ad_{42}\rangle$ & $0$ & $(-2,0,0)$ \\
\hline
$|\ad_{42}\rangle$ & $-i |\ad_{42}\rangle$ & $|\ad_{23}\rangle$ & $0$ & $(0,-2,0)$ \\
\hline
$|\at_{\hat{1}}\rangle$ & $-i\sqrt{-i} |\at_{\hat{1}}\rangle$ & $|\at_{\hat{1}}\rangle$ & $0$ & $(-1,-1,-1)$ \\
\hline
$|\at_{\hat{2}}\rangle$ & $-i\sqrt{-i} |\at_{\hat{2}}\rangle$ & $|\at_{\hat{3}}\rangle$ & $0$ & $(-1,1,1)$ \\
\hline
$|\at_{\hat{3}}\rangle$ & $-i\sqrt{-i} |\at_{\hat{3}}\rangle$ & $|\at_{\hat{4}}\rangle$ & $0$ & $(1,-1,1)$ \\
\hline
$|\at_{\hat{4}}\rangle$ & $-i\sqrt{-i} |\at_{\hat{4}}\rangle$ & $|\at_{\hat{2}}\rangle$ & $0$ & $(1,1,-1)$ \\
\hline
\end{tabular}
\caption{The 16 states, actions of operators $\hat{\Sigma}$ and $\hat{\Omega}$ on the states, energies and $U(1)^3$ charges of the Abelian KT model.}
\label{tab:energy-actions-etc}
\end{table}
We see that out of the 16 states, only 2 states, $|A_+\rangle$ and $|A_-\rangle$, are invariant under $U(1)^3$ symmetry. Thus they span the singlet-sector of the theory.

\subsection{An extension of the Abelian KT model to a $1d$ lattice: Two sites}
\label{ssec:KT2site}

\subsubsection{Hamiltonian}

We can extend the KT model to a $1d$ lattice in a manner similar to what was done in~\cite{Gu:2016oyy} for the SYK model.

Let us consider copies of the KT model on each site of a one-dimensional lattice with $L$ sites and introduce interactions between the nearest neighbors. The Hamiltonian is given by
\beq
\begin{split}
H = \frac{g}{4} \sum_{a=1}^L \psi^{(a)}_{ijk} \psi^{(a)}_{ilm} \psi^{(a)}_{njm} \psi^{(a)}_{nlk} + \frac{\lambda_r}{4} \sum_{a=1}^L\psi^{(a)}_{ijk} \psi^{(a)}_{ilm} \psi^{(a+1)}_{njm} \psi^{(a+1)}_{nlk} \\ + \frac{\lambda_g}{4} \sum_{a=1}^L \psi^{(a)}_{ijk} \psi^{(a+1)}_{ilm} \psi^{(a)}_{njm} \psi^{(a+1)}_{nlk} + \frac{\lambda_b}{4} \sum_{a=1}^L \psi^{(a)}_{ijk} \psi^{(a+1)}_{ilm} \psi^{(a+1)}_{njm} \psi^{(a)}_{nlk},
\end{split}
\eeq
where we impose periodic boundary conditions for fermions: $\psi^{(L+1)}=\psi^{(1)}$. As in the (one-site) KT model, the fermion $\psi^{(a)}$ ($a = 1, 2, \cdots, L$) in the KT chain model transform in the trifundamental representation of $U(1)^3$. In addition to $U(1)^3$ symmetry, it also has  $(\mathbb{Z}_2)^L$ symmetry under $\psi^{(a)} \rightarrow -\psi^{(a)}$ for any $a \in \{ 1, 2, ..., L \}$ where the corresponding $(\mathbb{Z}_2)^L$ charge is denoted by $(\eta_1, \eta_2, ..., \eta_L)$ ($\eta_i = \pm$).

The simplest case is when there are two lattice sites. Let us first consider the $L = 2$ case. The Hamiltonian for the $L = 2$ case is 
\bea
H &=& \frac{g}{4} \psi^{(1)}_{ijk} \psi^{(1)}_{ilm} \psi^{(1)}_{njm} \psi^{(1)}_{nlk} + \frac{g}{4} \psi^{(2)}_{ijk} \psi^{(2)}_{ilm}\psi^{(2)}_{njm} \psi^{(2)}_{nlk} + \frac{\lambda_r}{4} \psi^{(1)}_{ijk} \psi^{(1)}_{ilm}\psi^{(2)}_{njm} \psi^{(2)}_{nlk} \nn \\ 
&& ~~~~~~~~~~~~ + \frac{\lambda_g}{4} \psi^{(1)}_{ijk} \psi^{(2)}_{ilm}\psi^{(1)}_{njm}\psi^{(2)}_{nlk} + \frac{\lambda_b}{4} \psi^{(1)}_{ijk} \psi^{(2)}_{ilm}\psi^{(2)}_{njm}\psi^{(1)}_{nlk},
\eea
where $\psi^{(i)}$ is the fermionic field at the $i$th site and $\lambda_r$, $\lambda_g$ and $\lambda_b$ are the couplings of the three different types of interaction terms shown above. As mentioned, there is $\mathbb{Z}_2^2$ symmetry corresponding to $\psi^{(a)}\rightarrow -\psi^{(a)}$ for each value of $a \in \{1, 2\}$.

As before, we can define annihilation and creation operators $a_i$'s and $a_i^\dag$'s in terms of the components of $\psi^{(1)}$.
 
In a similar way, we can define annihilation and creation operators $b_i$'s and $b_i^\dag$'s by the same linear combinations of the corresponding components of $\psi^{(2)}$. As before, let us define operators $\hat{\Omega}$, $\hat{\Sigma}$ whose actions on the annihilation operators is as follows
\beq
\begin{split}
\hat{\Omega} \ a_i \ \hat{\Omega}^{-1} \equiv a_{(234)\cdot i}\;,\quad \hat{\Omega} \ b_i \ \hat{\Omega}^{-1} \equiv b_{(234)\cdot i} 
\label{def: omega}\ ,
\end{split}
\eeq

\beq
\hat{\Sigma} \ a_i \ \hat{\Sigma}^{-1} = e^{i\pi \over 4} a_i \;, \qquad \hat{\Sigma} \ b_i \ \hat{\Sigma}^{-1} = e^{i\pi \over 4} b_i \hspace{1cm} (i = 1, 2, 3, 4)\ .
\eeq
%
%
%
%
where $(234)\cdot i$ is the cyclic permutation of $i$ by $(234)\in S_3$ ($i = 1, 2, 3, 4$) [e.g. see \eqref{eq: example of omega action} and \eqref{def: spectral reflection}]. Furthermore, we define a lattice translation operator~$\hat{T}$:
\beq
\hat{T} \ a_i  \hat{T}^{-1} = b_i \;, \qquad \hat{T} \ b_i \hat{T}^{-1} = a_i \hspace{1cm} (i = 1, 2, 3, 4) 
\label{def: translation}\ .
\eeq
The Hamiltonian expressed in terms of these creation and annihilation operators has the form
\beq
H = g H_a^{(0)} + g H_b^{(0)} + \lambda_r H_{ab}^{(r)} + \lambda_g H_{ab}^{(g)} + \lambda_b H_{ab}^{(b)}\ ,
\eeq
where
\bea
H_a^{(0)} &\equiv& 2 \Bigl( a_1^\dag a_2^\dag a_3^\dag a_4^\dag + a_4 a_3 a_2 a_1 \Bigr)\;,\quad H_b^{(0)} \equiv \hat{T} H_a^{(0)} \hat{T}^{-1}\ , \\
H_{ab}^{(r)} &\equiv& - \frac{1}{2} (a_1^\dag a_1 - a_2^\dag a_2) (b_3^\dag b_3 - b_4^\dag b_4) - \frac{1}{2}  (b_1^\dag b_1 - b_2^\dag b_2) (a_3^\dag a_3 - a_4^\dag a_4) \cr
&&~~~~~~ + \frac{1}{2} (a_1^\dag b_1 - a_2^\dag b_2) (a_3^\dag b_3 - a_4^\dag b_4) + \frac{1}{2} (b_1^\dag a_1 - b_2^\dag a_2) (b_3^\dag a_3 - b_4^\dag a_4) \cr
&&~~~~~~ + \frac{1}{2} \Bigl( a_1^\dag b_2^\dag + b_1^\dag a_2^\dag \Bigr) \Bigl( a_3^\dag b_4^\dag + b_3^\dag a_4^\dag \Bigr) + \frac{1}{2} \Bigl( b_4 a_3 + a_4 b_3 \Bigr) \Bigl( b_2 a_1 + a_2 b_1 \Bigr)\ , \\
H_{ab}^{(g)} &\equiv& \hat{\Omega} H_{ab}^{(r)}\hat{\Omega}^{-1}\;,\quad H_{ab}^{(b)} \equiv \hat{\Omega}^2 H_{ab}^{(r)}\hat{\Omega}^{-2}\ . 
\eea

\subsubsection{Singlet sector}

The singlet sector of the theory is the subspace that is invariant under $U(1)^3$ transformations. For the generic $L$-site chain we can consider a basis of the total Hilbert space where each element has $k_1,\ k_2,\ k_3\ $ and $k_4$ creation operators with charges $(+, +, +), (+, -, -), (-, +, -)$ and $(-, -, +)$, respectively acting on the vacuum. Such a basis element has charges \[(k_1+k_2-k_3-k_4,\ k_1-k_2+k_3-k_4,\ k_1-k_2-k_3+k_4)\] of $U(1)^3$ and, hence, can belong to the singlet sector if and only if 
\begin{equation}
k_1= k_2= k_3=k_4\equiv k\ , \hspace{1cm} (k \in\{0,1,2,\cdots ,L\})\ .
\end{equation}
For any basis vector that does belong to the singlet sector the $k$ creation operators of any particular charge belong to $k$ of the $L$ sites. Thus for any particular charge we can choose $k$ out of the $L$ sites to place the corresponding creation operators to construct a basis vector that belongs to the singlet sector. Therefore, the total number of such basis vectors is
\begin{equation}
(\text{Dimension of the singlet sector}) = \sum_{k=0}^L{{L}\choose {k}}^4\ .
\end{equation} 
For example, there are $\sum_{k=0}^2{{2}\choose {k}}^4=18$ states in the singlet sector of the two-site KT chain model.

A convenient basis for the singlet sector of the theory is given in Table~\ref{tab:2-site-KT-singlet-sector}.
\begin{table}[h!]
\centering
\begin{tabular}{|c|c|}
\hline
$\text{Basis\hspace{2mm} vector}$ & $(\mathbb{Z}_2)^2 \hspace{2mm} \text{charges}$ \\   
\hline
\hline
$|A_{\sigma_1} B_{\sigma_2}\rangle$ & $(+,+)$ \\ 
$|\ad_{ij}\bd_{\widehat{ij}}\rangle$ & $(+,+)$ \\
$|\as_i\bt_{\hat{i}}\rangle$ & $(-,-)$ \\
$|\at_{\hat{i}}\bs_i\rangle$ & $(-,-)$ \\
\hline
\end{tabular}
\caption{The singlet-sector basis vectors and corresponding $(\mathbb{Z}_2)^2$ charges of the 2-site Abelian KT chain model. Here $\sigma_1,\sigma_2=\pm $.} 
\label{tab:2-site-KT-singlet-sector}
\end{table}
The states are constructed from the vacuum by the action of the appropriate operators in the following way.
\beq
\begin{split}
|A_{\sigma_1} B_{\sigma_2} \rangle &\equiv A_{\sigma_1}^{\dag} B_{\sigma_2}^{\dag} |\ \rangle\ , \\
|\ad_{ij} \bd_{\widehat{ij}}\rangle &\equiv (a^{\dag^2})_{ij}(b^{\dag^2})_{\widehat{ij}} |\ \rangle\ , \\
|\as_i \bt_{\hat{i}}\rangle &\equiv  a^{\dag}_i (b^{\dag^3})_{\hat{i}}|\ \rangle\ , \\
|\at_{\hat{i}} \bs_i \rangle &\equiv  a^{\dag^3}_{\hat{i}} b^{\dag}_i |\ \rangle\ . \\
\end{split}
\eeq
where $\sigma_1,\sigma_2= \pm $ and $i,j=1,2,3,4$. These states are related to each other by the operator $\hat{\Sigma}$ and $\hat{T}$:
\beq
\begin{split}
|A_- B_- \rangle &= \hat{\Sigma}|A_+ B_+ \rangle, \\
|A_- B_+ \rangle &= \hat{\Sigma}|A_+ B_- \rangle, \\
|\at_{\hat{i}} \bs_i \rangle &= -\hat{T}|\as_i \bt_{\hat{i}}\rangle.
\end{split}
\eeq
It is also convenient to define the following states in order to diagonalize the Hamiltonian.
\beq
\begin{split}
|ab^3\rangle_{\sigma_1\sigma_2\sigma_3} &\equiv \frac{1}{2} \Bigl(|\as_1 \bt_{\hat{1}}\rangle + \sigma_1 \ |\as_2 \bt_{\hat{2}}\rangle + \sigma_2 \ |\as_3 \bt_{\hat{3}}\rangle + \sigma_3 \ |\as_4 \bt_{\hat{4}}\rangle \Bigr), \\
|a^3b\rangle_{\sigma_1\sigma_2\sigma_3} &\equiv \hat{T}|ab^3\rangle_{\sigma_1\sigma_2\sigma_3}, \\
|a^2b^2\rangle_{\sigma} &\equiv \frac{1}{\sqrt{2}} \Bigl( |\ad_{12} \bd_{\widehat{12}}\rangle + \sigma \hat{T} |\ad_{12} \bd_{\widehat{12}}\rangle \Bigr), \\
\hat{\Omega}|a^2b^2\rangle_{\sigma} &\equiv \frac{1}{\sqrt{2}} \Bigl( |\ad_{13} \bd_{\widehat{13}}\rangle + \sigma\hat{T} |\ad_{13} \bd_{\widehat{13}}\rangle \Bigr), \\
\hat{\Omega}^2 |a^2b^2\rangle_{\sigma} &\equiv \frac{1}{\sqrt{2}} \Bigl( |\ad_{14} \bd_{\widehat{14}}\rangle + \sigma\hat{T} |\ad_{14} \bd_{\widehat{14}}\rangle \Bigr), \\
\end{split}
\eeq
where $\sigma$ is again $+1$ or $-1$, and $(\sigma_1, \sigma_2, \sigma_3)$ can take values only from the following set
\begin{gather*}
\{ (+, +, +), (+, -, -), (-, +, -), (-, -, +) \}.
\end{gather*}

\subsubsection{Spectrum}

There are 4 middle states (states with zero energy) in the two-site KT chain model given by 
\beq
\begin{split}
H|A_+ B_-\rangle &= H |A_- B_+\rangle = H |ab^3\rangle_{+++}= H |a^3b\rangle_{+++} = 0. \\
\end{split}
\eeq
We take $\Delta_{ij} \equiv \lambda_i - \lambda_j$ $(i, j = r, g, b)$. There are 12 states which become middle states only at the symmetric point of the couplings $\lambda_r = \lambda_g = \lambda_b = \lambda$. Nine of them are found to be
\beq
\begin{split}
H |a^2 b^2\rangle_{-} &= \Delta_{bg} |a^2b^2\rangle_{-}\ , \\
H \hat{\Omega}^2 |a^2 b^2\rangle_{-} &= \Delta_{gr} \hat{\Omega}^2 |a^2 b^2\rangle_{-}\ , \\
H \hat{\Omega}|a^2 b^2\rangle_{-} &= \Delta_{rb} \hat{\Omega}|a^2 b^2\rangle_{-}\ , \\
H\frac{1}{\sqrt{2}} \Bigl( |ab^3\rangle_{+--} \pm |a^3b\rangle_{+--} \Bigr) &= \pm \Delta_{bg} \frac{1}{\sqrt{2}}\Bigl( |ab^3\rangle_{+--} \pm |a^3b\rangle_{+--} \Bigr)\ , \\
H\frac{1}{\sqrt{2}}\Bigl( |ab^3\rangle_{--+} \pm |a^3b\rangle_{--+} \Bigr) &= \pm \Delta_{gr} \frac{1}{\sqrt{2}}\Bigl(|ab^3\rangle_{--+} \pm |a^3b\rangle_{--+}\Bigr)\ , \\
H\frac{1}{\sqrt{2}}\Bigl( |ab^3\rangle_{-+-} \pm |a^3b\rangle_{-+-} \Bigr) &= \pm\Delta_{rb} \frac{1}{\sqrt{2}}\Bigl(|ab^3\rangle_{-+-} \pm |a^3b\rangle_{-+-}\Bigr)\ . \\
\end{split}
\eeq
The Hamiltonian for the rest of the states can be written as
\beq
\begin{split}
H \left( \begin{array}{c} |A_+ B_+\rangle \\ |A_- B_-\rangle \\ |a^2b^2\rangle_{+} \\ \hat{\Omega}^2 |a^2b^2\rangle_{+} \\ \hat{\Omega}|a^2b^2\rangle_{+} \end{array} \right) =
\left( \begin{array}{ccccc} 
4g & 0 & \lambda_b + \lambda_g & \lambda_g + \lambda_r & \lambda_r + \lambda_b \\
0 & -4g & \Delta_{bg} & \Delta_{gr} & \Delta_{rb} \\
 \lambda_b + \lambda_g & \Delta_{bg} & \Delta_{bg} & 0 & 0 \\
 \lambda_g + \lambda_r &\Delta_{gr} & 0 & \Delta_{gr} & 0 \\
 \lambda_r + \lambda_b & \Delta_{rb} & 0 & 0 & \Delta_{rb} \\
 \end{array} \right)
\left( \begin{array}{c} |A_+ B_+ \rangle \\ |A_- B_- \rangle \\ |a^2b^2\rangle_{+} \\ \hat{\Omega}^2 |a^2b^2\rangle_{+} \\ \hat{\Omega}|a^2b^2\rangle_{+} \end{array} \right).
\end{split}
\eeq

\subsubsection{Large $g$ limit}

In the large $g$ limit (i.e., $g \gg \lambda_r, \lambda_g, \lambda_b$), we can treat the hopping interaction $\lambda_r H_{ab}^{(r)}$, $\lambda_g H_{ab}^{(g)}$, $\lambda_b H_{ab}^{(b)}$ as perturbations over the on-site interaction $H^{(0)} = g H_a^{(0)}+g H_b^{(0)}$.

In this limit, the states with absolute energies of order $\mathcal{O}(g)$ are located near the tail of the spectrum (spectral tail states) whereas the states with absolute energies of order $\mathcal{O}(\lambda)$ and less populate the central region of the spectrum (midspectral states).
 
When $g$ is very large, we get spectral tail states proportional to
\beq
|A_\pm B_\pm\rangle, 
\eeq
with energies $\pm 4g$ , respectively.  In this limit, the energies of the other three midspectral states [up to order $\mathcal{O}(\lambda)$] are given by 
\beq
|a^2b^2\rangle_{+}\;:\; \Delta_{bg}\;,\qquad \hat{\Omega}^2 |a^2b^2\rangle_{+}\;:\; \Delta_{gr}\;,\qquad \hat{\Omega}|a^2b^2\rangle_{+}\;:\; \Delta_{rb}. 
\eeq 
We see that up to first order in perturbation, the energies of the two spectral tail states are unaffected by the perturbation.

\subsubsection{Symmetric coupling for all the hopping terms}

When $\lambda_r = \lambda_g = \lambda_b = \lambda$, we have 13 middle states as mentioned earlier, and three more middle states:
\bea
 && 2g |a^2b^2\rangle_{+} - \frac{\lambda}{\sqrt{2}} (|A_+ B_+\rangle - \ |A_- B_-\rangle)\ , \nn \\
 && 2g \hat{\Omega}^2 |a^2b^2\rangle_{+} - \frac{\lambda}{\sqrt{2}} (|A_+ B_+\rangle - \ |A_- B_-\rangle)\ , \\
 && 2g \hat{\Omega}|a^2b^2\rangle_{+} - \frac{\lambda}{\sqrt{2}} (|A_+ B_+\rangle - \ |A_- B_-\rangle)\ .\nn  
\eea
Furthermore, there are two other states given by
\bea
&& \sqrt{2} g (|A_+ B_+\rangle - \ |A_- B_-\rangle) + \lambda \Bigl(|a^2b^2\rangle_{+} + \hat{\Omega}^2 |a^2b^2\rangle_{+} + \hat{\Omega}|a^2b^2\rangle_{+} \Bigr) \nn \\
&& ~~~~~~~~ \pm \sqrt{2g^2+\frac{3}{2}\lambda^2}\  (|A_+ B_+\rangle - \ |A_- B_-\rangle),
\eea
with energies $\pm 2\sqrt{4g^2+3\lambda^2}$, respectively. We summarize the spectrum of the symmetric hopping coupling case in Table~\ref{tab:eq-coupling}.
\begin{table}[h!]
\centering
\begin{tabular}{|c|c|}
\hline
Eigenvalue & Degeneracy \\   
\hline
\hline
$0$ & $16$ \\
\hline
$\pm 2\sqrt{4g^2+3\lambda^2}$ & $1$ \\
\hline
\end{tabular}
\caption{Spectrum for $\lambda_r = \lambda_g = \lambda_b = \lambda$ case for the two-site Abelian KT chain model.} 
\label{tab:eq-coupling}
\end{table}

\subsubsection{Comments}

\begin{itemize}
\item In the two-site case, we see that there are far more middle states when the three hopping couplings $\lambda_r$, $\lambda_g$ and $\lambda_b$ are symmetric. We see a similar behavior for the three-site and four-site cases as well. We expect that this would be true generally for the $L$-sites KT chain model.

\item We see that four states, i.e., $|A_+ B_-\rangle, |A_- B_+ \rangle,| ab^3\rangle_{+++}$ and $|a^3b\rangle_{+++}$, are middle states irrespective of the values of the couplings; i.e., their energies are protected under change of the couplings. In the four-site case, we see similar protected middle states. Although in the two-site case we do not find any other protected state with non-zero energy, we observe some such protected states in both the three-site and four-site cases with nonzero energies.

\item As expected, the Hamiltonian does not mix states with different $(\mathbb{Z}_2)^2$ charges. We will use this fact while studying the three-site and the four-site cases and look at the eigenstates and eigenvalues in each subsector with particular $(\mathbb{Z}_2)^L$ charges.

\end{itemize}

\section{Abelian KT Chain Model: Three Sites}
\label{sec:KT3site}

\subsection{Hamiltonian}

The Hamiltonian of the three-site KT chain model is defined by
\beq
\begin{split}
H = g\ H_a^{(0)} + g\ H_b^{(0)} + g\ H_c^{(0)} + \sum_{\col\in\{\R,\G,\B\}} (\lambda_\col H_{ab}^{(\col)} + \lambda_\col H_{bc}^{(\col)} +\lambda_\col H_{ca}^{(\col)}),
\end{split}
\eeq
where on-site interaction is given by
\beq
H_a^{(0)} \equiv 2 \Bigl(a_1^\dag a_2^\dag a_3^\dag a_4^\dag+a_4 a_3 a_2 a_1\Bigr)\;,\qquad H_b^{(0)}\equiv \hat{T} H_a^{(0)} \hat{T}^{-1}\;,\qquad H_c^{(0)}\equiv \hat{T} H_b^{(0)} \hat{T}^{-1},
\eeq
%
%
and the hopping interaction is
\begin{align}
H_{ab}^{(r)}\equiv &- \frac{1}{2} (a_1^\dag a_1-  a_2^\dag a_2)(b_3^\dag b_3-  b_4^\dag b_4)+ \frac{1}{2} (a_1^\dag b_1-  a_2^\dag b_2)(a_3^\dag b_3-  a_4^\dag b_4)\cr
& + \frac{1}{2} \Bigl(a_1^\dag b_2^\dag +b_1^\dag a_2^\dag \Bigr) \Bigl(a_3^\dag b_4^\dag +b_3^\dag a_4^\dag \Bigr)+\left(\;a\;\leftrightarrow \; b\;\right),
\end{align}
\beq
H_{ab}^{(g)} \equiv \hat{\Omega} H_{ab}^{(r)} \hat{\Omega}^{-1}\;,\quad H_{ab}^{(b)} \equiv \hat{\Omega} H_{ab}^{(g)}\hat{\Omega}^{-1},
\eeq
and similar for $H_{bc}^{(r)}$ etc.

\subsection{Singlet sector}

In the three-site KT chain model, the singlet sector has 164 states, and we define a basis for each subsector with definite $(\mathbb{Z}_2)^3$ charges below.

\subsubsection{The $(+,+,+)$ subsector}

The basis in this subsector are given in Table~\ref{tab:ppp-sector}.
\begin{table}[h!]
\centering
{\renewcommand{\arraystretch}{1.5}
	\begin{tabular}{|>{\centering\arraybackslash}m{3.5cm}|>{\centering\arraybackslash}m{2.7cm}|}
  \hline
  Form of basis &  Degeneracy \\       
  \hline
  \hline
  $|A_{\sigma_1}B_{\sigma_2}C_{\sigma_3}\rangle$ & $8$ \\
  $|A_\sigma \bd_{ij} \cd_{\widehat{ij}}\rangle $ & $12$ \\
  $|\ad_{ij}B_\sigma \cd_{\widehat{ij}}\rangle $ & $12$ \\
  $|\ad_{ij}\bd_{\widehat{ij}}C_\sigma \rangle $ & $12$ \\
  \hline
\end{tabular}
\caption{The basis vectors in the $(+, +, +)$ subsector. $\sigma, \sigma_1, \sigma_2, \sigma_3=\pm$. $(i, j)$ is a pair of distinct elements chosen from the set $\{ 1, 2, 3, 4 \}$.}
\label{tab:ppp-sector}}
\end{table}
where we define
\begin{align}
|A_{\sigma_1} B_{\sigma_2} C_{\sigma_3} \rangle &\equiv A^\dag_{\sigma_1}B^\dag_{\sigma_2}C^\dag_{\sigma_3}|\ \rangle, \\
|A_\sigma \bd_{ij}\cd_{\widehat{ij}}\rangle &\equiv A^\dag_{\sigma}(b^{\dag^2})_{ij} (c^{\dag^2})_{\widehat{ij}} |\ \rangle, \\
|\ad_{ij} B_\sigma \cd_{\widehat{ij}}\rangle &\equiv  (a^{\dag^2})_{ij}B^\dag_{\sigma} (c^{\dag^2})_{\widehat{ij}}|\ \rangle , \\
|\ad_{ij}\bd_{\widehat{ij}} C_\sigma \rangle &\equiv (a^{\dag^2})_{ij} (b^{\dag^2})_{\widehat{ij}}C^\dag_{\sigma}|\ \rangle.
\end{align}
%
%
%
%
Since $\sigma, \sigma_1, \sigma_2, \sigma_3$ can take the values $+1$ or $-1$, and $(i, j)$ is a pair of distinct elements chosen from the set $\{ 1, 2, 3, 4 \}$, the total number of states in $(+,+,+)$ subsector is 44.

Due to the lattice translational symmetry, it is useful to introduce a projection operator onto $\mathbb{Z}_3$ charge eigenspace:
\begin{equation}
\projt_p\equiv {1\over 3} \sum_{n=0}^2 e^{i{2\pi p\over 3} n}\trans^n\ ,\hspace{1cm} (p=0,1,2)\ .
\end{equation}
In addition, we find it convenient to define the following states:
\beq
\begin{split}
|(a^2b^2 + \sigma_1 b^2a^2)C_{\sigma_2}\rangle_{ij\widehat{ij}} &\equiv |(\ad_{ij}\bd_{\widehat{ij}} + \sigma_1 \bd_{ij}\ad_{\widehat{ij}})C_{\sigma_2} \rangle\ ,
\end{split}
\eeq
where $\sigma_1$ and $\sigma_2$ can take the values in $\{+,-\}$. We call the states of the form $|(a^2b^2 - b^2a^2)C_{\sigma}\rangle$ ``biquadratic difference states'', and the states of the form $|(a^2b^2 + b^2a^2)C_{\sigma}\rangle$ ``biquadratic sum states.''

\paragraph{Biquadratic difference states:} 

There are $18$ biquadratic difference eigenstates of the form 
\beq
\Psi^p_{\text{BiQdiff},\sigma}\equiv \sqrt{3} \projt_p \begin{pmatrix}
 |(a^2b^2-b^2a^2)C_\sigma\rangle_{12\widehat{12}}  \\ 
 |(a^2b^2-b^2a^2)C_\sigma\rangle_{13\widehat{13}}  \\ 
|(a^2b^2-b^2a^2)C_\sigma\rangle_{14\widehat{14}}   \\ 
\end{pmatrix} \ ,
\eeq
%
%
where $\sigma=\pm 1$ and $p=0,1,2$. Therefore, we have $18$ states in this subsector given by $\Psi^p_{\text{BiQdiff},\sigma}$, and the Hamiltonian is found to be
\begin{align}
H^p_{\text{BiQdiff},\sigma}\equiv&
2\sigma g
\begin{pmatrix}
1 & 0 & 0  \\
0 & 1 & 0  \\
0 & 0 & 1 \\
\end{pmatrix} - \sigma \left({ \omega^2+\omega+1\over 2} - {1\over 2} \right)
\begin{pmatrix}
\lambda_g + \lambda_b & 0 & 0 \\
0 & \lambda_b + \lambda_r & 0 \\
0 & 0 & \lambda_r + \lambda_g \\
\end{pmatrix} 
\cr
& + \left({\omega^2+\omega+1\over 2} - {3\over 2}\right) 
\begin{pmatrix}
\Delta_{gb} & 0 & 0 \\
0 & \Delta_{br} & 0 \\
0 & 0 & \Delta_{rg} \\
\end{pmatrix},
\end{align}
%
%
where $\omega = e^{i\frac{2\pi p}{3}}$ and $\Delta_{ij} \equiv \lambda_i-\lambda_j$. 
%

\paragraph{Biquadratic sum states and their partners:} 

The remaining 26 states can be decomposed into blocks of states as described below such that the action of the Hamiltonian is closed within each block. Each of these blocks contain Bloch states obtained out of biquadratic sum states and their partners which are of the form $ |A_\sigma B_\sigma C_\sigma\rangle$ and $ |A_\sigma B_{-\sigma} C_{-\sigma}\rangle $ where $\sigma =\pm $. 

There are two blocks of five states with zero Bloch momentum:
\beq
\Psi^0_{\text{BiQsum},\sigma}\equiv \sqrt{3} \projt_0 
\begin{pmatrix}
 {1\over \sqrt{3}} \ |A_\sigma B_\sigma C_\sigma\rangle \\ 
 |A_\sigma B_{-\sigma} C_{-\sigma}\rangle \\ 
 |(a^2b^2+b^2a^2)C_\sigma\rangle_{12\widehat{12}} \\
 |(a^2b^2+b^2a^2)C_\sigma\rangle_{13\widehat{13}} \\ 
 |(a^2b^2+b^2a^2)C_\sigma\rangle_{14\widehat{14}} \\ 
 \end{pmatrix}\ ,
\eeq
%
%
where $\sigma=\pm $. The Hamiltonian in these blocks is
\beq
H^0_{\text{BiQsum},\sigma} = 
\begin{pmatrix} 
6\sigma g & 0 & -\sqrt{6}f_{+,r} & -\sqrt{6}f_{+,g} & -\sqrt{6}f_{+,b} \\
0 & -2\sigma g & -{\sqrt{2}}f_{-,r} & -{\sqrt{2}}f_{-,g} & -{\sqrt{2}}f_{-,g} \\
-\sqrt{6}f_{+,r}& -{\sqrt{2}}f_{-,r} & 2\sigma g+f_{r} & 0 & 0 \\
-\sqrt{6}f_{+,g} & -{\sqrt{2}}f_{-,g} & 0 & 2\sigma g+f_{g} & 0 \\
-\sqrt{6}f_{+,b} & -{\sqrt{2}}f_{-,b} & 0 & 0 &2\sigma g+f_{b} 
\end{pmatrix}\ ,
\eeq
%
%
%
where we define
\begin{alignat}{2}
f_{\pm,r} \equiv \frac{1}{2}\Bigl[ -(\lambda_g+\lambda_b) \pm \sigma(\lambda_g-\lambda_b) \Bigr]\;,  &&\hspace{0.5cm} f_{r} \equiv -2(\lambda_g-\lambda_b)+\sigma(\lambda_g+\lambda_b)\ , \cr
f_{\pm,g} \equiv \frac{1}{2}\Bigl[ -( \lambda_b+\lambda_r) \pm \sigma(\lambda_b-\lambda_r)  \Bigr]\;,  &&\hspace{0.5cm} f_{g} \equiv -2(\lambda_b-\lambda_r)+\sigma(\lambda_b+\lambda_r)\ , \cr
f_{\pm,b} \equiv \frac{1}{2}\Bigl[ -( \lambda_r+\lambda_g) \pm \sigma(\lambda_r-\lambda_g)  \Bigr]\;,  &&\hspace{0.5cm} f_{b} \equiv -2(\lambda_r-\lambda_g)+\sigma(\lambda_r+\lambda_g) \ .
\end{alignat}

In addition, we found 4 blocks of $4$ states with Bloch momentum $p=1,2$:
\beq
\Psi^p_{\text{BiQsum},\sigma}\equiv \sqrt{3} \projt_p 
\begin{pmatrix} 
 |A_\sigma B_{-\sigma} C_{-\sigma}\rangle \\ 
 |(a^2b^2+b^2a^2)C_\sigma\rangle_{12\widehat{12}} \\
 |(a^2b^2+b^2a^2)C_\sigma\rangle_{13\widehat{13}} \\ 
 |(a^2b^2+b^2a^2)C_\sigma\rangle_{14\widehat{14}} \\
\end{pmatrix}\ ,
\eeq
%
%
%
where $\sigma = \pm$ and $p = 1, 2$. The Hamiltonian in such blocks is given by
\beq
H^p_{\text{BiQsum},\sigma} \equiv -\sigma 
\begin{pmatrix}
2 g &  h_{r}\sqrt{2}\omega^{-1} \ &  h_{g}\sqrt{2}\omega^{-1} \ &  h_{b}\sqrt{2}\omega^{-1} \  \\
 h_{r}\sqrt{2}\omega \ & -2g - \sigma h_{r} & 0 & 0 \\
 h_{g}\sqrt{2}\omega \ & 0 &  -2g - \sigma h_{g} & 0 \\
 h_{b}\sqrt{2}\omega \ & 0 & 0 &  -2g - \sigma h_{b} \\
\end{pmatrix},
\eeq
%
%
where $\omega = e^{i\frac{2\pi p}{3}}$ and we define
\begin{align}
h_{r} &\equiv \frac{1}{2}\Bigl[ -(\lambda_g-\lambda_b)-\sigma(\lambda_b+\lambda_g) \Bigr], \cr 
h_{g} &\equiv \frac{1}{2}\Bigl[ -(\lambda_b-\lambda_r)-\sigma(\lambda_b+\lambda_r)  \Bigr],  \cr
h_{b} &\equiv \frac{1}{2}\Bigl[ -(\lambda_r-\lambda_g)-\sigma(\lambda_r+\lambda_g)  \Bigr].  
\end{align}
In total this accounts for $26$ biquadratic sum states and their partners. 

\paragraph{Large $g$ spectrum:} 

At large $g$, we have only spectral tail states that is, states with energies of $\mathcal{O}(g)$ in this sector.

First of all, there are two states with energies $\pm 6g$:
\beq
|A_\pm B_\pm C_\pm\rangle, 
\eeq
In addition, we have six states, 
\begin{align}
|A_\pm B_\pm C_\mp \rangle \equiv & A^\dag_\pm B^\dag_\pm C^\dag_\mp |\ \rangle\ , \\
|A_\pm B_\mp C_\pm\rangle \equiv & A^\dag_\pm B^\dag_\mp C^\dag_\pm |\ \rangle\ , \\
|A_\mp B_\pm C_\pm\rangle \equiv & A^\dag_\mp B^\dag_\pm C^\dag_\pm |\ \rangle\ , 
\end{align}
where the upper signs give  an energy $+2g$ at large $g$ whereas lower signs give states with an energy $-2g$ at large $g$, respectively. Finally, let us consider $36$ states of the form 
\beq
\begin{split}
|b^2_{ij}c^2_{\widehat{ij}} A_\pm \rangle &\equiv b^{\dag^2}_{ij}c^{\dag^2}_{\widehat{ij}} A^\dag_\pm |\ \rangle \ , \\
|c^2_{ij}a^2_{\widehat{ij}} B_\pm\rangle  &\equiv  \hat{T}|b^2_{ij}c^2_{\widehat{ij}}  A_\pm \rangle \ , \\
|a^2_{ij}b^2_{\widehat{ij}} C_\pm \rangle &\equiv \hat{T}^2|b^2_{ij}c^2_{\widehat{ij}}  A_\pm \rangle \ . 
\end{split}
\eeq
Among them $18$ states (with plus signs) have energy $+2g$ and the other $18$ states (with minus signs) have energy $-2g$ at large $g$. In total, we have $21$ states with energy $+2g$ and $21$ states with energy $-2g$ in this sector.

\paragraph{Symmetric couplings:} 

We summarize the spectrum of symmetric hopping couplings (i.e., $\lambda_r = \lambda_g = \lambda_b = \lambda$) in Table~\ref{tab:symmetric-3-site-KT-model},
\begin{table}[h!]
\centering
{\renewcommand{\arraystretch}{1.3}
\begin{tabular}{| >{\centering\arraybackslash}m{6cm} | >{\centering\arraybackslash}m{2cm} |}
\hline
Eigenvalue & {\small Degeneracy} \\
\hline
\hline
$2\sigma(g-\lambda)$ & $3$ \\
$\sigma(2 g+\lambda)$ & $6$  \\
$2\sigma (g+\lambda) $ & $2$  \\
$\sigma (2 g-\lambda) $ & $4$  \\
$\frac{1}{2}\sigma (\pm\lambda + \sqrt{16 g^2 - 8 g \lambda + 25 \lambda^2})$ &$2$\\
$\sigma R_1[P]$ & $1$\\
$\sigma R_2[P]$ & $1$\\
$\sigma R_3[P]$ & $1$\\
\hline
\end{tabular}}
\caption{The spectrum in the subsector $(+, +, +)$ for $\lambda_r = \lambda_g = \lambda_b = \lambda$ of the 3-site Abelian KT chain model. Here $\sigma \in \{ +, - \}$} 
\label{tab:symmetric-3-site-KT-model}
\end{table}
where $R_1[P], R_2[P], R_3[P]$ are the three roots of the polynomial
\beq 
P(\alpha) \equiv 24 g^3 + 24 g^2 \lambda + (-4 g^2 + 8 g \lambda - 24 \lambda^2) \alpha + (-6 g - 2 \lambda) \alpha^2 + \alpha^3 = 0\ . 
\label{def:polynomial 3sites}
\eeq
The general solution of the equation $P(\alpha) = 0$ is found to be
\begin{equation}
R_n[P]\equiv\alpha_n \equiv \frac{2}{3}(3 g+\lambda) \nn  + \frac{\omega_n(48 g^2+76 \lambda^2)}{3(U + 
  V )^{\frac{1}{3}}} \nn   +\frac{(U + V )^{\frac{1}{3}}}{3\omega_n}, \hspace{1cm} (n=0,1,2) \ ,
\end{equation}
where $\omega=e^{{2\pi i\over 3 }n }$ $(n=0,1,2)$ and
\beq
U \equiv 8\lambda(-36 g^2  + 81 g \lambda + 28 \lambda^2)\; ,\;\; V\equiv \sqrt{(-48 g^2 - 76 \lambda^2)^3 +     U^2}\ .
\eeq
%

\subsubsection{The $(+, -, -)$ subsector}

The basis of the $(+,-,-)$ subsector is given in table \ref{tab:pmm-sector-basis}.
\begin{table}[h!]
\centering
{\renewcommand{\arraystretch}{1.5}
\begin{tabular}{ | >{\centering\arraybackslash}m{3.5cm} | >{\centering\arraybackslash}m{3cm} |}
\hline
Form of basis & {\small Number of states} \\       
\hline
\hline
$|A_\sigma \bt_{\hat{i}}\cs_i\rangle$ & $8$ \\
$|A_\sigma \bs_i\ct_{\hat{i}}\rangle $ & $8$ \\
$|\ad_{ij}\bs_k \cs_l\rangle $ & $12$ \\
$|\ad_{ij}\bt_{\hat{i}}\ct_{\hat{j}}\rangle $ & $12$ \\
\hline
\end{tabular}}
\caption{The basis vectors in the $(+, -, -)$ subsector of three-site Abelian KT chain model. Here $\sigma = \pm$} 
\label{tab:pmm-sector-basis}
\end{table}
Here, the states are defined by
\begin{equation}
\begin{split}
|A_\sigma \bt_{\hat{i}}\cs_i\rangle &\equiv  A^\dag
_\sigma (b^{\dag^3})_{\hat{i}} c^\dag_i|\ \rangle\ , \\
| A_\sigma \bs_{i}\ct_{\hat{i}}\rangle &\equiv  A^\dag
_\sigma  b^\dag_i (c^{\dag^3})_{\hat{i}}|\ \rangle\ , \\
|\ad_{ij}\bs_k \cs_l\rangle &\equiv (a^{\dag^2})_{ij} b^\dag_k c^\dag_l|\Omega \rangle\ , \\
|\ad_{ij}\bt_{\hat{i}} \ct_{\hat{j}}\rangle &\equiv (a^{\dag^2})_{ij} (b^{\dag^3})_{\hat{i}}(c^{\dag^3})_{\hat{i}}\rangle|\ \rangle\ . \\
\end{split}
\end{equation}
where $\sigma=\pm$ and $i,j,k$ and $l$ are distinct elements chosen from the set $\{1,2,3,4\}$.

There are 40 states in the $(+,-,-)$ subsector which can be divided into eight blocks of five states:
\begin{align}
&\Psi^{(a)}_{\text{\tiny Cubic},\sigma_0\sigma_r\sigma_g\sigma_b} = \frac{1}{2}
\begin{pmatrix}
 |A_{\sigma_0}b^3 c\rangle  \\
 | A_{\sigma_0} c^3b \rangle  \\
  |a^2bc-\sigma_0 (b^3a)(c^2ac)\rangle \\
 |abac-\sigma_0 (b^3a)(cac^2)\rangle \\
 |ba^2c-\sigma_0 (b^3a)(ac^3)\rangle \\
\end{pmatrix}+ \frac{1}{2}\sigma_r\begin{pmatrix}
 |A_{\sigma_0} b^2 cb\rangle\\
 | A_{\sigma_0} c^2bc\rangle \\
|a^2cb-\sigma_0 (b^2ab)(c^3a) \rangle\\
 |baca-\sigma_0 (b^2ab)(ac^3) \rangle\\
 |abca-\sigma_0 (b^2ab)(cac^2) \rangle\\
\end{pmatrix}\cr
&\hspace{2cm}+\frac{1}{2} \sigma_g\begin{pmatrix}
 |A_{\sigma_0} bcb^2\rangle  \\
 | A_{\sigma_0} cbc^2 \rangle  \\
 |bca^2-\sigma_0 (bab^2) (ac^3)\rangle  \\
|acab-\sigma_0 (bab^2) (c^3a)\rangle  \\
 |acba-\sigma_0 (bab^2)(c^2ac) \rangle  \\
\end{pmatrix}+\frac{1}{2}\sigma_b\begin{pmatrix}
  |A_{\sigma_0} cb^3 \rangle \\
 |A_{\sigma_0} b c^3\rangle  \\
|cba^2-\sigma_0 (ab^3)(cac^2) \rangle  \\
  |caba-\sigma_0 (ab^3)(c^2ac) \rangle  \\
  |ca^2b-\sigma_0 (ab^3)(c^3a) \rangle  \\
\end{pmatrix}\ ,
\end{align}
where $\sigma_0, \sigma_r, \sigma_g, \sigma_b =\pm $ and $\sigma_r\sigma_g\sigma_b = 1$. Note that we define
\begin{equation}
\begin{split}
|abac\rangle &\equiv a_1^\dag b_2^\dag a_3^\dag c_4^\dag |\; \rangle\ ,\\
|(b^3a)(cac^2)\rangle &\equiv b_1^\dag b_2^\dag b_3^\dag a_4^\dag  c_1^\dag a_2^\dag c_3^\dag c_4^\dag |\;\rangle\ ,
\end{split}
\end{equation}
and so on. The Hamiltonian in such blocks is given by
\begin{equation}
\begin{split}
H^{\text{Cubic}}_{\sigma_0\sigma_1\sigma_2\sigma_3} &\equiv
 \left( \begin{array}{ccccc} 
 2\sigma_0 g & h_{0}&   f_{r}   &   f_{g}  &  f_{b}   \\
 h_{0} & 2\sigma_0 g &    \sigma_r f_{r}   &  \sigma_g f_{g}  &   \sigma_b f_{b} \\
 f_{r}  & \sigma_r f_{r} &  h_{r}& 0 & 0 \\
 f_{g} & \sigma_g f_{g} &  0 &  h_{g} & 0\\
 f_{b}  & \sigma_b f_{b} & 0 & 0 &  h_{b}  \\
\end{array} \right) \\
\end{split}\ ,
\end{equation}
where
\begin{align}
h_{0} &\equiv \frac{1}{2}\Bigl[ \lambda_r(\sigma_g-\sigma_b)+\lambda_g(\sigma_b-\sigma_r)+\lambda_b(\sigma_r-\sigma_g ) \Bigr] \ , \cr
f_{r} &\equiv \frac{1}{2}\Bigl[ -(\lambda_g-\lambda_b)+\sigma_0(\lambda_b+\lambda_g) \Bigr]\ ,  \cr
f_{g} &\equiv \frac{1}{2}\Bigl[ -(\lambda_b-\lambda_r)+\sigma_0(\lambda_r+\lambda_b)  \Bigr]\ ,  \cr
f_{b} &\equiv \frac{1}{2}\Bigl[ -(\lambda_r-\lambda_g)+\sigma_0(\lambda_g+\lambda_r)  \Bigr]\ ,  \cr
h_{r} &\equiv  \frac{1}{2}\Bigl[ -(\lambda_g-\lambda_b)-\sigma_0(-\sigma_r+\sigma_g+\sigma_b)\lambda_r -\sigma_0(\sigma_r \lambda_r+\sigma_g \lambda_g+\sigma_b \lambda_b)  \Bigr]\ , \cr
h_{g} &\equiv \frac{1}{2}\Bigl[ -(\lambda_b-\lambda_r) -\sigma_0(\sigma_r-\sigma_g+\sigma_b)\lambda_g -\sigma_0(\sigma_r \lambda_r+\sigma_g \lambda_g+\sigma_b \lambda_b) \Bigr]\ , \cr
h_{b} &\equiv \frac{1}{2}\Bigl[ -(\lambda_r-\lambda_g) -\sigma_0(\sigma_r+\sigma_g-\sigma_b)\lambda_b -\sigma_0(\sigma_r \lambda_r+\sigma_g \lambda_g+\sigma_b \lambda_b)  \Bigr]\ .
\end{align}

\paragraph{Large $g$:} 

At large $g$, we have the 16 spectral tail states [i.e., states with energy of $\mathcal{O}(g)$] in this sector:
\begin{align}
|\bs_i\ct_{\hat{i}} A_\pm \rangle\equiv& b^\dag_i (c^{\dag^3})_{\hat{i}} A^\dag_\pm |\ \rangle \ , \\
|\bt_{\hat{i}}\cs_i A_\pm \rangle   \equiv& (b^{\dag^3})_{\hat{i}} c^\dag_i A^\dag_\pm |\ \rangle \ ,
\end{align}
where $i = 1, 2, 3, 4 $. Moreover, there are 24 midspectral states in this sector of the following form.
\begin{equation}
\begin{split}
 |[b,c]b^2c^2a^2 \pm a^2 [b,c] \rangle_{ijkl}  &\equiv 
\frac{1}{2} (b_i^\dag c^\dag_j-  c_i^\dag b^\dag_j)b_k^\dag b_l^\dag c^\dag_k c_l^\dag a_i^\dag a^\dag_j |\ \rangle \pm \frac{1}{2}a_i^\dag a^\dag_j (b_k^\dag c^\dag_l-  c_k^\dag b^\dag_l)  |\ \rangle\ , \\
 |\{b,c\}b^2c^2a^2 \pm a^2 \{b,c\} \rangle_{ijkl}  &\equiv 
\frac{1}{2} (b_i^\dag c^\dag_j+  c_i^\dag b^\dag_j)b_k^\dag b_l^\dag c^\dag_k c_l^\dag a_i^\dag a^\dag_j |\ \rangle \pm \frac{1}{2}a_i^\dag a^\dag_j (b_k^\dag c^\dag_l+  c_k^\dag b^\dag_l)  |\ \rangle\ ,
\end{split}
\end{equation}
and we summarize their energies in Table \ref{tab:pmm-sector}.
\begin{table}[h!]
\centering
{\renewcommand{\arraystretch}{1.2}
\begin{tabular}{ | >{\centering\arraybackslash}m{0.4cm} | >{\centering\arraybackslash}m{0.4cm} | >{\centering\arraybackslash}m{4cm} |}
\hline
$ij$ & $kl$  & {\small $|[b,c]b^2c^2a^2 - a^2 [b,c] \rangle_{ijkl} $ } \\       
\hline
\hline
$12$ & $34$ & $0$ \\
$34$ & $12$ & $0$ \\
$13$ & $24$ & $0$ \\
$24$ & $13$ & $0$ \\
$14$ & $23$ & $0$ \\
$23$ & $14$ & $0$ \\
\hline   
\end{tabular}
}

{\renewcommand{\arraystretch}{1.2}
\begin{tabular}{ | >{\centering\arraybackslash}m{0.4cm} | >{\centering\arraybackslash}m{0.4cm} | >{\centering\arraybackslash}m{4.2cm} |>{\centering\arraybackslash}m{4.2cm} |>{\centering\arraybackslash}m{4.2cm} |}
\hline
$ij$ & $kl$   & {\small $|\{b,c\}b^2c^2a^2 + a^2 \{b,c\} \rangle_{ijkl} $} & {\small $ |\{b,c\}b^2c^2a^2 - a^2 \{b,c\} \rangle_{ijkl} $} & {\small $ |[b,c]b^2c^2a^2 + a^2 [b,c] \rangle_{ijkl} $ } \\
\hline
\hline
$12$ & $34$  & $ (\lambda_r+\lambda_b)$  & $ -(\lambda_g+\lambda_r)$& $ (\lambda_b-\lambda_g)$ \\
$34$ & $12$  & $ (\lambda_r+\lambda_b)$  & $ -(\lambda_g+\lambda_r)$& $ (\lambda_b-\lambda_g)$ \\
$13$ & $24$  & $ (\lambda_g+\lambda_r)$  & $ -(\lambda_b+\lambda_g)$& $ (\lambda_r-\lambda_b)$ \\
$24$ & $13$  & $ (\lambda_g+\lambda_r)$  & $ -(\lambda_b+\lambda_g)$& $(\lambda_r-\lambda_b)$ \\
$14$ & $23$  & $ (\lambda_b+\lambda_g)$  & $ -(\lambda_r+\lambda_b)$ & $ (\lambda_g-\lambda_r)$ \\
$23$ & $14$  & $ (\lambda_b+\lambda_g)$  & $ -(\lambda_r+\lambda_b)$& $ (\lambda_g-\lambda_r)$ \\
\hline  
\end{tabular}}
\caption{The $6 \times 4 = 24$ midspectral states in the $(+, -, -)$ subsector of three-site KT chain model and its energy levels as $g\to \infty$.} 
\label{tab:pmm-sector}
\end{table}

\paragraph{Symmetric couplings:} 

The spectrum of the symmetric hopping couplings (i.e., $\lambda_r = \lambda_g = \lambda_b = \lambda$) is given in Table~\ref{tab:pmm-symmetric-case}.
\begin{table}[h!]
\centering
{\renewcommand{\arraystretch}{1.3}
\begin{tabular}{ | >{\centering\arraybackslash}m{6cm} | >{\centering\arraybackslash}m{2cm} |}
\hline
Eigenvalues ($\sigma \in \{ +, - \}$) & {\small Degeneracy} \\       
\hline
\hline
$0$ & $6$ \\
$2\sigma g$ & $1$ \\
$2\sigma \lambda $ & $2$ \\
$\sigma (g+\sqrt{g^2+4\lambda^2}) $ & $3$ \\
$\sigma (g-\sqrt{g^2+4\lambda^2}) $ & $3$ \\
$\sigma ( g + \lambda + \sqrt{g^2 - 2 g \lambda + 3 \lambda^2})$ & 3 \\
$\sigma (- g -  \lambda + \sqrt{g^2 - 2 g \lambda + 3 \lambda^2})$ & 3 \\
$\sigma ( g -  \lambda + \sqrt{g^2 + 2 g \lambda + 7 \lambda^2})$ & 1 \\
$\sigma ( -g + \lambda + \sqrt{g^2 + 2 g \lambda + 7 \lambda^2})$ & 1 \\ 
\hline
\end{tabular}}
\caption{Spectrum in the sector $(+, -, -)$ for $\lambda_r = \lambda_g = \lambda_b = \lambda$ case of the 3-site Abelian KT chain model.} 
\label{tab:pmm-symmetric-case}
\end{table}

\subsubsection{Other subsectors}

\paragraph{The $(-, +, -)$ subsector:}

The energy eigenstates in the $(-, +, -)$ subsector are obtained by acting the single translation $\hat{T}$ on the states in the $(+, -, -)$ subsector. The corresponding energy eigenvalues are the same as those in the $(+, -, -)$ subsector.

\paragraph{The $(-, -, +)$ subsector:}

Similarly, the energy eigenstates in  $(-, -, +)$ subsector can be found by acting the translation operator $\hat{T}^2$ on the states in the $(+, -, -)$ subsector. The corresponding energy eigenvalues are also the same as those in the $(+, -, -)$ subsector.

\subsubsection{A comparison of the spectra in different sectors}

\paragraph{Large $g$:}

The energy eigenvalues (up to first order in perturbation) and their corresponding degeneracies in different sectors are shown in Table \ref{tab:large-g-3-site-KT}.
\begin{table}[h!]
\centering
\begin{tabular}{|c|c|c|c|c|c|}
\hline
\multirow{2}{*}{Eigenvalue} & \multicolumn{5}{ c| }{Degeneracy}  \\ \cline{2-6}
 & {\tiny $(+, +, +)$} & {\tiny $(+, -, -)$}  & {\tiny $(-, +, -)$} & {\tiny $(-, -, +)$} & Total \\ 
 \hline
 \hline
 $+6g$ & 1 & 0 & 0 & 0 & 1 \\
 \hline
 $-6g$ & 1 & 0 & 0 & 0 & 1 \\
 \hline
 $+2g$ & 21 & 8 & 8 & 8 & 45 \\
 \hline
 $-2g$ & 21 & 8 & 8 & 8 & 45 \\
 \hline
 $\lambda_r+\lambda_b$ & 0 & 2 & 2 & 2 & 6 \\
 \hline
 $-(\lambda_r+\lambda_b)$ & 0 & 2 & 2 & 2 & 6 \\
 \hline
 $\lambda_g+\lambda_r$ & 0 & 2 & 2 & 2 & 6 \\
 \hline
 $-(\lambda_g+\lambda_r)$ & 0 & 2 & 2 & 2 & 6 \\
 \hline
 $\lambda_b+\lambda_g$ & 0 & 2 & 2 & 2 & 6 \\
 \hline
 $-(\lambda_b+\lambda_g)$ & 0 & 2 & 2 & 2 & 6 \\
 \hline
 $\lambda_r-\lambda_b$ & 0 & 2 & 2 & 2 & 6 \\
 \hline
 $\lambda_g-\lambda_r$ & 0 & 2 & 2 & 2 & 6 \\
 \hline
 $\lambda_b-\lambda_g$ & 0 & 2 & 2 & 2 & 6 \\
 \hline
 $0$ & 0 & 6 & 6 & 6 & 18 \\
 \hline
\end{tabular}
\caption{The spectrum of the 3-site Abelian KT chain model in different subsectors at large $g$.}
\label{tab:large-g-3-site-KT}
\end{table}

\paragraph{The spectrum at symmetric hopping:}

We compare the spectrum of subsectors for the case of symmetric hopping couplings. See Table~\ref{tab:spec-symm-hopp-3-site-KT}.
\begin{table}[h!]
\centering
\begin{tabular}{|c|c|c|c|c|c|}
\hline
\multirow{2}{*}{Eigenvalue} & \multicolumn{5}{ c| }{Degeneracy}  \\ \cline{2-6}
& {\tiny $(+,+,+)$}   & {\tiny $(+,-,-)$}  & {\tiny $(-,+,-)$}  & {\tiny $(-,-,+)$} & Total \\ 
\hline
\hline
0 & 0 & 6 & 6 & 6 & 18 \\
\hline
$\pm 2g$ & 0 & 1 & 1 & 1 & 3 \\
\hline
$\pm 2(g-\lambda)$ & 3 & 0 & 0 & 0 & 3 \\
\hline
$\pm (2g-\lambda)$ & 4 & 0 & 0 & 0 & 4 \\
\hline
$\pm 2(g+\lambda)$ & 2 & 0 & 0 & 0 & 2 \\
\hline
$\pm (2g+\lambda)$ & 6 & 0 & 0 & 0 & 6 \\
\hline
$\pm 2\lambda$ & 0 & 2 & 2 & 2 & 6 \\
\hline
$\pm \frac{1}{2} (\lambda + \sqrt{16 g^2 - 8 g \lambda + 25 \lambda^2})$ & 2 & 0 & 0 & 0 & 2 \\
\hline
$\pm  \frac{1}{2}(-\lambda + \sqrt{16 g^2 - 8 g \lambda + 25 \lambda^2})$ & 2 & 0 & 0 & 0 & 2 \\
\hline
$\pm (g+\sqrt{g^2+4\lambda^2})$ & 0 & 3 & 3 & 3 & 9 \\
\hline
$\pm (-g+\sqrt{g^2+4\lambda^2})$ & 0 & 3 & 3 & 3 & 9 \\
\hline
$\pm ( g + \lambda + \sqrt{g^2 - 2 g \lambda + 3 \lambda^2})$ & 0 & 3 & 3 & 3 & 9 \\
\hline
$\pm (- g -  \lambda + \sqrt{g^2 - 2 g \lambda + 3 \lambda^2})$ & 0 & 3 & 3 & 3 & 9 \\
\hline
$\pm ( g -  \lambda + \sqrt{g^2 + 2 g \lambda + 7 \lambda^2})$ & 0 & 1 & 1 & 1 & 3 \\
\hline
$\pm ( -g + \lambda + \sqrt{g^2 + 2 g \lambda + 7 \lambda^2})$ & 0 & 1 & 1 & 1 & 3 \\ 
\hline
$\pm R_1[P]$ & 1 & 0 & 0 & 0 & 1 \\
\hline
$\pm R_2[P]$ & 1 & 0 & 0 & 0 & 1 \\
\hline
$\pm R_3[P]$ & 1 & 0 & 0 & 0 & 1 \\
\hline
\end{tabular}
\caption{The Spectrum in the different subsectors for the case $\lambda_r = \lambda_g = \lambda_b = \lambda$ of the three-site Abelian KT chain model.}
\label{tab:spec-symm-hopp-3-site-KT} 
\end{table}

\subsection{Spectral properties of the three-site Abelian KT model}

\subsubsection{Band diagrams for eigenvalues}

\begin{figure}[htp]

\subfloat[Asymmetric case.]{\includegraphics[width=8.0cm,angle=0]{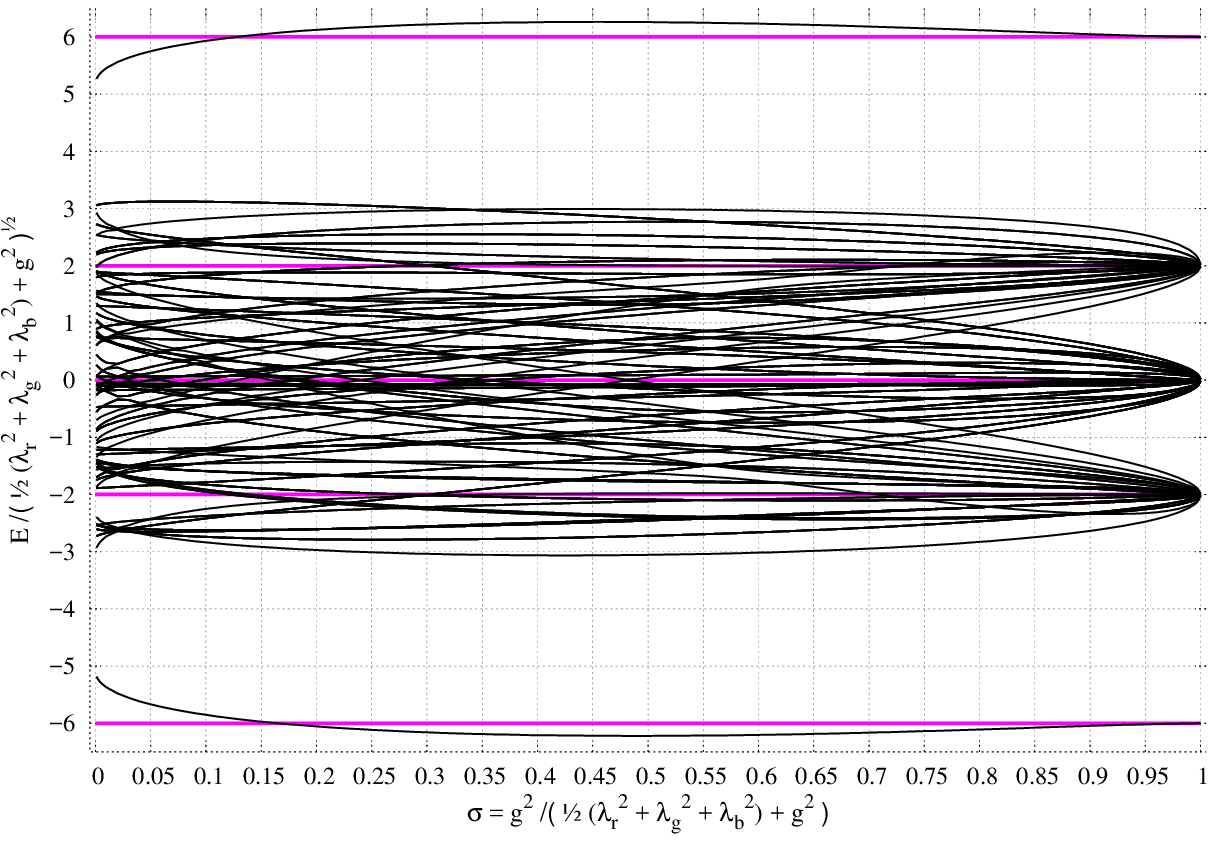}}
\subfloat[Symmetric case.]{\includegraphics[width=8.0cm,angle=0]{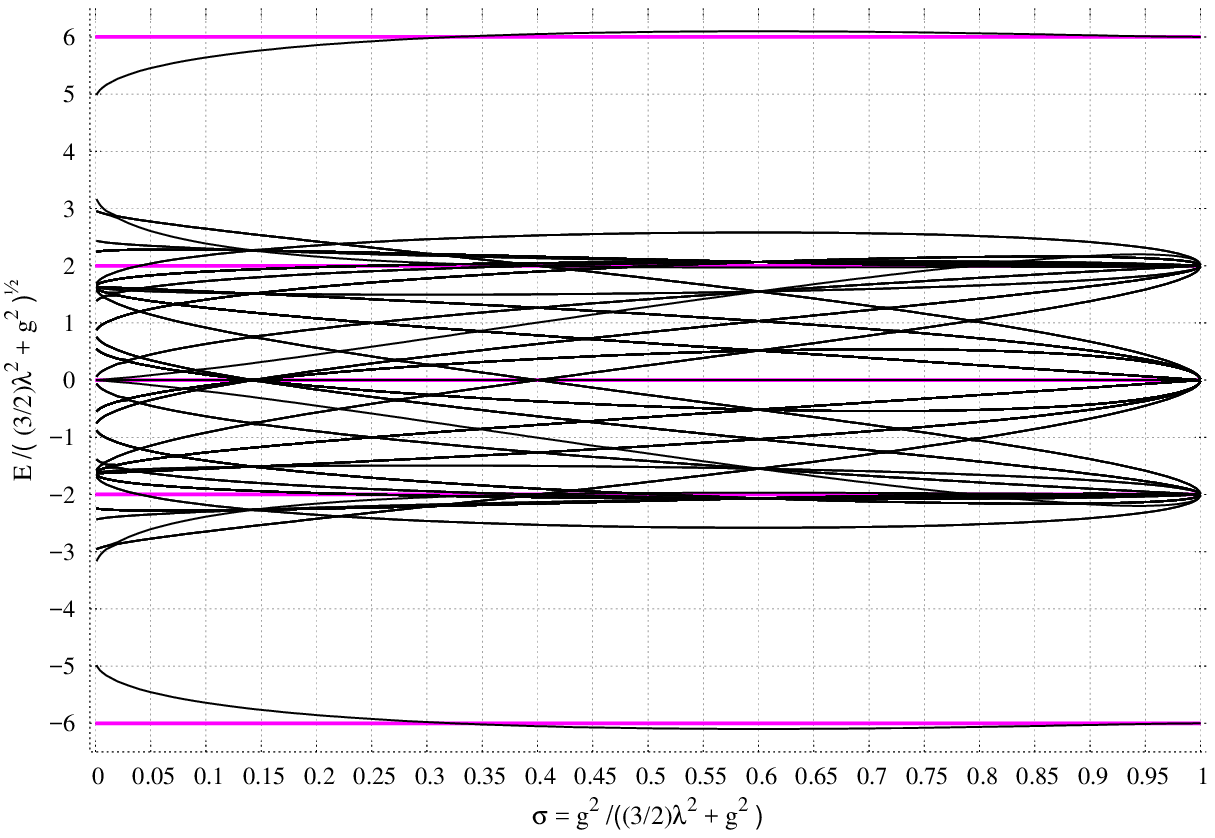}}

\caption{(a) $U(1)^3$ singlet spectrum : Rescaled eigenvalues $E/\sqrt{\hf(\lambda_r^2 + \lambda_g^2 + \lambda_b^2) + g^2}$ of the three-site Abelian KT chain model Hamiltonian with asymmetric hopping couplings against the coupling ratio $\sigma$ (degeneracies not shown). Here $\sigma \equiv \frac{g^2}{\hf(\lambda_r^2 + \lambda_g^2 + \lambda_b^2) + g^2}$. (b) $U(1)^3$ singlet spectrum :  Rescaled eigenvalues $E/\sqrt{\frac{3}{2}\lambda^2 + g^2}$ of the three-site Abelian KT chain model Hamiltonian with symmetric hopping couplings against the coupling ratio $\sigma$ (degeneracies not shown). Here $\sigma \equiv \frac{g^2}{\frac{3}{2}\lambda^2 + g^2}$.}
\label{fig:band-diag-3-site-KT-model}

\end{figure}

In Fig.~\ref{fig:band-diag-3-site-KT-model}, we show the band diagrams of rescaled eigenvalues $E/\sqrt{\hf(\lambda_r^2 + \lambda_g^2 + \lambda_b^2) + g^2}$ of the 3-site Abelian KT chain model with symmetric and asymmetric hopping couplings against the coupling ratio $\sigma \equiv \frac{g^2}{\hf(\lambda_r^2 + \lambda_g^2 + \lambda_b^2) + g^2}$.

We see that when the couplings of the hopping terms vanish (i.e., in the limit $\lambda_r = \lambda_g = \lambda_b=0$ for the asymmetric case, and $\lambda = 0$ in the symmetric case) the bands collapse to give one level with energy $0$ and degeneracy $72$; two levels with energies $2g$ and $-2g$, each having a degeneracy $45$; and two more non-degenerate levels with energies $6g$ and $-6g$.
%

\subsubsection{Cumulative spectral function, level spacing distribution and $r$-parameter statistics}

In the preceding sections, we have given a detailed description of the singlet spectrum of the $L = 2, 3$ site KT chain models. We will now turn to an analysis of the general characteristics of the spectrum. We will begin by reviewing some general notions about the nature of the spectrum and what they reveal about the Hamiltonian.

It is an essential insight due to Wigner and Dyson \cite{1967SIAMR...9....1W, 1962JMP.....3..140D, 1962JMP.....3..157D} that the spectrum of any sufficiently generic (i.e., nonintegrable) Hamiltonian can be modeled by a spectrum of a random matrix. Thus, in a system which is ergodic, i.e., a system where eigenstate thermalization hypothesis (ETH) holds, the Hamiltonian effectively behaves like a random matrix. This, in turn, the spectrum shows a characteristic level repulsion; i.e., the adjacent energies in an ergodic system tend not to cluster together but rather feel an effective repulsion resulting in a specific structure in the energy spectrum. Such a level repulsion is familiar from, say, the perturbation theory of two-level systems where the off-diagonal entries of the interaction Hamiltonian mixes the levels resulting in a level repulsion. Thus, if we denote the level spacing between two adjacent energy levels in an ergodic system as $\delta$, the probability of $\delta$ taking a value near zero is vanishingly small.

In contrast, in a many-body localized state, the off-diagonal entries are very much suppressed thus resulting in a breakdown of ergodicity. Thus, the eigenvalues corresponding to localized states are essentially uncorrelated random numbers without any spectral rigidity and hence they fall into a Poisson distribution. Thus, an examination of the statistics of the spectrum gives us crucial clues as to the nature of the Hamiltonian and its ergodicity.

We can apply a statistical measure known as nearest neighbor spacing distribution in order to extract this information. The first step is to perform an unfolding procedure on the eigenvalues of the model. We need to define the spectral staircase function, also known as the cumulative spectral function, \cite{Bohigas:1983er, Guhr:1997ve} for the unfolding procedure.  

The spectral staircase function $N(E)$ is defined as
\beq
N(E) \equiv \sum_n \Theta(E - E_n),
\eeq
where $\Theta$ is the Heaviside step function and $E_n$ represents the $n$th energy level from the ordered set of energy levels $\{E_1, E_2, \cdots, E_n\}$ of the model. It is easy to see that $N(E)$ is a counting function; it jumps by one unit each time an energy level $E_n$ is encountered. Thus $N(E)$ gives the number of energy levels $E_n$ with energy less than $E$. 

\begin{figure}[htp]

\centering
\includegraphics[width=4.0in]{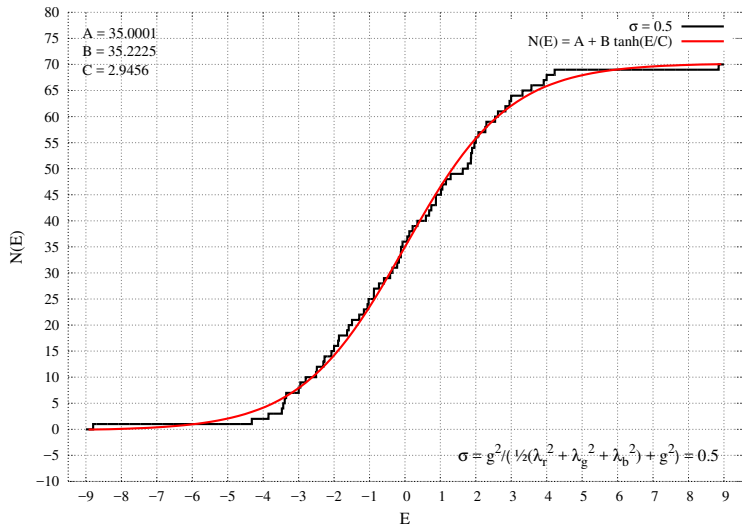}
\caption{The spectral staircase function $N(E)$ of the three-site KT chain model for asymmetric case of the hopping couplings. The plot is for $\sigma = 0.5$, which corresponds to $\lambda_r/g = 0.8255, \lambda_g/g = 0.3005$ and $\lambda_b/g = 1.1083$. Note that here $E$ is measured in units of the on-site coupling $g$.}
\label{fig:spec-staircase-3-site-KT-model}

\end{figure}

In Fig.~\ref{fig:spec-staircase-3-site-KT-model}, we show the spectral staircase function $N(E)$ of the three-site Abelian KT chain model for asymmetric case of the hopping couplings. The plot is for the coupling ratio $\sigma = 0.5$, which corresponds to the hopping couplings $\lambda_r/g = 0.8255, \lambda_g/g = 0.3005$ and $\lambda_b/g = 1.1083$. Note that in the plot energy $E$ is given in units of the on-site coupling $g$.

The next step is to define a function $\bar{N}(E)$, which is the mean staircase function interpolating $N(E)$. We can now map the energies $\{E_1, E_2, \cdots, E_n\}$ onto numbers $\{\xi_1, \xi_2, \cdots, \xi_n\}$ by
\beq
\xi_k = \xi(E_k),~~k = 1, \cdots, n.
\eeq
This would give us a new spectrum, $\{\xi_1, \xi_2, \cdots, \xi_n\}$, which we call an ordered unfolded spectrum of energy levels. This spectrum has a constant mean spacing of unity \cite{Bohigas:1983er}.

We are now in a position to calculate the nearest neighbor spacing distribution. Let us define a quantity
\beq
s \equiv \xi_{k+1} - \xi_k,
\eeq
which is the spacing between two neighboring energy levels. The nearest neighbor spacing distribution $P(s)$ gives the probability that the spacing between two neighboring energy levels is $s$. The unfolding procedure mentioned above ensures that both $P(s)$ and its mean are normalized to unity.

We can use the nearest neighbor spacing distribution $P(s)$ to study the short-range fluctuations in the spectrum. We also note that there is another statistical measure of the energy level spacings known as the spectral rigidity. It measures the long-range correlations in the model. We do not diagnose the spectral rigidity properties of our models in this paper.

We have a strong indication that the three-site Abelian KT chain model we consider here is integrable. The probability distribution $P(s)$ behaves like
\beq
P(s) \sim e^{-s},
\eeq
which is the characteristic of a Poisson process. This in turn indicates that the energy levels are uncorrelated, that is, they are distributed at random. We also see that the maximum value of the distribution occurs at $s=0$, indicating a level clustering in the model.  

\begin{figure}[htp]

\subfloat[]{\includegraphics[width=3.1in]{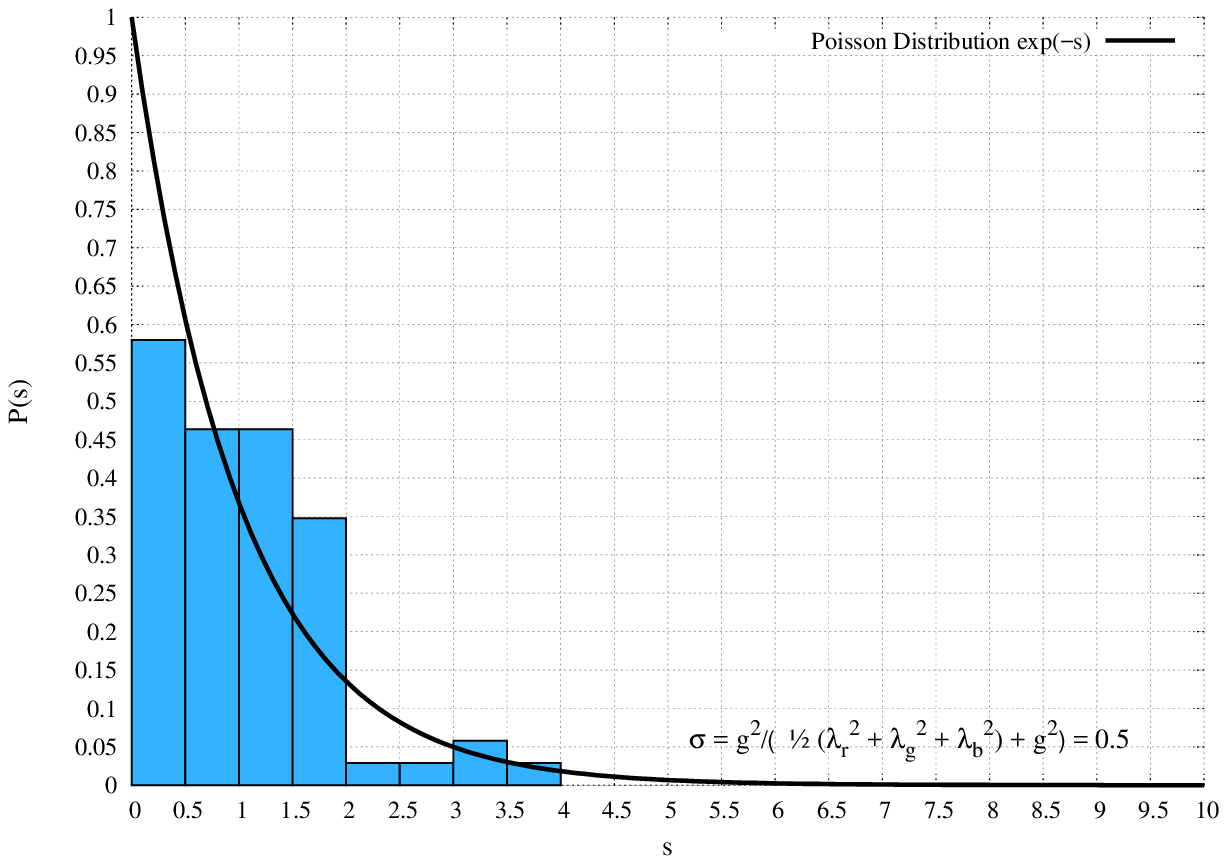}}
\subfloat[]{\includegraphics[width=3.1in]{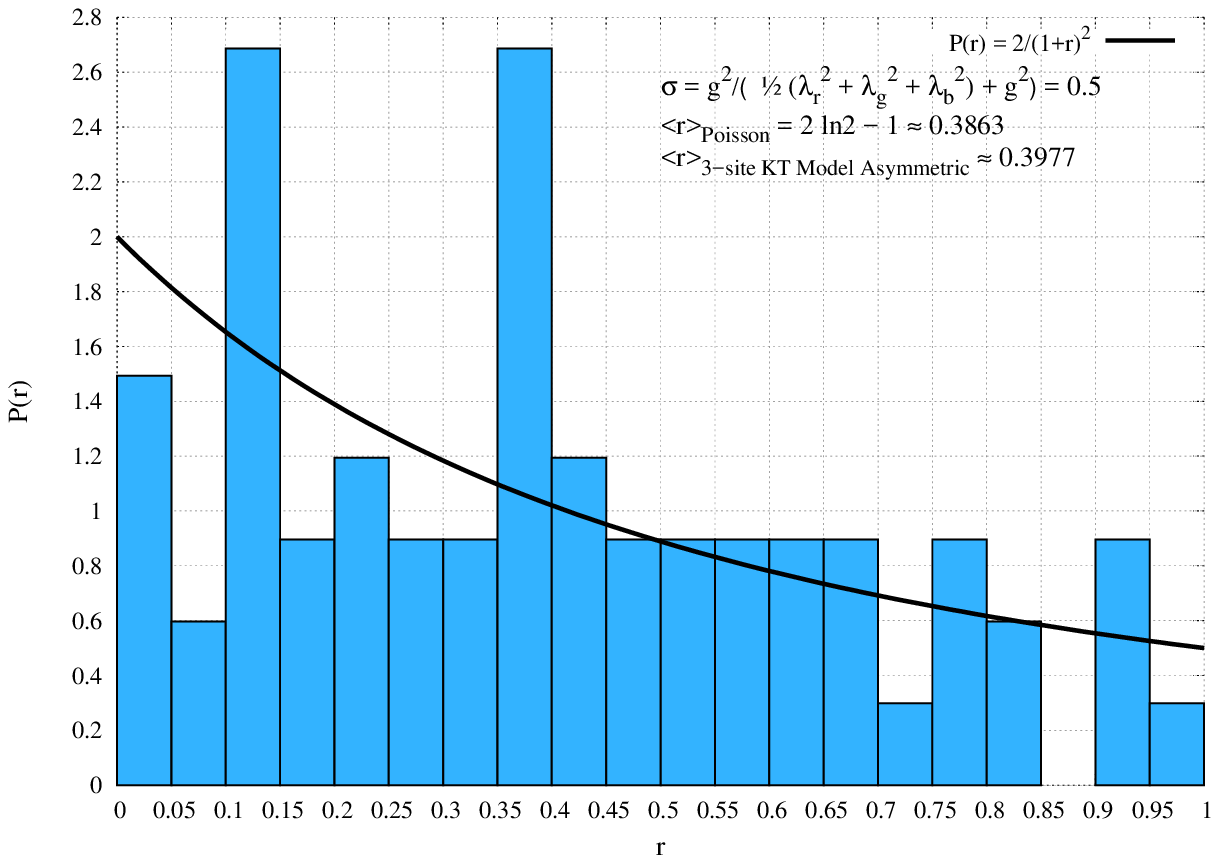}}

\caption{(a) The histogram of nearest neighbor spacing distribution $P(s)$ against $s$ for the three-site Abelian KT chain model with asymmetric hopping couplings. Level clustering is evident in the model, indicating the the system is integrable. (b) The histogram of $r$-parameter distribution $P(r)$ against $r$ for the three-site Abelian KT chain model with asymmetric hopping couplings. In both cases the plots are for the coupling ratio $\sigma = 0.5$, which corresponds to $\lambda_r/g = 0.8255, \lambda_g/g = 0.3005$ and $\lambda_b/g = 1.1083$; and the fits are to Poisson distribution.
\label{fig:s-Ps-r-Pr-3-site-KT-model}}

\end{figure}

In Fig.~\ref{fig:s-Ps-r-Pr-3-site-KT-model} (left) we give the nearest neighbor spacing distribution $P(s)$ against $s$ for the three-site Abelian KT chain model with asymmetric hopping couplings. Level clustering is evident in the figure. It becomes more and more apparent as we go to four- and five-site models.

We also note that chaotic systems generally exhibit level repulsion. That is, the difference between neighboring eigenvalues is statistically unlikely to be small compared to the mean eigenvalue spacing.

We can also diagnose the ergodicity of the system using the statistics of a dimensionless quantity called the $r$-parameter \cite{Oganesyan:2007aa}. This parameter characterizes the correlations between adjacent gaps in the energy spectrum. It is defined as the ratio
\beq
r_n \equiv \frac{{\rm min}(s_n, s_{n-1})}{{\rm max}(s_n, s_{n-1})},
\eeq
where
\beq
s_n \equiv E_{n+1} - E_n,
\eeq
and $E_n$ the ordered set of energy levels of the Hamiltonian. 

The $r$-parameter takes values between $[0,1]$. When spectrum is Poisson the probability distribution of this ratio is 
\beq
P(r) = \frac{2}{(1+r)^2},
\eeq
with the mean value 
\beq
\langle r \rangle_{\rm Poisson} = 2 \ln 2 - 1 \approx 0.3863.
\eeq
For comparison, we also note that for large Gaussian orthogonal ensemble (GOE) random matrices the mean value is 
\beq
\langle r \rangle_{\rm GOE} \approx 0.5295.
\eeq
In Fig.~\ref{fig:s-Ps-r-Pr-3-site-KT-model} (right), we give the $r$-parameter distribution $P(r)$ against $r$ for the 3-site Abelian KT chain model with asymmetric hopping couplings. The fit is to a Poisson distribution.

\subsection{Comments}

\begin{itemize}
\item The $r$-parameter statistics (shown in Fig.~\ref{fig:s-Ps-r-Pr-3-site-KT-model}) do not show a clear fit for Poisson (or for that matter random matrix) behavior. Due to the small number of states going into the fit, $r$-parameter statistics is inconclusive in this case. But as we will see in the following, with an increase in the number of sites, a better fit to Poisson-like behavior and level statistics can be obtained.  

\item Here, we see that there are no middle states when the three couplings of the hopping terms are all different. This has also been seen in the five-site case. We expect this to be a generic behavior whenever the number of sites is odd.

\item As in the two-site case we see that there are many middle states when $\lambda_r = \lambda_g = \lambda_b = \lambda$. In particular, we have 18 middle states in this case.
\item There are six protected states three of which have energy $2g$ and the other three have energy $-2g$. These states are  
\begin{equation}
\begin{split}
\sum_{i=1}^4 \Big(|A_\pm \bs_i \ct_{\hat{i}}\rangle +|A_\pm \bt_{\hat{i}} \cs_i\rangle\Big),\\
\hat{T}\Big[\sum_{i=1}^4 \Big(|A_\pm \bs_i \ct_{\hat{i}}\rangle +|A_\pm \bt_{\hat{i}} \cs_i\rangle\Big)\Big],\\
\hat{T}^2\Big[\sum_{i=1}^4 \Big(|A_\pm \bs_i \ct_{\hat{i}}\rangle +|A_\pm \bt_{\hat{i}} \cs_i\rangle\Big)\Big].
\end{split}
\end{equation}
The states with the $(+)$ have energy $2g$ and those with the $(-)$ have energy $-2g$.
\end{itemize}

\section{Abelian KT Chain Model: Four Sites}
\label{sec:KT4site}

\subsection{Hamiltonian}

The Hamiltonian is 
\begin{equation}
H =  g(H_a^{(0)}+ H_b^{(0)} + H_c^{(0)}+ H_d^{(0)}) + \sum_{\col\in \{r,g,b\}}\lambda_{\col} \left[ H_{ab}^{(\col)}+H_{bc}^{(\col)}+H_{cd}^{(\col)}+H_{da}^{(\col)}\right]\ ,
\end{equation}
with similar definitions for hopping term $H_{bc}^{(r)}$ {\it etc}. as that of the three-site case.

The singlet sector can be further divided into eight subsectors having definite $(\mathbb{Z}_2)^4$ charges (See Table~\ref{tab:dim-charges-4-site-KT}). We find that there are a large number of middle states in the spectrum. Thus we perform a detailed analysis of these middle states in appendix~\ref{app:4site middle asym} for the case of asymmetric hopping couplings and appendix~\ref{app:4site middle sym} for the case of symmetric hopping couplings. We enumerate the dimensions of these subsectors and the number of middle states in each of them in Table \ref{tab:dim-charges-4-site-KT}. We provide the spectra in the $(+, +, +, +)$, $(-, -, -, -)$, $(+, -, +, -)$ and $(+, +, -, -)$ subsectors in Tables \ref{tab:spectrum pppp sector}, \ref{tab:spectrum mmmm sector}, \ref{tab:spectrum pmpm sector} and \ref{tab:spectrum ppmm sector}, respectively.
\begin{center}
\begin{table}[h!]
\begin{center}
\begin{tabular}{|c|c|c|c|}\hline
\multirow{2}{*}{$(\mathbb{Z}_2)^4$ charges} & \multirow{2}{*}{Dimension} & \multicolumn{2}{c|}{No. of middle states}\\\cline{3-4}
 & & Asymmetric  & Symmetric \\
 \hline
 $(+,+,+,+)$  & 250 & 26 & 132\\
 \hline
 $(+,-,+,-)$ & 224 & 26 & 112\\
 \hline
 $(-,+,-,+)$  & 224 & 26 & 112\\
 \hline
 $(+,+,-,-)$  & 224 & 0 & 112\\
 \hline
 $(-,+,+,-)$  & 224 & 0 & 112\\
 \hline
 $(-,-,+,+)$  & 224 & 0 & 112\\
 \hline
 $(+,-,-,+)$  & 224 & 0 & 112\\
 \hline
 $(-,-,-,-)$ & 216 & 29 & 134\\
 \hline
 Total & 1810 & 107 & 938\\
 \hline
\end{tabular}
\end{center}
\caption{The dimensions of different subsectors with definite $(Z_2)^4$ charges and the number of middle states in each of them.}
\label{tab:dim-charges-4-site-KT}  
\end{table}
\end{center}

\subsection{The spectrum at symmetric hopping}

We summarize the spectrum in each sector in the following table.


\begin{table}[h!]
\centering
{\bf The $(+, +, +, +)$ Sector}\\
\vspace{0.2cm}
\begin{tabular}{|c|c|}
\hline
Energy eigenvalues  & Degeneracy\\
\hline
\hline
0  & 132 \\
 $\pm 4 g$  & 3 \\
 $\pm 2 \lambda  $ & 10 \\
 $\pm 2 \sqrt{2} \lambda $  & 5 \\
 $\pm 4 \sqrt{g^2+\lambda ^2}$  & 3 \\
 $\pm 2 \sqrt{2} \sqrt{2 g^2+\lambda ^2}$  & 5 \\
 $\pm 4 \sqrt{g^2+2 \lambda ^2} $ & 1 \\
$ \pm 2 \sqrt{2} \sqrt{2 g^2+3 \lambda ^2}  $& 1 \\
$ \pm 2 \sqrt{4 g^2+3 \lambda ^2}  $& 6 \\
$ \pm 2 \sqrt{2} \sqrt{2 g^2+2 \lambda ^2-\sqrt{\lambda ^2 \left(8 g^2+\lambda ^2\right)}}  $& 2 \\
$ \pm 2 \sqrt{2} \sqrt{2 g^2+2 \lambda ^2+\sqrt{\lambda ^2 \left(8 g^2+\lambda ^2\right)}}  $& 2 \\
$ \pm\sqrt{2} \sqrt{8 g^2+5 \lambda ^2-\sqrt{\lambda ^2 \left(64 g^2+9 \lambda ^2\right)}}  $& 6 \\
$ \pm\sqrt{2} \sqrt{8 g^2+5 \lambda ^2+\sqrt{\lambda ^2 \left(64 g^2+9 \lambda ^2\right)}} $ & 6 \\
$ \pm\sqrt{2} \sqrt{4 g^2+7 \lambda ^2-\sqrt{16 g^4+40 \lambda ^2 g^2+\lambda ^4}} $&  2 \\
$ \pm\sqrt{2} \sqrt{4 g^2+7 \lambda ^2+\sqrt{16 g^4+40 \lambda ^2 g^2+\lambda ^4}} $ & 2 \\
$ \pm 2 \sqrt{2} \sqrt{g^2+2 \lambda ^2-\sqrt{g^4+2 \lambda ^2 g^2+4 \lambda ^4}} $&  1 \\
$ \pm2 \sqrt{2} \sqrt{g^2+2 \lambda ^2+\sqrt{g^4+2 \lambda ^2 g^2+4 \lambda ^4}} $&  1 \\
$ \pm\sqrt{R_1[Q_1]} $& 1 \\
$\pm\sqrt{R_2[Q_1]} $&  1 \\
$\pm\sqrt{R_3[Q_1]} $&  1 \\
\hline
 Total no. of states & 250\\
 \hline
\end{tabular}
\caption{The spectrum in the (+, +, +, +) subsector.}
\label{tab:spectrum pppp sector}  
\end{table}
Here $R_i[Q_1]$ is the $i$th root of the polynomial: 
\bea
Q_1(\alpha) &=& \alpha ^3 + \alpha ^2 \left(-96 g^2-104 \lambda ^2\right) + \alpha \left(2304 g^4+3200 g^2 \lambda ^2+2304 \lambda ^4\right) \nn \\
&&~~~~~~~~~~~~~~~-61440 g^2 \lambda ^4 - 16384 g^6.
\eea


\begin{table}[h!]
\centering
{\bf The $(-, -, -, -,)$ sector}\\
\vspace{0.2cm}
\begin{tabular}{|c|c|}
\hline
Energy eigenvalues & Degeneracy\\
\hline
\hline
0  & 134 \\
$\pm 2 \lambda$   & 18 \\
$\pm 2 \sqrt{2} \lambda$   & 14 \\
$\pm 2 \sqrt{5} \lambda $  & 6 \\
$\pm 2 \sqrt{6} \lambda $  & 2 \\
$\pm 4 \sqrt{2} \lambda $  & 1 \\
\hline 
Total no. of states & 216\\
\hline
\end{tabular}
\caption{Spectrum in the $(-,-,-,-)$ subsector}
\label{tab:spectrum mmmm sector}
\end{table}


\begin{table}[h!]
\centering
{\bf The $(+,-,+,-,)$ sector}\\
\vspace{0.2cm}
\begin{tabular}{|c|c|}
\hline
Energy eigenvalues  & Degeneracy\\
\hline
0  & 112 \\
$\pm 2 g $ & 3 \\
$\pm 4 g $ & 1 \\
$\pm 2 \sqrt{g^2+\lambda ^2} $ & 17 \\
$\pm 4 \sqrt{g^2+\lambda ^2}  $& 3 \\
$\pm 2 \sqrt{g^2+2 \lambda ^2} $ & 17 \\
$\pm 2 \sqrt{g^2+3 \lambda ^2}  $& 3 \\
$\pm 2 \sqrt{g^2+5 \lambda ^2}  $& 3 \\
$\pm 2 \sqrt{g^2+8 \lambda ^2}  $& 1 \\
$\pm \sqrt{2} \sqrt{5 g^2+7 \lambda ^2-\sqrt{9 g^4+102 \lambda ^2 g^2+\lambda ^4}} $ & 1 \\
$\pm \sqrt{2} \sqrt{5 g^2+7 \lambda ^2+\sqrt{9 g^4+102 \lambda ^2 g^2+\lambda ^4}} $ & 1 \\
$\pm \sqrt{2} \sqrt{5 g^2+6 \lambda ^2-\sqrt{9 g^4+60 \lambda ^2 g^2+4 \lambda ^4}} $ & 3\\
$\pm \sqrt{2} \sqrt{5 g^2+6 \lambda ^2+\sqrt{9 g^4+60 \lambda ^2 g^2+4 \lambda ^4}} $ & 3 \\
\hline
\text{Total number of states} & 224\\
\hline
\end{tabular}
\caption{The spectrum in the $(+, -, +, -)$ subsector.}
\label{tab:spectrum pmpm sector}
\end{table}

%
\begin{table}[h!]
\centering
{\bf The $(+, +, -, -)$ sector}\\
\vspace{0.2cm}
\begin{tabular}{|c|c|}
\hline
Energy eigenvalues  & {\small Degeneracy}\\
\hline
0 &  112 \\
$\pm 2 \sqrt{g^2 - g \lambda + \lambda^2}$  & 14 \\
$\pm  2 \sqrt{g^2 + g \lambda + \lambda^2}  $& 14 \\
$\pm 2 \sqrt{g^2 -3 g \lambda + 3\lambda^2} $ & 3 \\
$\pm 2 \sqrt{g^2 +3 g \lambda + 3\lambda^2}  $& 3  \\
$\pm \sqrt{10 g^2+2 g \lambda +14 \lambda ^2-2\sqrt{9 g^4 - 6 g^3 \lambda + 31 g^2 \lambda^2 -  10 g \lambda^3 + \lambda^4}}  $ & 1 \\
$\pm \sqrt{10 g^2+2 g \lambda +14 \lambda ^2+2\sqrt{9 g^4 - 6 g^3 \lambda + 31 g^2 \lambda^2 -  10 g \lambda^3 + \lambda^4}}$  & 1 \\
$\pm \sqrt{10 g^2-2 g \lambda +14 \lambda ^2-2\sqrt{9 g^4 + 6 g^3 \lambda + 31 g^2 \lambda^2 +  10 g \lambda^3 + \lambda^4}}  $& 1 \\
$\pm \sqrt{10 g^2-2 g \lambda +14 \lambda ^2+2\sqrt{9 g^4 + 6 g^3 \lambda + 31 g^2 \lambda^2 +  10 g \lambda^3 + \lambda^4}} $&  1 \\
$\pm\frac{g}{2}R_1[Q(\lambda/g)]  $ & 3 \\
$\pm\frac{g}{2}R_2[Q(\lambda/g)]  $ & 3 \\
$\pm\frac{g}{2}R_3[Q(\lambda/g)] $  & 3 \\
$\pm\frac{g}{2}R_1[Q(-\lambda/g)] $  & 3 \\
$\pm\frac{g}{2}R_2[Q(-\lambda/g)] $  & 3 \\
$\pm\frac{g}{2}R_3[Q(-\lambda/g)]  $ & 3 \\
\hline
\text{Total no. of states} & 224\\
\hline
\end{tabular}
\caption{The spectrum in the $(+, +, -, -)$ subsector. $R_i [Q(\eta; \alpha)]$ denotes the $i${\it\tiny th} solution of the polynomial $Q(\eta; \alpha)$.}
\label{tab:spectrum ppmm sector}
\end{table}
{\noindent Here $R_i[Q(\eta)]$ is the $i$th root of the polynomial:}
\begin{align}
Q (\eta; \alpha) \equiv 16384 + 32768 \eta + 40960  \eta^2 + 
 49152 \eta^3 + 180224 \eta^4 + 
 90112 \eta^5 +
 81920 \eta^6 - \cr  (2304 + 2560 \eta + 
    7424 \eta^2 + 3072\eta^3 + 
    7424 \eta^4)\alpha + (96 + 32 \eta + 
    160\eta^2) \alpha^2 -\alpha^3.
\end{align} 

%


{\noindent {\bf The $(-, +, -, +)$ sector} : The eigenvalues $(-, +, -, +)$ sector are the same as those of the $(+, -, +, -)$ subsector due to the translational symmetry of the system.}

\vspace{0.2cm}


{\noindent {\bf The $(-, +, +, -)$, $(-,-,+,+)$ and $(+,-,-,+)$ sectors}: The eigenvalues in these sectors are the same as those in the $(+, +, -, -)$ sector due to the translational symmetry of the system.}

\subsection{A comparison of the spectra in different sectors at symmetric hopping}

By translational symmetry, the spectra in the $(+, -, +, -)$ and $(-, +, -, +)$ subsectors are the same. Similarly, the spectrum in the $(+, +, -, -)$, $(-, +, +, -)$, $(-, -, +, +)$ and $(+, -, -, +)$ subsectors are also the same. We show the energy eigenvalues and their corresponding degeneracies for the $(+, +, +, +)$, $(-, -, -, -)$, $(+, -, +, -)$ and $(+, +, -, -)$ subsectors in Tables~\ref{tab:symm-4-site-KT-b} and~\ref{tab:symm-4-site-KT-a}. It is important to bear in mind that the degeneracies in the $(+, -, +, -)$ subsector should be multiplied by 2 to get the total degeneracies of the same in the singlet sector. Similarly, the degeneracies in the $(+, +, -, -)$ subsector also should be multiplied by 4 to get the total degeneracies of the same in the singlet sector.

\begin{table}[h!]
\centering
\begin{tabular}{|c|c|c|c|c|}
\hline
\multirow{2}{*}{Eigenvalue} & \multicolumn{4}{ c| }{Degeneracy}  \\ \cline{2-5}
& {\tiny $(\pm ,\pm ,\pm, \pm)$}   & {\tiny $(+,-,+,-)$}  & {\tiny $(+,+,-,-)$} & Total\\ 
\hline
\hline
 $\pm \sqrt{2} (5 g^2+7 \lambda ^2-$  & \multirow{2}{*}{0} & \multirow{2}{*}{1\;(2)} & \multirow{2}{*}{0 }& \multirow{2}{*}{$ 2$} \\
 $\sqrt{9 g^4+102 \lambda ^2 g^2+\lambda ^4})^\frac{1}{2}$   &  &  &  & \\
 \hline 
 $\pm \sqrt{2} (5 g^2+7 \lambda ^2+$ & \multirow{2}{*}{0} & \multirow{2}{*}{1 \;(2)} &\multirow{2}{*}{ 0} &\multirow{2}{*}{ $2$} \\
 $\sqrt{9 g^4+102 \lambda ^2 g^2+\lambda ^4})^\frac{1}{2}$  &   &  &  & \\
 \hline
 $\pm \sqrt{2} (5 g^2+6 \lambda ^2-$  & \multirow{2}{*}{0 }& \multirow{2}{*}{3 \;(2)} & \multirow{2}{*}{0 }& \multirow{2}{*}{$ 6$ }\\
 $\sqrt{9 g^4+60 \lambda ^2 g^2+4 \lambda ^4})^\frac{1}{2}$    &  &  &  & \\
 \hline
 $\pm \sqrt{2} (5 g^2+6 \lambda ^2+$  & \multirow{2}{*}{0} & \multirow{2}{*}{3 \;(2)} &\multirow{2}{*}{ 0 }& \multirow{2}{*}{$6$ }\\
 $\sqrt{9 g^4+60 \lambda ^2 g^2+4 \lambda ^4})^\frac{1}{2}$    &  &  &  & \\
 \hline
 $\pm 2 \sqrt{g^2 - g \lambda + \lambda^2}$  & 0 & 0 & $14 \;(4)$ & $56$ \\
 \hline
 $\pm  2 \sqrt{g^2 + g \lambda + \lambda^2}$  & 0 & 0 & $14 \;(4)$ & $56$ \\
 \hline
 $\pm 2 \sqrt{g^2 -3 g \lambda + 3\lambda^2}$  & 0 & 0 & $3 \;(4)$ & $12$ \\
 \hline
 $\pm 2 \sqrt{g^2 +3 g \lambda + 3\lambda^2}$  & 0 & 0 & $3 \;(4)$ & $12$ \\
 \hline
 {\small $\pm \Bigl[10 g^2+2 g \lambda +14 \lambda ^2-$ }& \multirow{2}{*}{0} & \multirow{2}{*}{0} & \multirow{2}{*}{$1 \;(4)$} & \multirow{2}{*}{$4$} \\
 {\small  $2\sqrt{9 g^4 - 6 g^3 \lambda + 31 g^2 \lambda^2 -  10 g \lambda^3 + \lambda^4}\Bigr]^\frac{1}{2}$   } &  &  &  & \\
 \hline
 {\small $\pm\Bigl[10 g^2+2 g \lambda +14 \lambda ^2+$  } & \multirow{2}{*}{0} & \multirow{2}{*}{0} & \multirow{2}{*}{$1 \;(4)$} & \multirow{2}{*}{$4$} \\
 {\small $2\sqrt{9 g^4 - 6 g^3 \lambda + 31 g^2 \lambda^2 -  10 g \lambda^3 + \lambda^4}\Bigr]^\frac{1}{2}$ }   &  &  &  & \\
 \hline
{\small  $\pm\Bigl[10 g^2-2 g \lambda +14 \lambda ^2-$ }  & \multirow{2}{*}{0 }& \multirow{2}{*}{0 }& \multirow{2}{*}{$1 \;(4)$} &\multirow{2}{*}{ $ 4$ } \\
{\small  $2\sqrt{9 g^4 + 6 g^3 \lambda + 31 g^2 \lambda^2 +  10 g \lambda^3 + \lambda^4}\Bigr]^\frac{1}{2}$ }   &  &  &  & \\
 \hline
{\small  $\pm \Bigl[10 g^2-2 g \lambda +14 \lambda ^2+$ } &\multirow{2}{*}{ 0} & \multirow{2}{*}{0 }& \multirow{2}{*}{$1 \;(4)$} & \multirow{2}{*}{$4$} \\
{\small  $2\sqrt{9 g^4 + 6 g^3 \lambda + 31 g^2 \lambda^2 +  10 g \lambda^3 + \lambda^4}\Bigr]^\frac{1}{2}$   } &  &  &  & \\
 \hline
 $\pm\frac{g}{2}R_1[Q(\lambda/g)]$   & 0 & 0 & $3 \;(4)$ & 12 \\
 \hline
 $\pm\frac{g}{2}R_2[Q(\lambda/g)]$  & 0 & 0 &$ 3\;(4)$ & 12 \\
 \hline
 $\pm\frac{g}{2}R_3[Q(\lambda/g)]$   & 0 & 0 & $ 3\;(4)$ & 12 \\
 \hline
 $\pm\frac{g}{2}R_1[Q(-\lambda/g)]$  & 0 & 0 & $ 3\;(4)$ & 12 \\
 \hline
 $\pm\frac{g}{2}R_2[Q(-\lambda/g)]$   & 0 & 0 &$ 3\;(4)$ &12 \\
 \hline
 $\pm\frac{g}{2}R_3[Q(-\lambda/g)]$  & 0 & 0 & $ 3\;(4)$ & 12 \\
 \hline
\end{tabular}
\caption{The spectrum in different subsectors for the case $\lambda_r = \lambda_g = \lambda_b = \lambda$ of four-site Abelian KT chain model. The number in parenthesis denotes the degeneracy of the corresponding states.} 
\label{tab:symm-4-site-KT-b}
\end{table}
\begin{table}[h!]
\centering
\begin{tabular}{| >{\centering\arraybackslash}m{6.4cm} | >{\centering\arraybackslash}m{1.3cm} | >{\centering\arraybackslash}m{1.3cm} | >{\centering\arraybackslash}m{1.3cm} | >{\centering\arraybackslash}m{1.3cm} | >{\centering\arraybackslash}m{1.1cm} |}
\hline
\multirow{2}{*}{Eigenvalue} & \multicolumn{5}{ c| }{Degeneracy}  \\ \cline{2-6}
& {\tiny $(+,+,+,+)$}  & {\tiny $(-,-,-,-)$}  & {\tiny $(+,-,+,-)$}  & {\tiny $(+,+,-,-)$} & Total\\ 
\hline
\hline
 0 & $132\; (1)$ & $134\;(1)$ & $112 \; (2)$ & $112 \; (4)$ & $938$ \\
 \hline
 $\pm 2 \lambda$ & $10 \; (1)$ & $18\; (1)$ & 0 & 0 & $28$ \\
 \hline
 $\pm 2 \sqrt{2} \lambda$ & $5 \; (1)$ & $14 \; (1)$ & 0 & 0 & $ 19$ \\
 \hline
 $\pm 2 \sqrt{5} \lambda$ & 0 & $6\; (1)$ & 0 & 0 & 6 \\
 \hline
 $\pm 2 \sqrt{6} \lambda$ & 0 & $2 \; (1)$ & 0 & 0 & 2 \\
 \hline
 $\pm 4 \sqrt{2} \lambda$ & 0 & $1\; (1)$ & 0 & 0 & 1 \\
 \hline
 $ \pm 2 g$ & 0 & 0 & $3\; (2)$ & 0 & $ 6$ \\
 \hline
 $ \pm 4 g$ & $3 \; (1)$ & 0 & $1 \; (2)$ & 0 & $ 5$ \\
 \hline
 $ \pm 2 \sqrt{g^2+\lambda ^2}$ & 0 & 0 & $17\; (2)$ & 0 & $34$ \\
 \hline
 $ \pm 4 \sqrt{g^2+\lambda ^2}$ & $3 \; (1)$ & 0 & $3 \; (2)$ & 0 & $ 9$ \\
 \hline
 $ \pm 2 \sqrt{2} \sqrt{2 g^2+\lambda ^2}$ & $5 \; (1)$ & 0 & 0 & 0 & $5$ \\
 \hline
 $ \pm 4 \sqrt{g^2+2 \lambda ^2}$ & $1 \; (1)$ & 0 & 0 & 0 & $1$ \\
 \hline
 $\pm 2 \sqrt{2} \sqrt{2 g^2+3 \lambda ^2}$ & $1 \; (1)$ & 0 & 0 & 0 & $1$ \\
 \hline
 $ \pm 2 \sqrt{4 g^2+3 \lambda ^2}$ & $6 \; (1)$ & 0 & 0 & 0 & $6$ \\
 \hline
 $\pm 2 \sqrt{g^2+2 \lambda ^2}$ & 0 & 0 & $17 \; (2)$ & 0 & $  34$ \\
 \hline
 $\pm 2 \sqrt{g^2+3 \lambda ^2}$ & 0 & 0 & $3 \; (2)$ & 0 & $ 6$ \\
 \hline
 $\pm 2 \sqrt{g^2+5 \lambda ^2}$ & 0 & 0 & $3 \; (2)$ & 0 & $  6$ \\
 \hline
 $\pm 2 \sqrt{g^2+8 \lambda ^2}$ & 0 & 0 & $1\; (2)$ & 0 & $  2$ \\
 \hline 
 {\small $\pm 2 \sqrt{2} \left[2 g^2+2 \lambda ^2-\sqrt{\lambda ^2 \left(8 g^2+\lambda ^2\right)}\right]^\frac{1}{2}$} & $2 \; (1)$ & 0 & 0 & 0 & $2$ \\
 \hline
  {\small $\pm 2 \sqrt{2} \left[2 g^2+2 \lambda ^2+\sqrt{\lambda ^2 \left(8 g^2+\lambda ^2\right)}\right]^\frac{1}{2}$} & 2\; (1) & 0 & 0 & 0 & $2$ \\
 \hline
 {\small   $\pm\sqrt{2} \left[8 g^2+5 \lambda ^2-\sqrt{\lambda ^2 \left(64 g^2+9 \lambda ^2\right)}\right]^\frac{1}{2}$ }& $6\; (1)$ & 0 & 0 & 0 & $6$ \\
 \hline
 {\small    $\pm\sqrt{2} \left[8 g^2+5 \lambda ^2+\sqrt{\lambda ^2 \left(64 g^2+9 \lambda ^2\right)}\right]^\frac{1}{2}$ }& $6\; (1)$ & 0 & 0 & 0 & $6$ \\
 \hline
 {\small    $\pm\sqrt{2} \left[4 g^2+7 \lambda ^2-\sqrt{16 g^4+40 \lambda ^2 g^2+\lambda ^4}\right]^\frac{1}{2}$} & $2\; (1)$ & 0 & 0 & 0 & $2$ \\
 \hline
 {\small    $\pm\sqrt{2} \left[4 g^2+7 \lambda ^2+\sqrt{16 g^4+40 \lambda ^2 g^2+\lambda ^4}\right]^\frac{1}{2}$} & $2\; (1)$ & 0 & 0 & 0 & $2$ \\
 \hline
 {\small    $\pm 2 \sqrt{2} \left[g^2+2 \lambda ^2-\sqrt{g^4+2 \lambda ^2 g^2+4 \lambda ^4}\right]^\frac{1}{2}$} & $1\; (1)$ & 0 & 0 & 0 & $1$ \\
 \hline
 {\small    $ \pm  2 \sqrt{2} \left[g^2+2 \lambda ^2+\sqrt{g^4+2 \lambda ^2 g^2+4 \lambda ^4}\right]^\frac{1}{2}$ }& $1\; (1)$ & 0 & 0 & 0 & $1$ \\
 \hline
 $\pm \sqrt{R_1[Q_1(\alpha)]}$ & $1\; (1)$ & 0 & 0 & 0 & $1$ \\
 \hline
 $\pm \sqrt{R_2[Q_1(\alpha)]}$ & $1\; (1)$ & 0 & 0 & 0 & $1$ \\
 \hline
 $\pm \sqrt{R_3[Q_1(\alpha)]}$ & $1\; (1)$ & 0 & 0 & 0 & $1$ \\
 \hline
\end{tabular}
\caption{The spectrum in different subsectors for the case $\lambda_r = \lambda_g = \lambda_b = \lambda$ of four-site Abelian KT chain model. The number in parenthesis denotes the degeneracy of the corresponding states.} 
\label{tab:symm-4-site-KT-a}
\end{table}

\subsection{Spectral properties of the four-site Abelian KT chain model}

\subsubsection{Band diagrams for eigenvalues}

\begin{figure}[htp]

\subfloat[]{\includegraphics[width=3.1in]{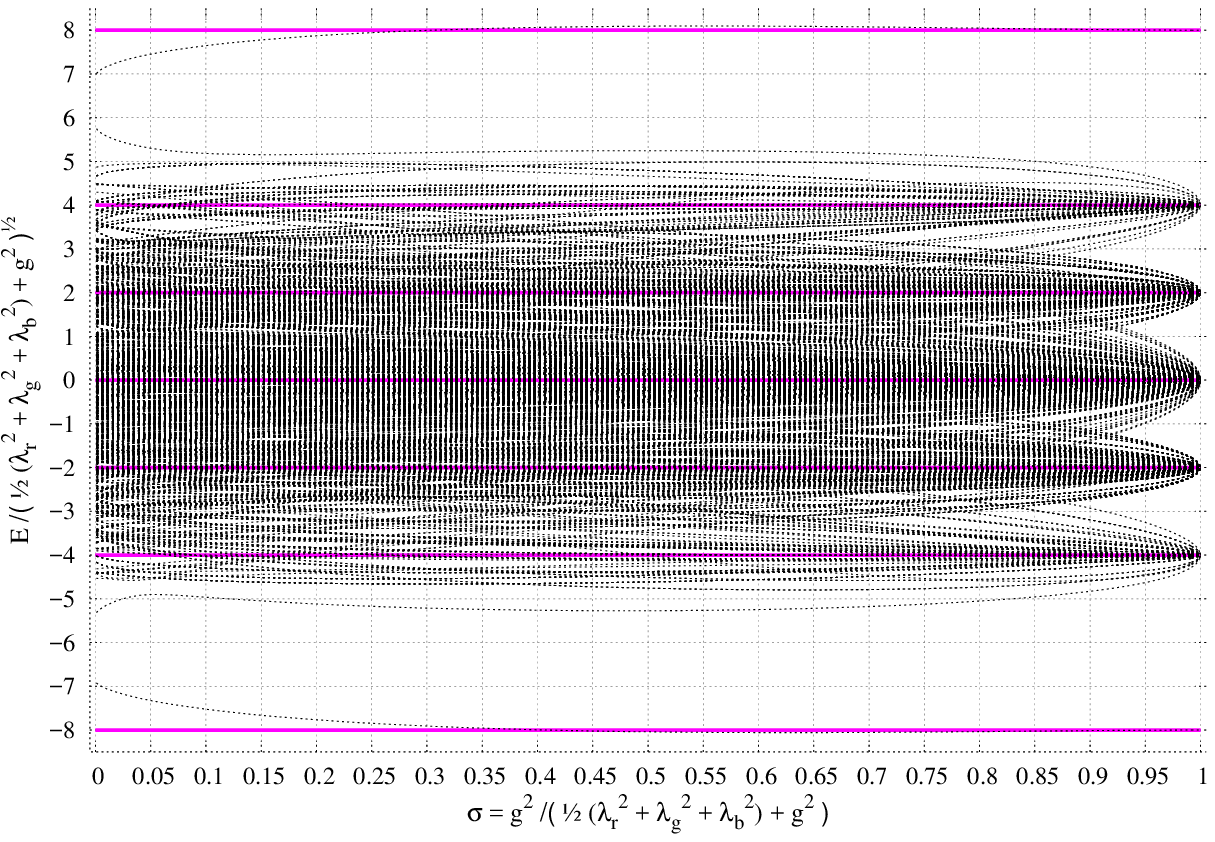}}
\subfloat[]{\includegraphics[width=3.1in]{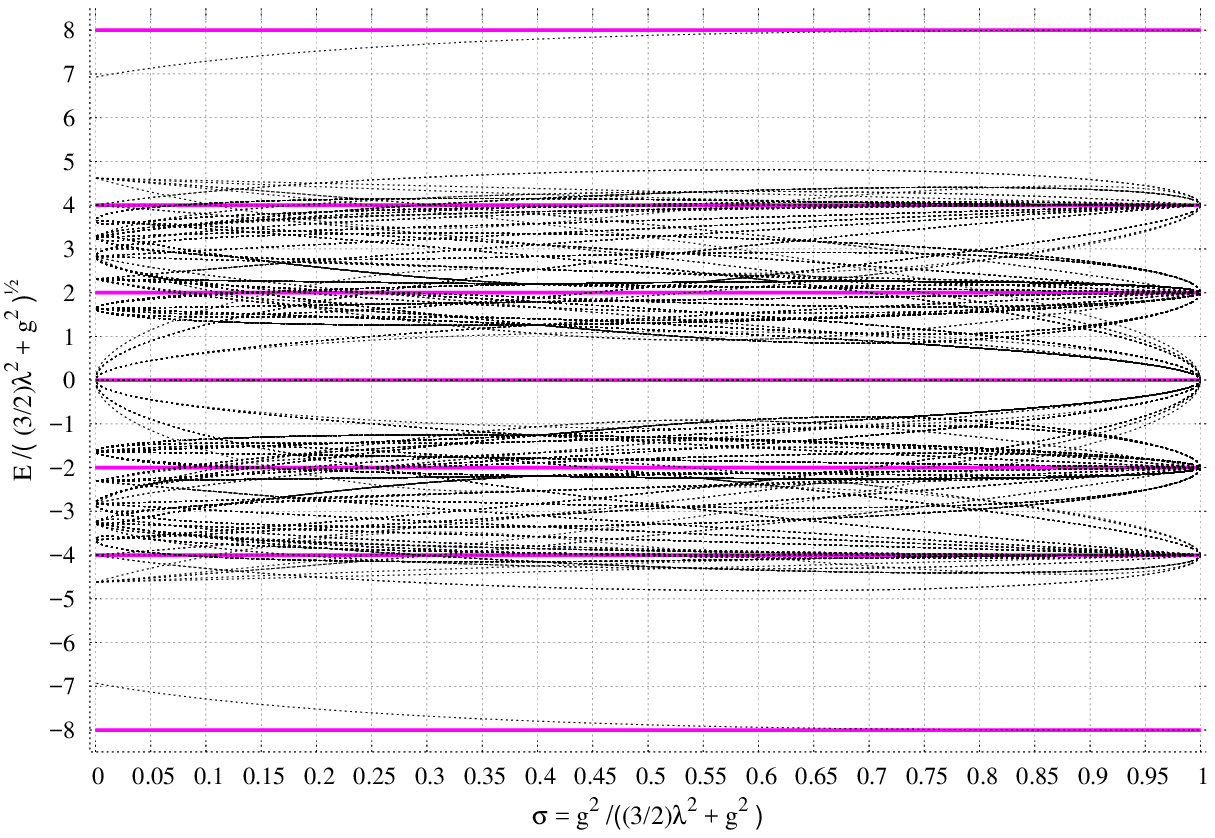}}

\caption{(a) $U(1)^3$ singlet spectrum :  Rescaled eigenvalues $E/\sqrt{\hf(\lambda_r^2 + \lambda_g^2 + \lambda_b^2) + g^2}$ of the four-site Abelian KT chain model Hamiltonian with asymmetric hopping couplings against the coupling ratio $\sigma$ (degeneracies not shown). Here $\sigma \equiv \frac{g^2}{\hf(\lambda_r^2 + \lambda_g^2 + \lambda_b^2) + g^2}$. (b) $U(1)^3$ singlet spectrum : Rescaled eigenvalues $E/\sqrt{\frac{3}{2}\lambda^2 + g^2}$ of the four-site Abelian KT chain model Hamiltonian with symmetric hopping couplings against the coupling ratio $\sigma$(degeneracies not shown). Here $\sigma \equiv \frac{g^2}{\frac{3}{2}\lambda^2 + g^2}$.}
\label{fig:band-diag-4-site-KT-model}
\end{figure}

In Fig.~\ref{fig:band-diag-4-site-KT-model}, we show the band diagrams of rescaled eigenvalues $E/\sqrt{\hf(\lambda_r^2 + \lambda_g^2 + \lambda_b^2) + g^2}$ of the four-site Abelian KT chain model Hamiltonian with symmetric and asymmetric hopping couplings against the coupling ratio $\sigma \equiv \frac{g^2}{\hf(\lambda_r^2 + \lambda_g^2 + \lambda_b^2) + g^2}$.

We see that when the couplings of the hopping terms vanish (that is, in the limit $\lambda_r = \lambda_g = \lambda_b = 0$ for the asymmetric case, and $\lambda = 0$ in the symmetric case) the bands collapse to give one level with energy $0$ and degeneracy $1056$; two levels with energies $2g$ and $-2g$, each with degeneracy $288$; two levels with energies $4g$ and $-4g$, each with degeneracy $88$; and two more nondegenerate levels with energies $8g$ and $-8g$. 

\subsubsection{Cumulative spectral function, level spacing distribution and $r$-parameter statistics}

\begin{figure}[htp]

\includegraphics[width=4.0in]{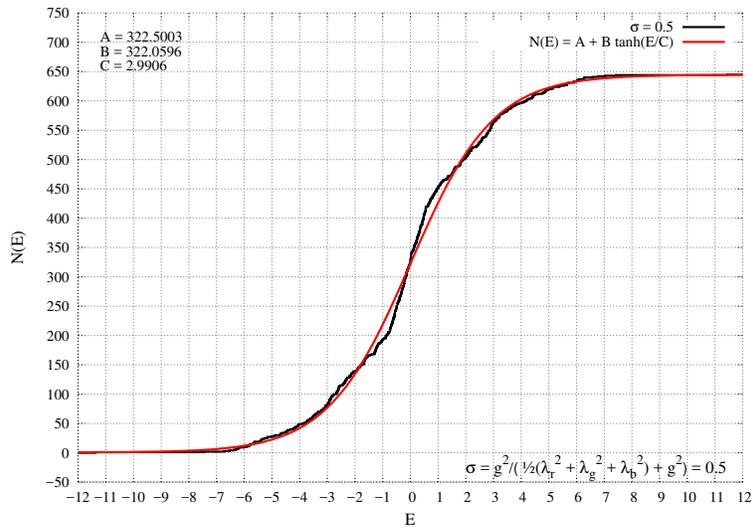}

\caption{The spectral staircase function $N(E)$ of the four-site Abelian KT chain model for asymmetric case of the hopping couplings. The plot is for $\sigma = 0.5$, which corresponds to $\lambda_r/g = 0.8255, \lambda_g/g = 0.3005$ and $\lambda_b/g = 1.1083$. Note that here $E$ is measured in units of the on-site coupling $g$.}
\label{fig:spec-staircase-4-site-KT-model}

\end{figure}

In Fig.~\ref{fig:spec-staircase-4-site-KT-model}, we show the spectral staircase function $N(E)$ of the four-site Abelian KT chain model for asymmetric case of the hopping couplings. The plot is for $\sigma = 0.5$, which corresponds to $\lambda_r/g = 0.8255, \lambda_g/g = 0.3005$ and $\lambda_b/g = 1.1083$. Note that here $E$ is measured in units of the on-site coupling $g$.

\begin{figure}[htp]

\subfloat[]{\includegraphics[width=3.1in]{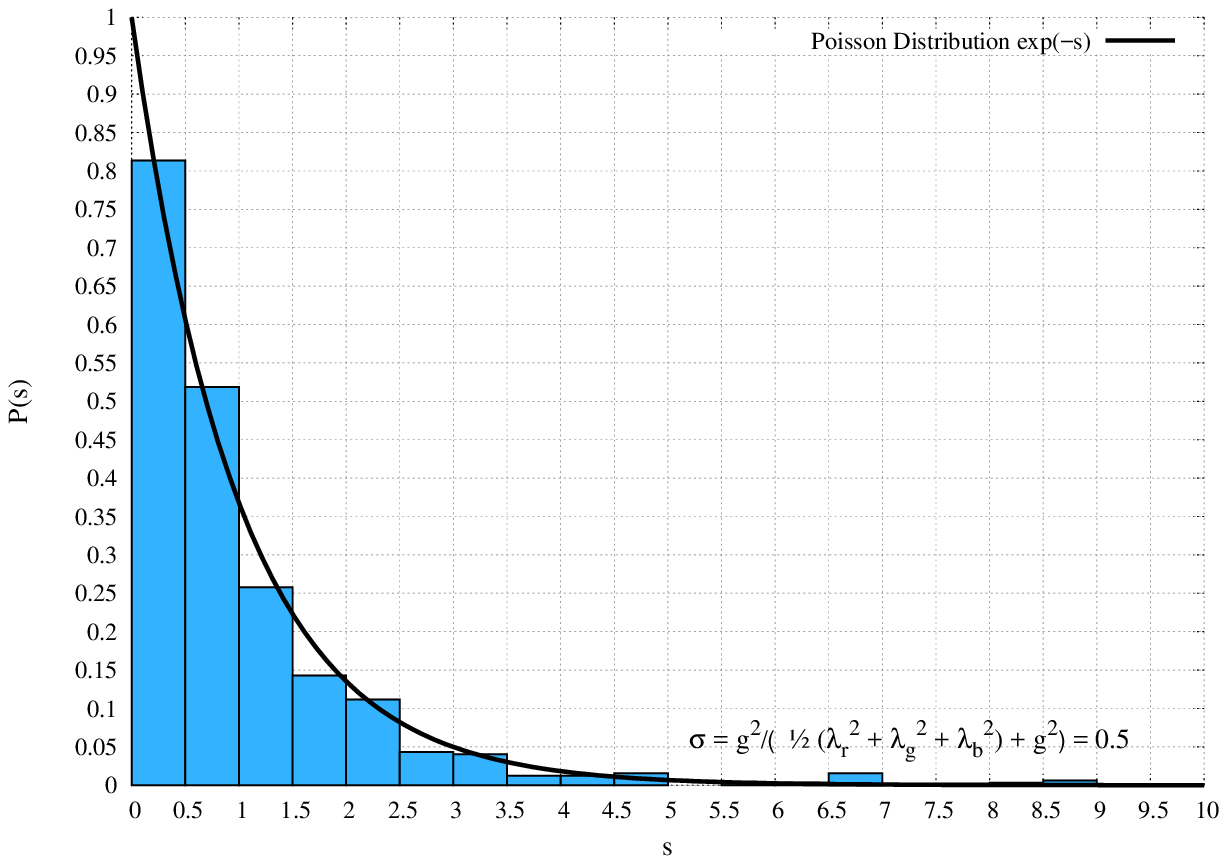}}
\subfloat[]{\includegraphics[width=3.1in]{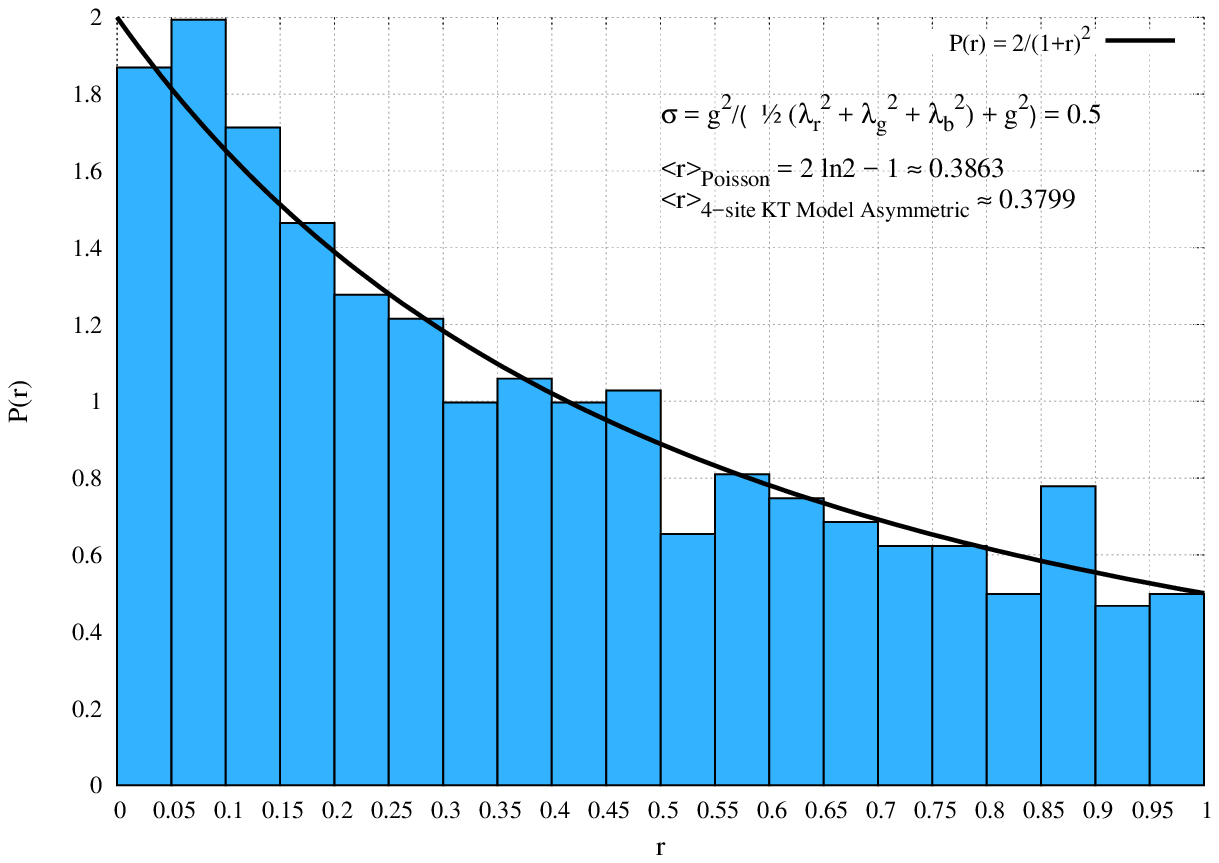}}

\caption{(a) The histogram of nearest neighbor spacing distribution $P(s)$ against $s$ for the four-site Abelian KT chain model with asymmetric hopping couplings. (b) The histogram of $r$-parameter distribution $P(r)$ against $r$ for the four-site Abelian KT chain model with asymmetric hopping couplings. In both cases the plots are for the coupling ratio $\sigma = 0.5$, which corresponds to $\lambda_r/g = 0.8255, \lambda_g/g = 0.3005$ and $\lambda_b/g = 1.1083$; and the fits are to Poisson distribution.}
\label{fig:s-Ps-r-Pr-4-site-KT-model}

\end{figure}

In Fig.~\ref{fig:s-Ps-r-Pr-4-site-KT-model} (left) we show the histogram of nearest neighbor spacing distribution $P(s)$ against $s$ for the four-site Abelian KT chain model with asymmetric hopping couplings. In Fig.~\ref{fig:s-Ps-r-Pr-4-site-KT-model} (right) we show the histogram of $r$-parameter distribution $P(r)$ against $r$ for the four-site Abelian KT chain model with asymmetric hopping couplings. In both cases the plots are for the coupling ratio $\sigma = 0.5$, which corresponds to $\lambda_r/g = 0.8255, \lambda_g/g = 0.3005$ and $\lambda_b/g = 1.1083$; and the fits are to Poisson distribution. From figure~\ref{fig:s-Ps-r-Pr-4-site-KT-model} it is evident that the model exhibits the characteristics of an integrable system.

\subsection{Comments}

\begin{itemize}

\item We see from the graphs of the unfolded level spacing distribution and the $r$-parameter statistics that the four-site KT chain has a spectrum that shows a more clear quasi many body localized behavior than the three-site chain.

\item The spectrum in this case has a much larger degeneracy in the middle when the hopping couplings are symmetric. In fact, more than half of the states ($938$ out of $1810$ singlet states) are exactly degenerate with zero energy in the symmetric hopping case. This should be contrasted with the generic asymmetric hopping where only $107$ states are degenerate. Both these degeneracies are a dramatic demonstration of the lack of level repulsion in these models.

\item The above behavior is broadly similar to the two-site and the three-site cases, except for the huge degeneracies. We expect a very fast growing degeneracy in the middle part of the spectrum to persist in the case of even number of sites with symmetric hopping. 

\item There are five protected states described in Appendix~\ref{app:4gprotected} with energy $4g$; three of these are in the $(+, +, +, +)$ subsector; 1 is in the $(+, -, +, -)$ subsector and the remaining one is in the $(-, +, -, +)$ subsector. Similarly, there are protected states with energy $-4g$ distributed in the different subsectors in a similar way.

\item In addition to the above protected states, in the case where the couplings of the hopping terms are equal, we have six protected states with energy $2g$ and six protected states with energy $-2g$. Out of the six states with energy $2g$, three are in the $(+, -, +, -)$ subsector and the remaining three are in the $(-, +, -, +)$ subsector. The distribution of the states with energy $-2g$ into the different subsectors is similar.

\end{itemize}

\section{Spectral Properties of $5$-site Abelian KT Chain}
\label{sec:KT5site}

In the five-site chain, the Hamiltonian is 
\begin{equation}
H = g (H_a^{(0)} +  H_b^{(0)}  +   H_c^{(0)}+ H_d^{(0)}+ H_e^{(0)})  + \sum_{\col\in \{r,g,b\} }\lambda_{\col}\left[ H_{ab}^{(\col)}+H_{bc}^{(\col)}+H_{cd}^{(\col)}+H_{de}^{(\col)}+H_{ea}^{(\col)} \right]\ .
\end{equation}

\begin{figure}[htp]

\centering
\includegraphics[width=4.0in]{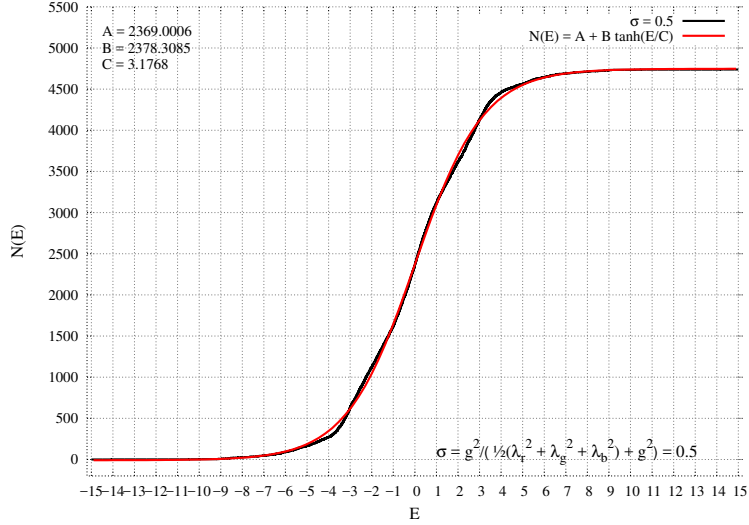}
\caption{The spectral staircase function $N(E)$ of the five-site Abelian KT chain model for asymmetric case of the hopping couplings. The plot is for $\sigma = 0.5$, which corresponds to $\lambda_r/g = 0.8255, \lambda_g/g = 0.3005$ and $\lambda_b/g = 1.1083$. Note that here $E$ is measured in units of the on-site coupling $g$.}
\label{fig:spec-staircase-5-site-KT-model}

\end{figure}

\begin{figure}[htp]

\subfloat[]{\includegraphics[width=3.1in]{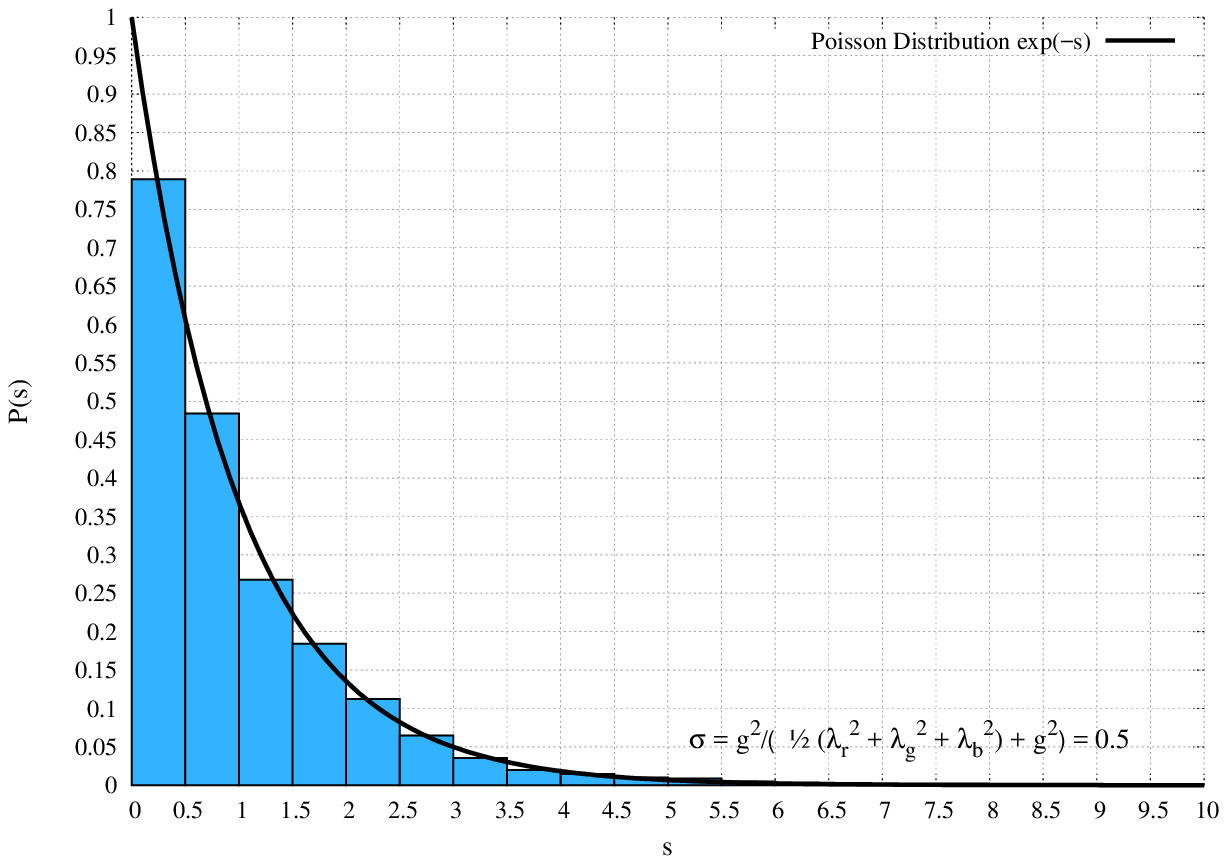}}
\subfloat[]{\includegraphics[width=3.1in]{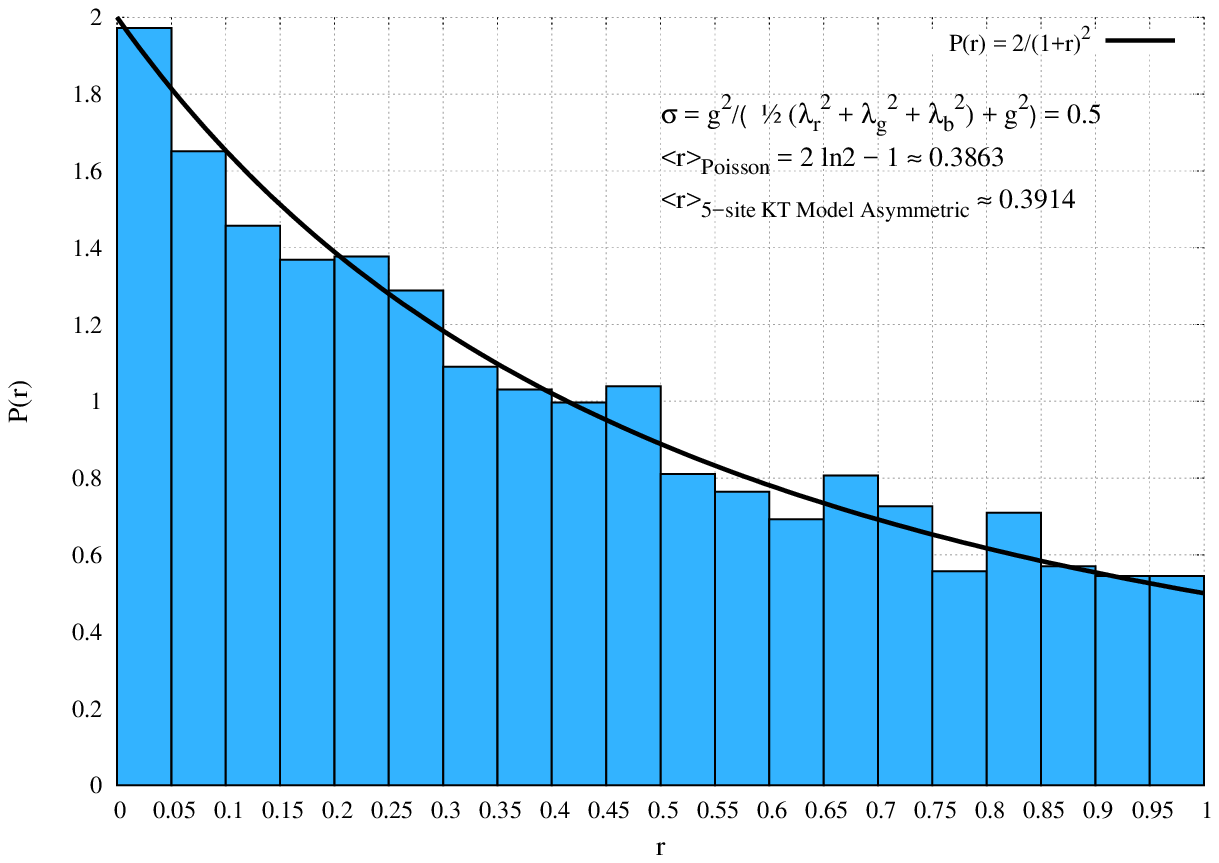}}

\caption{(a) The histogram of nearest neighbor spacing distribution $P(s)$ against $s$ for the five-site Abelian KT chain model with asymmetric hopping couplings. (b) The histogram of $r$-parameter distribution $P(r)$ against $r$ for the five-site Abelian KT chain model with asymmetric hopping couplings. In both cases the plots are for the coupling ratio $\sigma = 0.5$, which corresponds to $\lambda_r/g = 0.8255, \lambda_g/g = 0.3005$ and $\lambda_b/g = 1.1083$; and the fits are to Poisson distribution.}
\label{fig:s-Ps-r-Pr-5-site-KT-model}

\end{figure}

There are $21,252$ states in the singlet sector. We note that when the hopping couplings are unequal, there are no states at zero energy as expected from an odd site KT chain.  In Fig.~\ref{fig:spec-staircase-5-site-KT-model}, we show the spectral staircase function $N(E)$ of the five-site Abelian KT chain model for asymmetric case of the hopping couplings. The plot is for $\sigma = 0.5$, which corresponds to $\lambda_r/g = 0.8255, \lambda_g/g = 0.3005$ and $\lambda_b/g = 1.1083$. Note that here $E$ is measured in units of the on-site coupling $g$. It is clear from the $r$-parameter statistics (shown in Fig.~\ref{fig:s-Ps-r-Pr-5-site-KT-model}) that this model is close to being quasi-many-body localized.

We see from the graphs of the unfolded level spacing distribution and the $r$-parameter statistics that the five-site chain has a spectrum that shows a more clear quasi-many-body localized behavior than the three-site and the four-site chain. We expect this qMBL behavior to become even clearer in larger number of sites implying that as $L \rightarrow \infty $ this model is indeed quasi-many-body localized. 

\section{Spectral Form Factors and Thermodynamics of KT Model}
\label{sec:spectra}

\subsection{Spectral form factors}
\label{sec:Spectral form factor}

The spectral form factor was proposed in~\cite{Papadodimas:2015xma} to study the black hole information paradox, and it has been extensively used in measuring late-time discrete spectrum and in capturing the random matrix behavior of systems. (e.g. SYK model~\cite{Cotler:2016fpe}, tensor model~\cite{Krishnan:2016bvg, Krishnan:2017ztz}, 2D CFT~\cite{Dyer:2016pou} and D1-D5 system~\cite{Balasubramanian:2016ids}). This simple quantity could reveal the random matrix behavior of the SYK model at late times. The spectral form factors in that case can come with a dip, ramp and plateau~\cite{Cotler:2016fpe}. At earlier times, the spectral form factor decreases until it reaches the minimum value at the dip, at a time $t_d$, which is followed by a linear growth, the so-called ramp, until it reaches a plateau at a time $t_p$. In large $N_{\text{\tiny SYK}}$, the ramp and plateau can be easily distinguishable since the ratio of the plateau and dip time ${t_p\over t_d}$ is exponentially large (i.e., $e^{N_{\text{\tiny SYK}}\over 2}$). However, in $N=2$ KT model, the ratio of those two times is not large enough [i.e., ${t_p\over t_d}\sim \exp[ {2^3\over 2}]\sim \mathcal{O}(10^2)$], and it would be difficult to capture the clear linear growth between $t_d$ and $t_p$.

The spectral form factor is defined by
\beq
f(\beta,t) = \Big| \frac{Z(\beta+it)}{Z(\beta)} \Big|^2\ ,
\eeq
where $Z(\beta) = \Tr(e^{-\beta H})$ is the partition function (Here, $\beta = \frac{1}{k_B T}$ is the inverse temperature, and we consider the trace only over states in the singlet sector). It can be understood as analytic continuation of the partition function. Note that the long time average of the spectral form factor is bounded below, and it saturates the bound when there is no degeneracy in each energy level~\cite{Dyer:2016pou}
\begin{equation}
\overline{f}(\beta)\equiv \lim_{t\rightarrow \infty } {1\over t} \int_{0}^{t} dt'\; g(\beta,t')\geqq {Z(2\beta)\over |Z(\beta)|^2}\ .
\end{equation}

\begin{figure}[htp]

\subfloat[3-site asymmetric]{\includegraphics[width=3.1in]{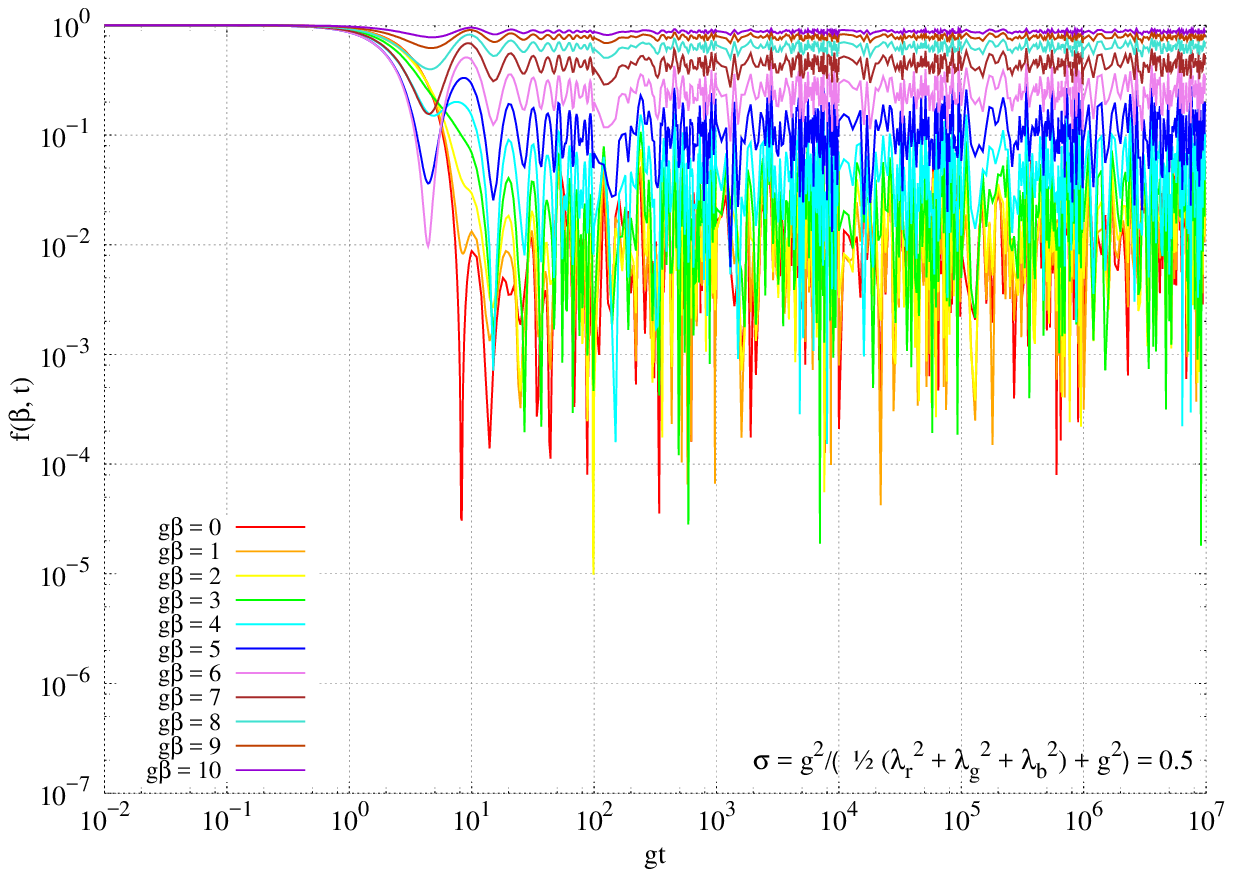}}
\subfloat[3-site symmetric]{\includegraphics[width=3.1in]{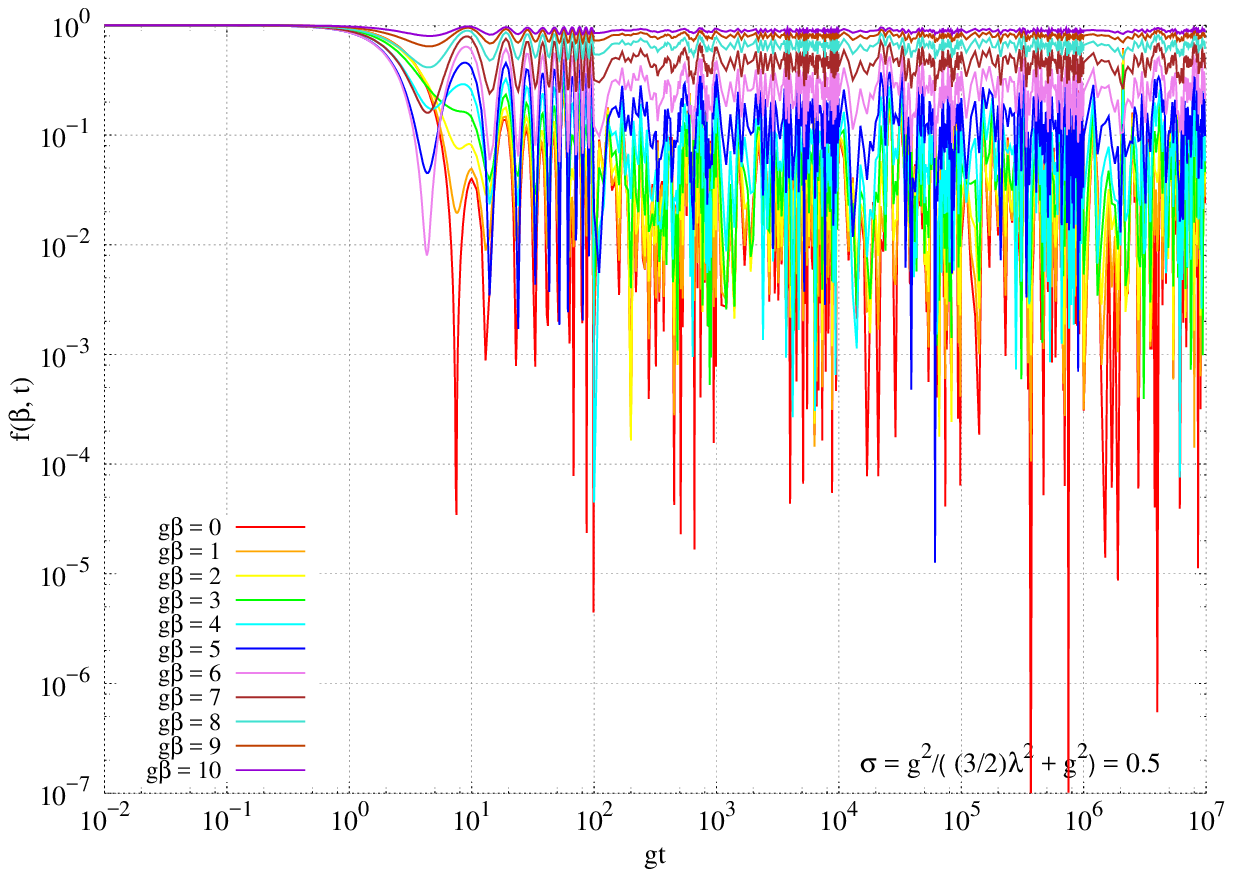}}

\caption{The spectral form factor $f(\beta, t)$ against time $gt$ for the three-site Abelian KT chain model at fixed coupling ratio $\sigma = 0.5$ and for various values of $g \beta$. Here $\sigma$ is defined as $\sigma \equiv \frac{g^2}{\hf (\lambda_r^2 + \lambda_g^2 + \lambda_b^2) + g^2}$. We take $\lambda_r/g = 0.8255, \lambda_g/g = 0.3005$ and $\lambda_b/g = 1.1083$ for the asymmetric hopping coupling case and $\lambda/g = 0.8165$ for the symmetric hopping coupling case, respectively. In both cases $g\beta$ runs from $0$ to $10$.}
\label{fig:sff-3-site-KT}

\end{figure}

%
%
\paragraph{Three-site KT Chain Model:} In Fig.~\ref{fig:sff-3-site-KT}, we show the spectral form factors $f(\beta, t)$ against time $tg$ for the three-site Abelian KT chain model at fixed coupling ratio $\sigma = 0.5$. Here $\sigma$ is the effective coupling which appears in large $N$ case
\begin{equation}
\frac{g^2}{\hf (\lambda_r^2 + \lambda_g^2 + \lambda_b^2) + g^2}\ ,
\end{equation}
for both the asymmetric and symmetric cases of the hopping couplings.

\begin{figure}[htp]

\subfloat[4-site asymmetric]{\includegraphics[width=3.1in]{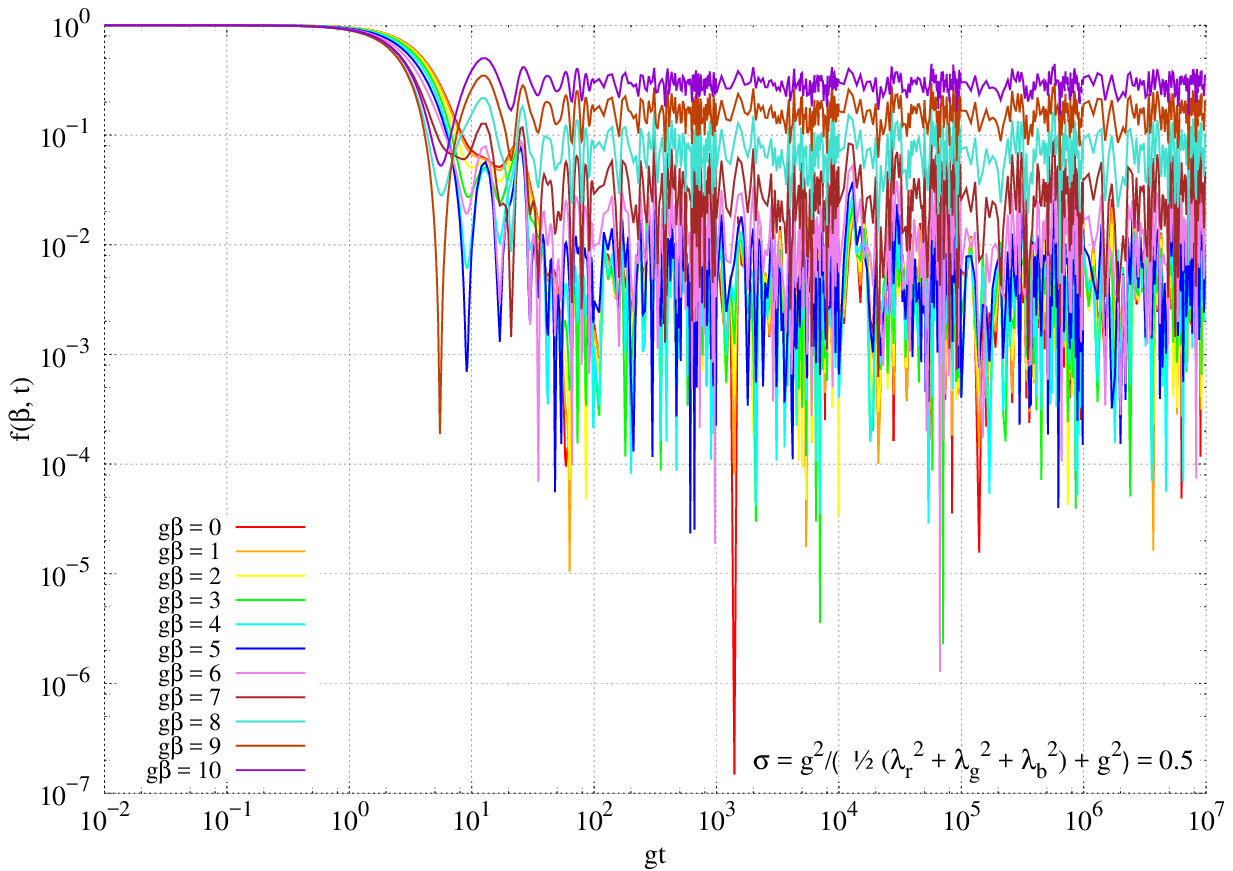}}
\subfloat[4-site symmetric]{\includegraphics[width=3.1in]{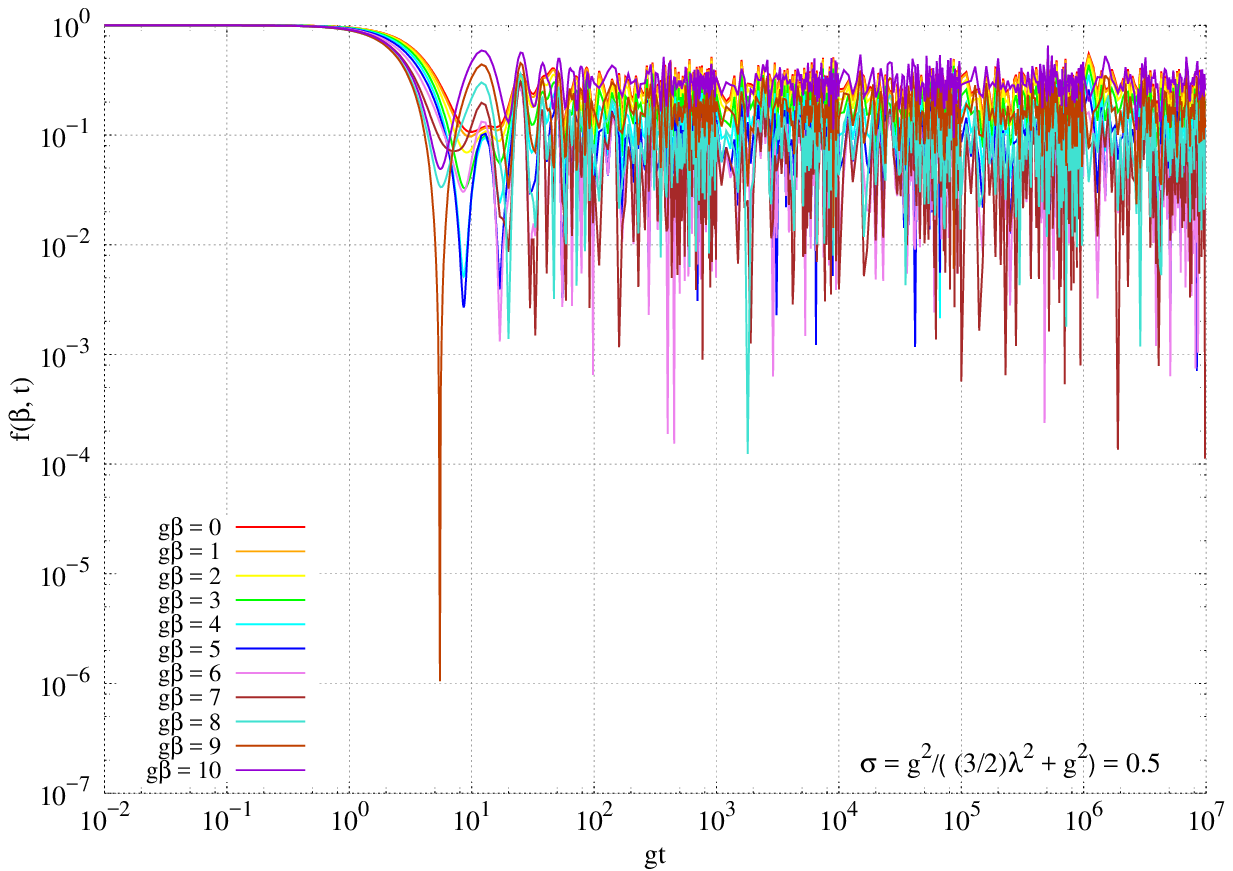}}

\caption{The spectral form factor $f(\beta, t)$ against time $gt$ for the four-site Abelian KT chain model at fixed coupling ratio $\sigma = 0.5$ and for various values of $g \beta$. Here $\sigma$ is defined as $\sigma \equiv \frac{g^2}{\hf (\lambda_r^2 + \lambda_g^2 + \lambda_b^2) + g^2}$. We take $\lambda_r/g = 0.8255, \lambda_g/g = 0.3005$ and $\lambda_b/g = 1.1083$ for the asymmetric hopping coupling case and $\lambda/g = 0.8165$ for the symmetric hopping coupling case, respectively. In both cases $g\beta$ runs from $0$ to $10$.}
\label{fig:sff-4-site-KT}

\end{figure}

\begin{figure}[htp]

\subfloat[5-site asymmetric]{\includegraphics[width=3.1in]{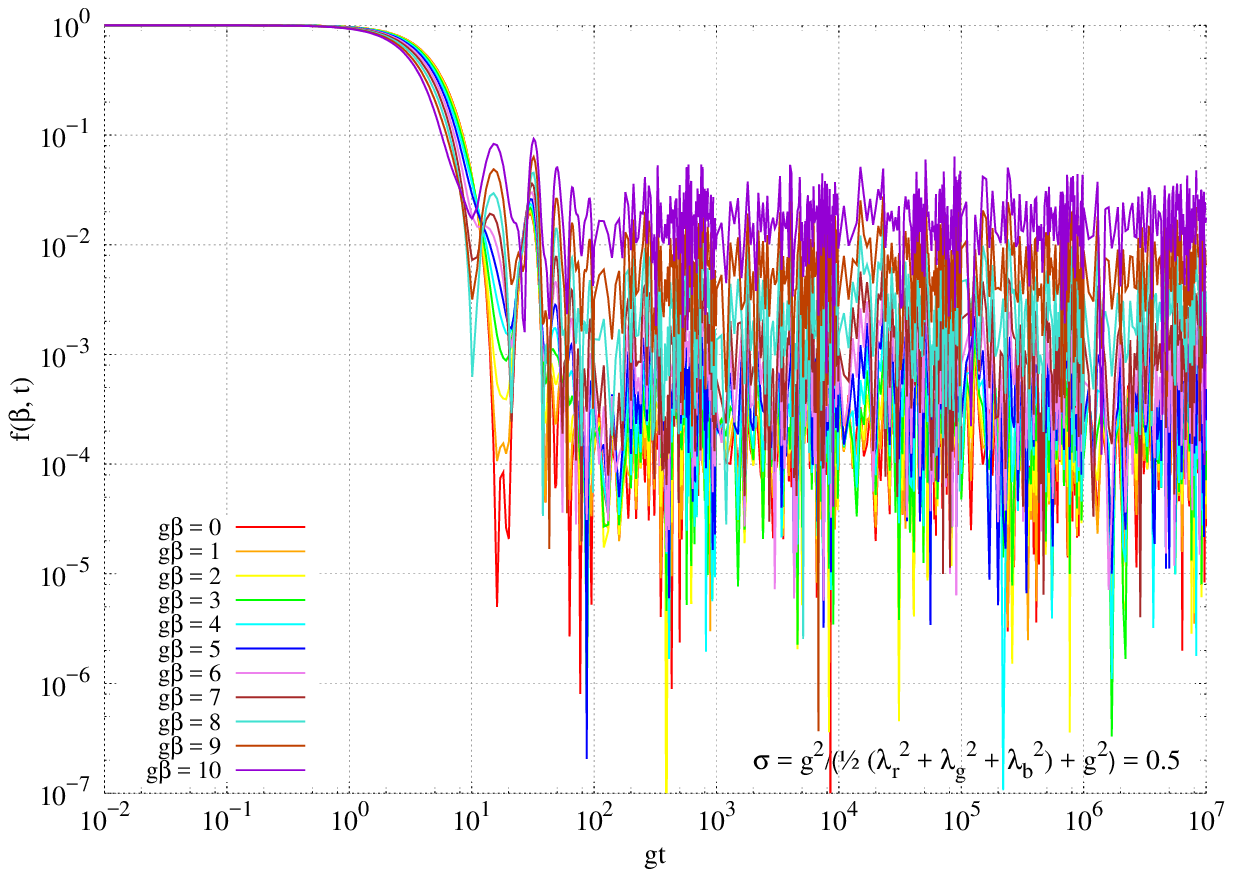}}
\subfloat[5-site symmetric]{\includegraphics[width=3.1in]{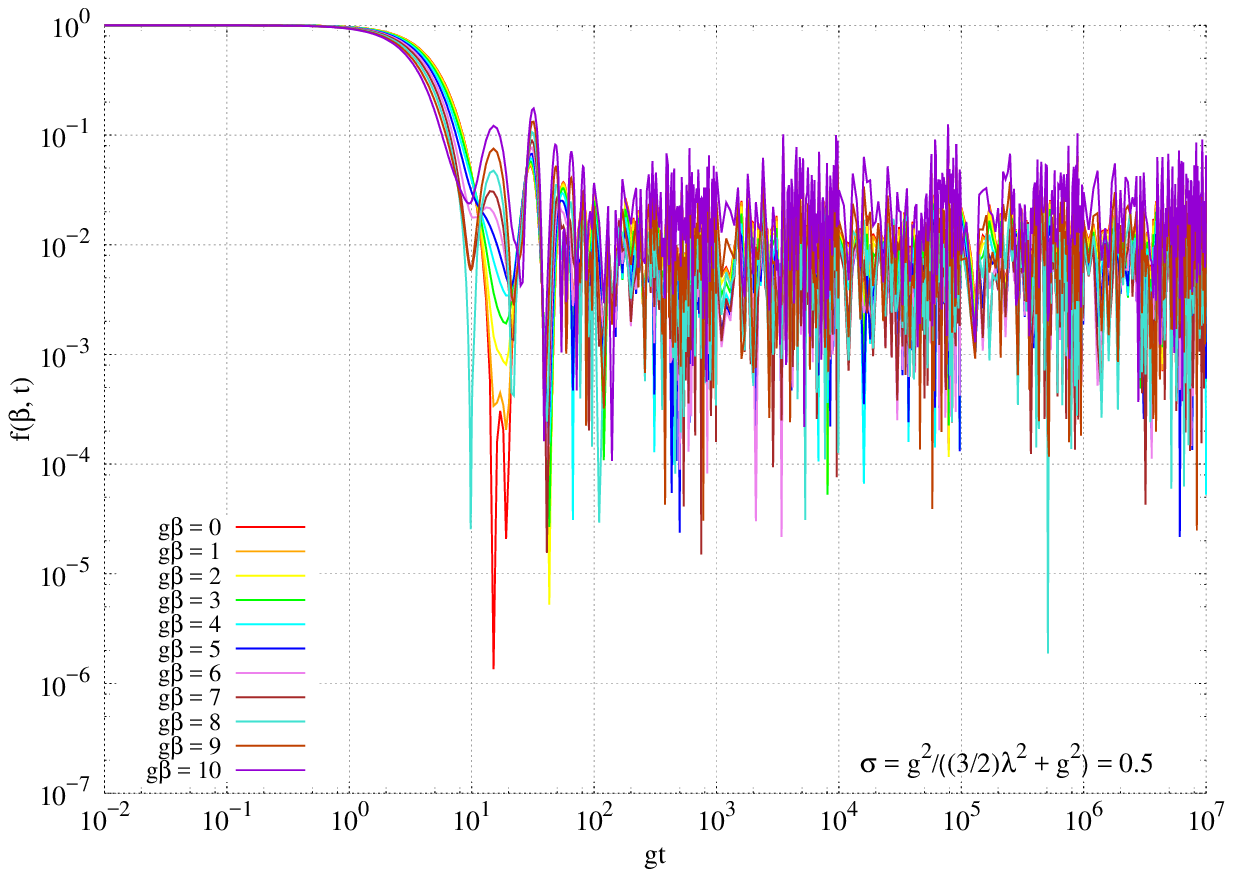}}

\caption{The spectral form factor $f(\beta, t)$ against time $gt$ for the five-site Abelian KT chain model at fixed coupling ratio $\sigma = 0.5$ and for various values of $g \beta$. Here $\sigma$ is defined as $\sigma \equiv \frac{g^2}{\hf (\lambda_r^2 + \lambda_g^2 + \lambda_b^2) + g^2}$. We take $\lambda_r/g = 0.8255, \lambda_g/g = 0.3005$ and $\lambda_b/g = 1.1083$ for the asymmetric hopping coupling case and $\lambda/g = 0.8165$ for the symmetric hopping coupling case, respectively. In both cases $g\beta$ runs from $0$ to $10$.}
\label{fig:sff-5-site-KT}

\end{figure}

The spectral from factor clearly exhibits a ballistic regime identified as the early-time plateau, a diffusive regime where the curve approaches a dip, an ergodic regime where it tries to climb back up and finally a quantum regime where it fluctuates around a mean value \cite{Liu:2016rdi}.   
\paragraph{Four-site KT Chain Model:} In Fig.~\ref{fig:sff-4-site-KT}, we show the spectral form factors $f(\beta, t)$ against time $tg$ for the $4$-site Abelian KT chain model at fixed coupling ratio $\sigma$.
\paragraph{Five-site KT Chain Model:} In Fig.~\ref{fig:sff-5-site-KT}, we show the spectral form factors $f(\beta, t)$ against time $tg$ for the five-site Abelian KT chain model at fixed coupling ratio $\sigma$.
%

\subsection{Thermodynamic properties}
\label{sec:thermodynamic properties}

\begin{figure}[htp]

\subfloat[Energy, 3-site asymmetric]{\includegraphics[width=3.1in]{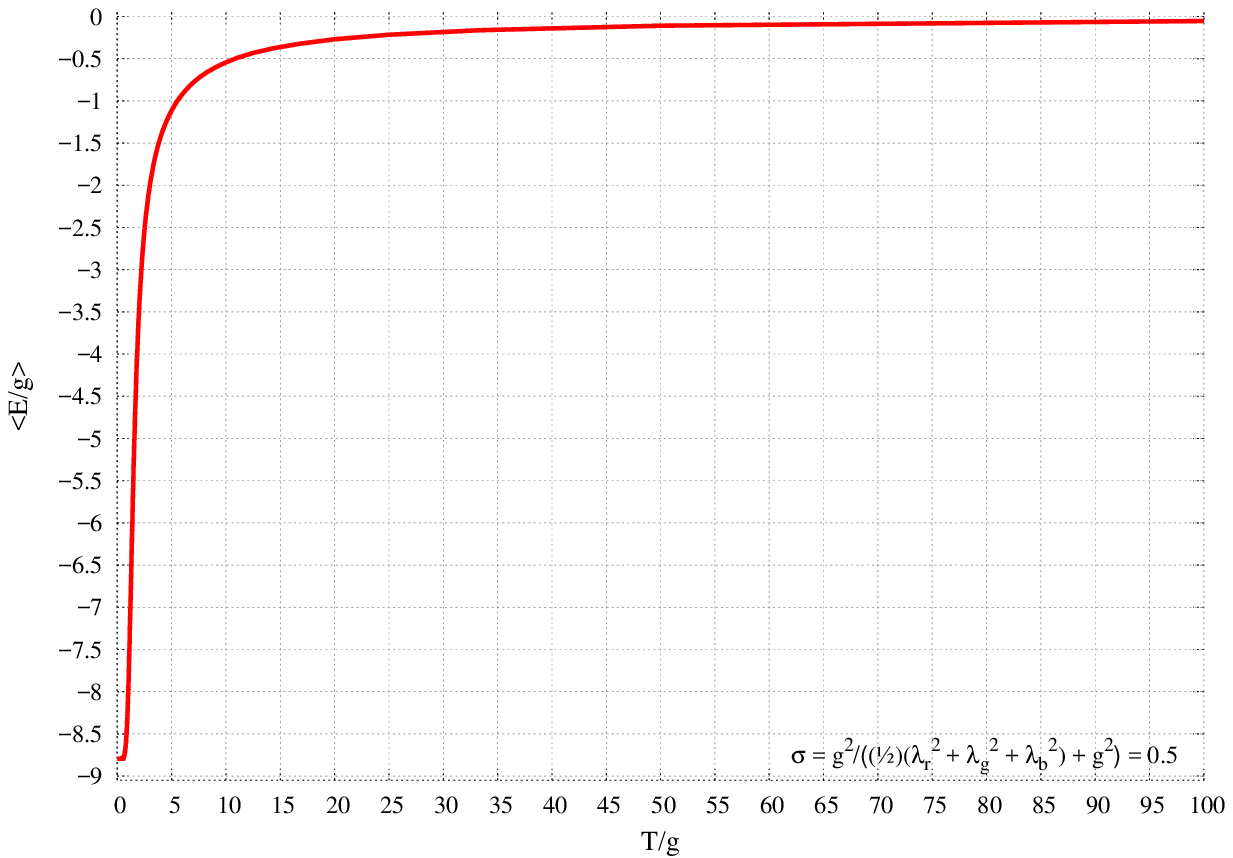}}
\subfloat[Entropy, 3-site asymmetric]{\includegraphics[width=3.1in]{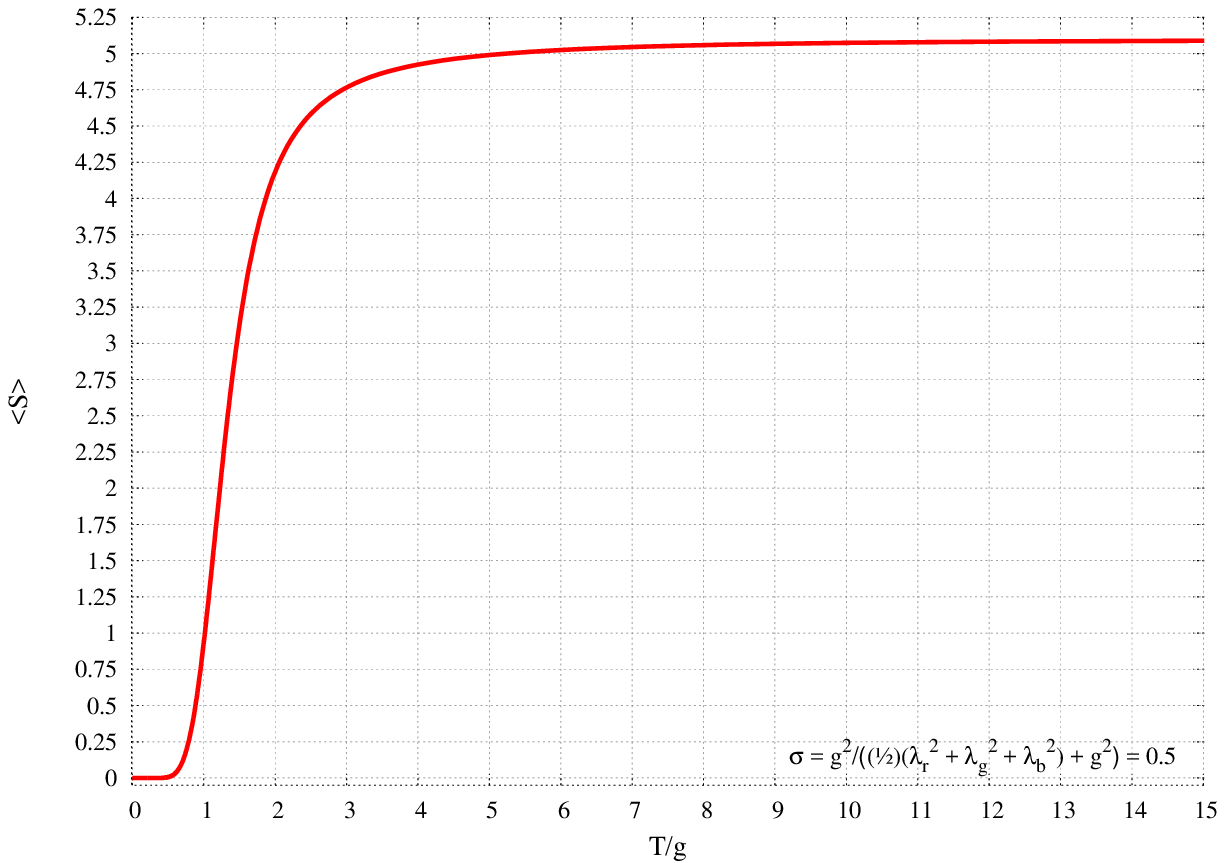}}

\caption{The plots above are for the three-site Abelian KT chain at fixed coupling ratio $\sigma = 0.5$ at asymmetric hopping couplings. Here $\sigma$ is defined as $\sigma \equiv \frac{g^2}{\hf (\lambda_r^2 + \lambda_g^2 + \lambda_b^2) + g^2}$. The ratio of the couplings are taken as $\lambda_r/g = 0.8255, \lambda_g/g = 0.3005$ and $\lambda_b/g = 1.1083$.  (a) The mean energy $\langle E/g \rangle$ against temperature $T/g$. (b) The mean entropy $\langle S \rangle$ against temperature $T/g$.}
\label{fig:3-site-KT-E-S-asym}

\end{figure}

\begin{figure}[htp]

\subfloat[Energy, 3-site symmetric]{\includegraphics[width=3.1in]{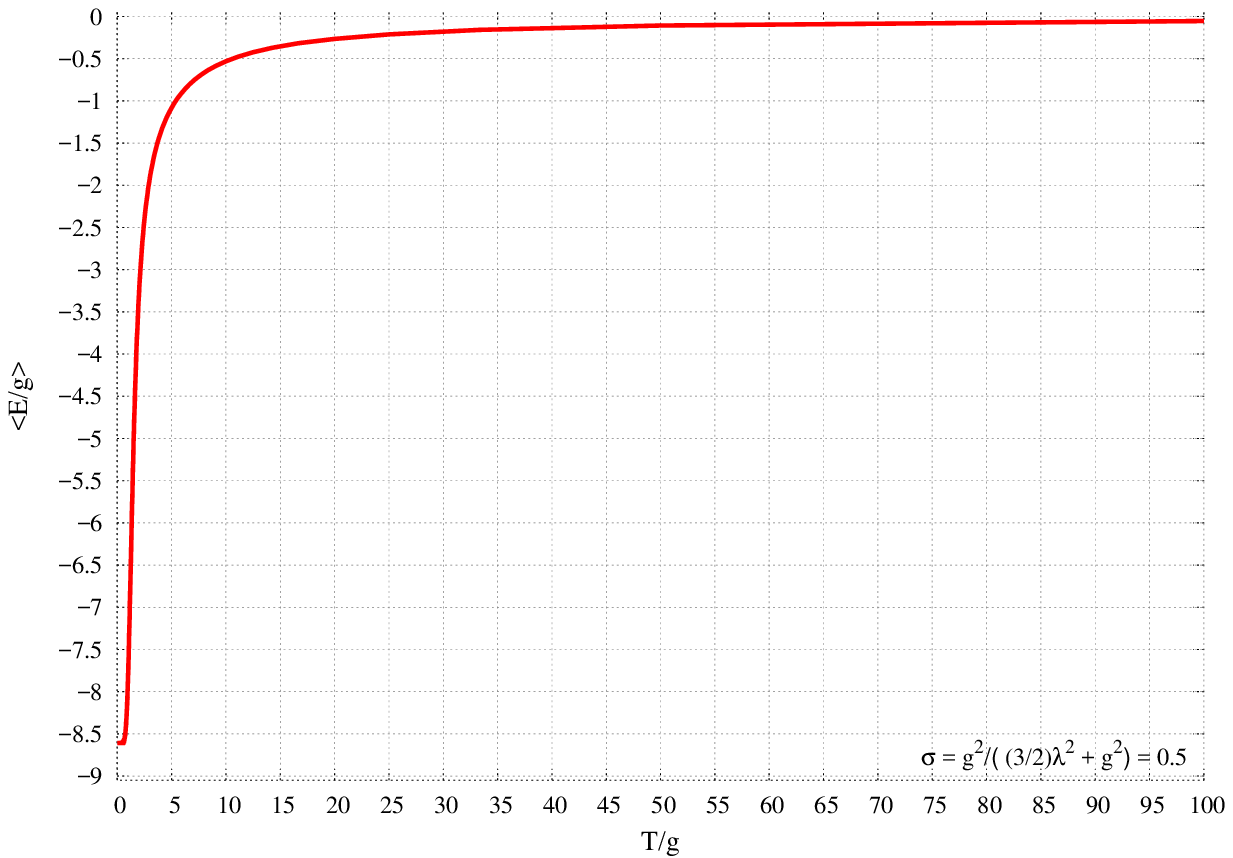}}
\subfloat[Entropy, 3-site symmetric]{\includegraphics[width=3.1in]{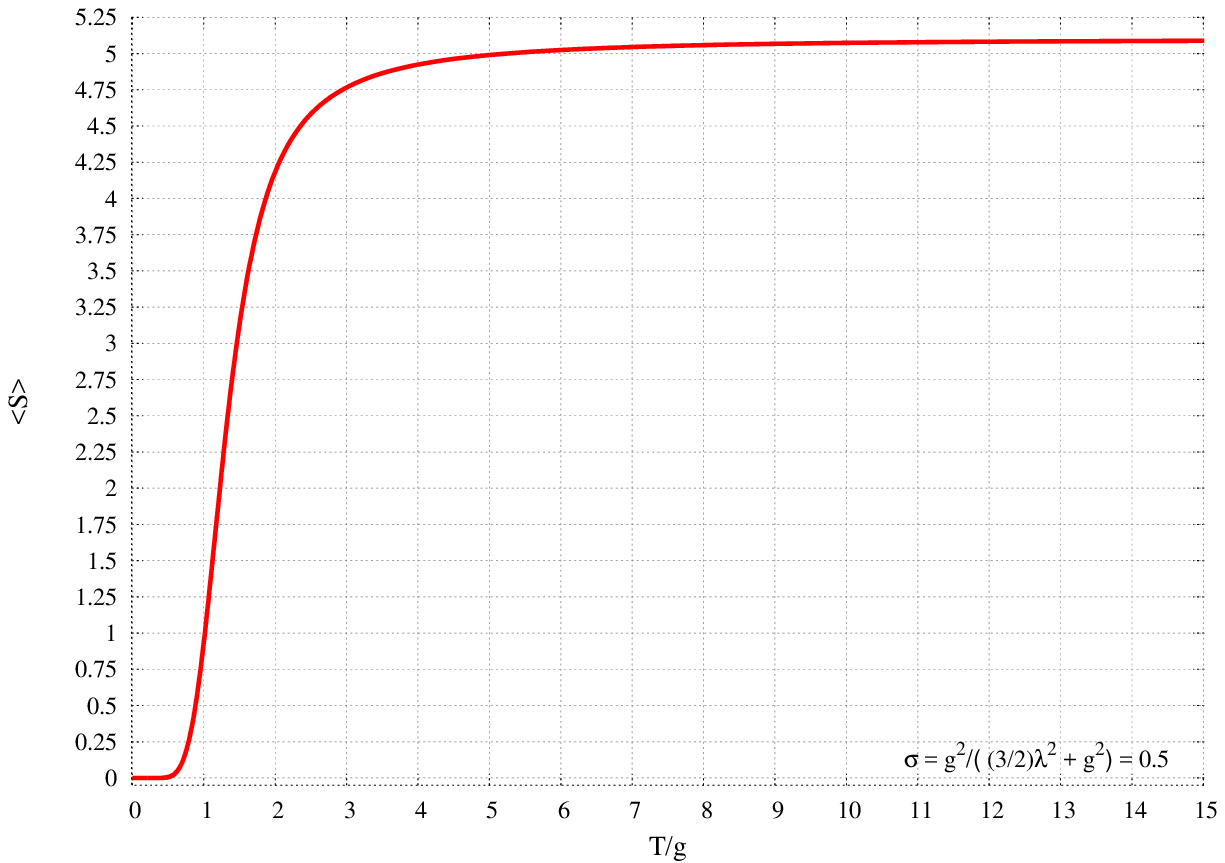}}

\caption{The plots above are for the three-site KT chain at fixed coupling ratio $\sigma = 0.5$ at symmetric hopping coupling. The ratio of the coupling $\sigma$ is defined as $\sigma \equiv \frac{g^2}{\frac{3}{2}\lambda^2 + g^2}$. This corresponds to $\lambda/g = 0.8165$. (a) The mean energy $\langle E/g \rangle$ against temperature $T/g$. (b) The mean entropy $\langle S \rangle$ against temperature $T/g$.}
\label{fig:3-site-KT-E-S-sym}

\end{figure}

\begin{figure}[htp]

\subfloat[Energy, 4-site asymmetric]{\includegraphics[width=3.1in]{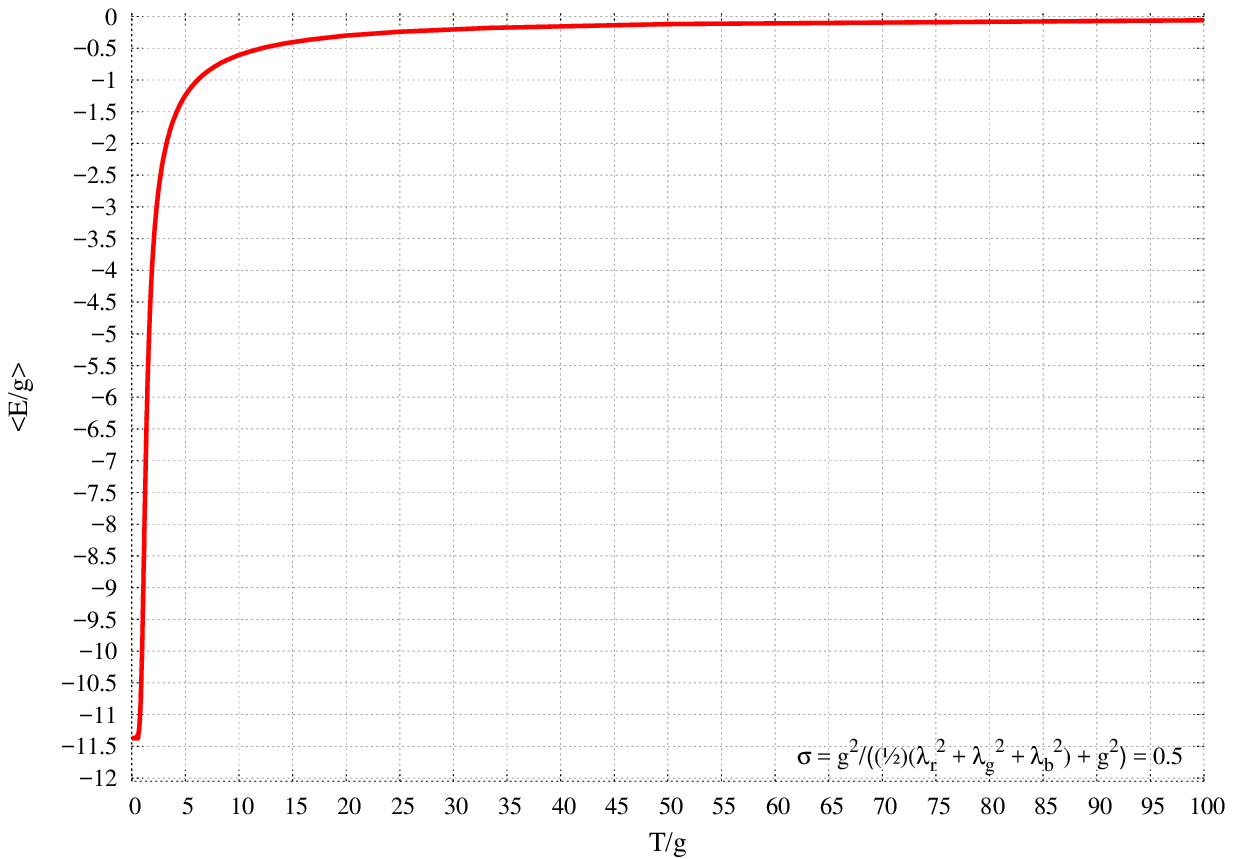}}
\subfloat[Entropy, 4-site asymmetric]{\includegraphics[width=3.1in]{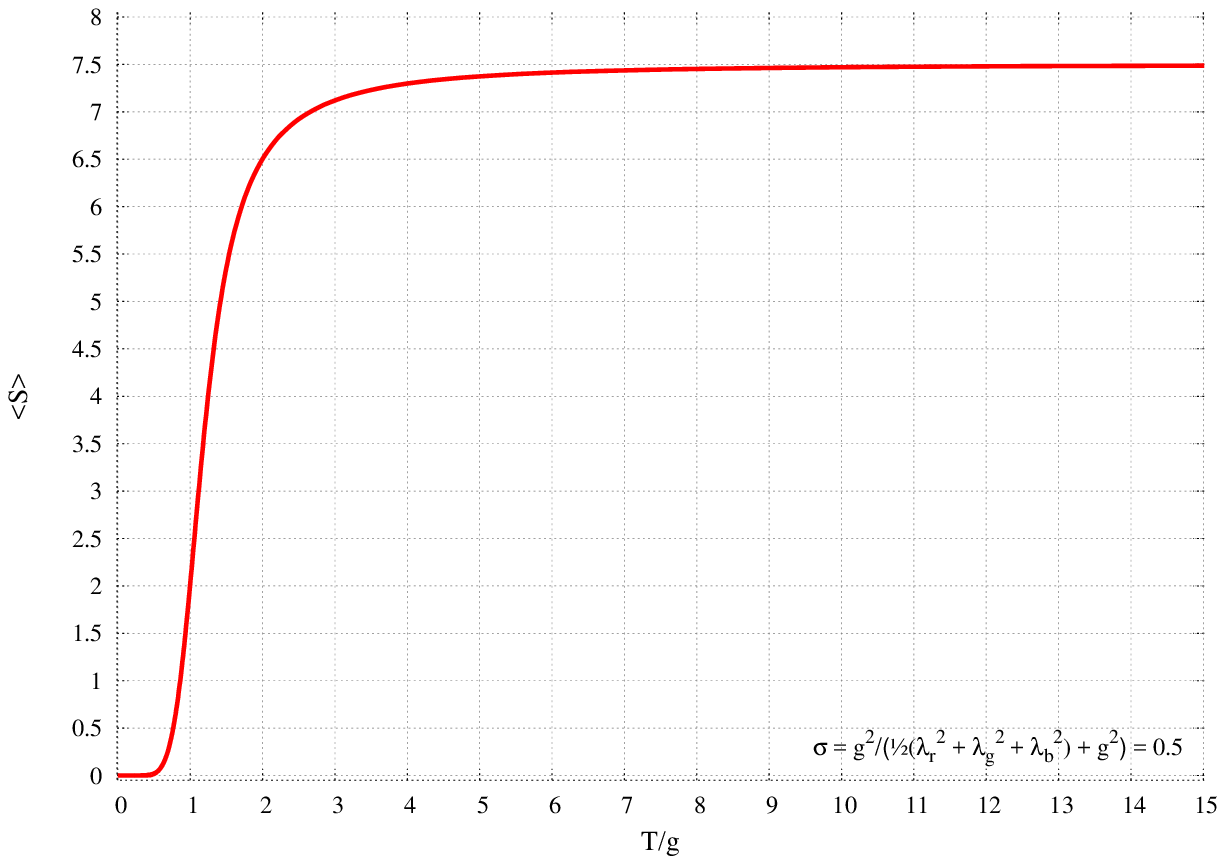}}

\caption{The plots above are for the four-site Abelian KT chain model at fixed coupling ratio $\sigma = 0.5$ at asymmetric hopping couplings. Here $\sigma$ is defined as $\sigma \equiv \frac{g^2}{\hf (\lambda_r^2 + \lambda_g^2 + \lambda_b^2) + g^2}$. The ratio of the couplings are taken as $\lambda_r/g = 0.8255, \lambda_g/g = 0.3005$ and $\lambda_b/g = 1.1083$. (a) The mean energy $\langle E/g \rangle$ against temperature $T/g$. (b) The mean entropy $\langle S \rangle$ against temperature $T/g$.}
\label{fig:4-site-KT-E-S-asym}

\end{figure}

\begin{figure}[htp]

\subfloat[Energy, 4-site symmetric]{
\includegraphics[width=3.1in]{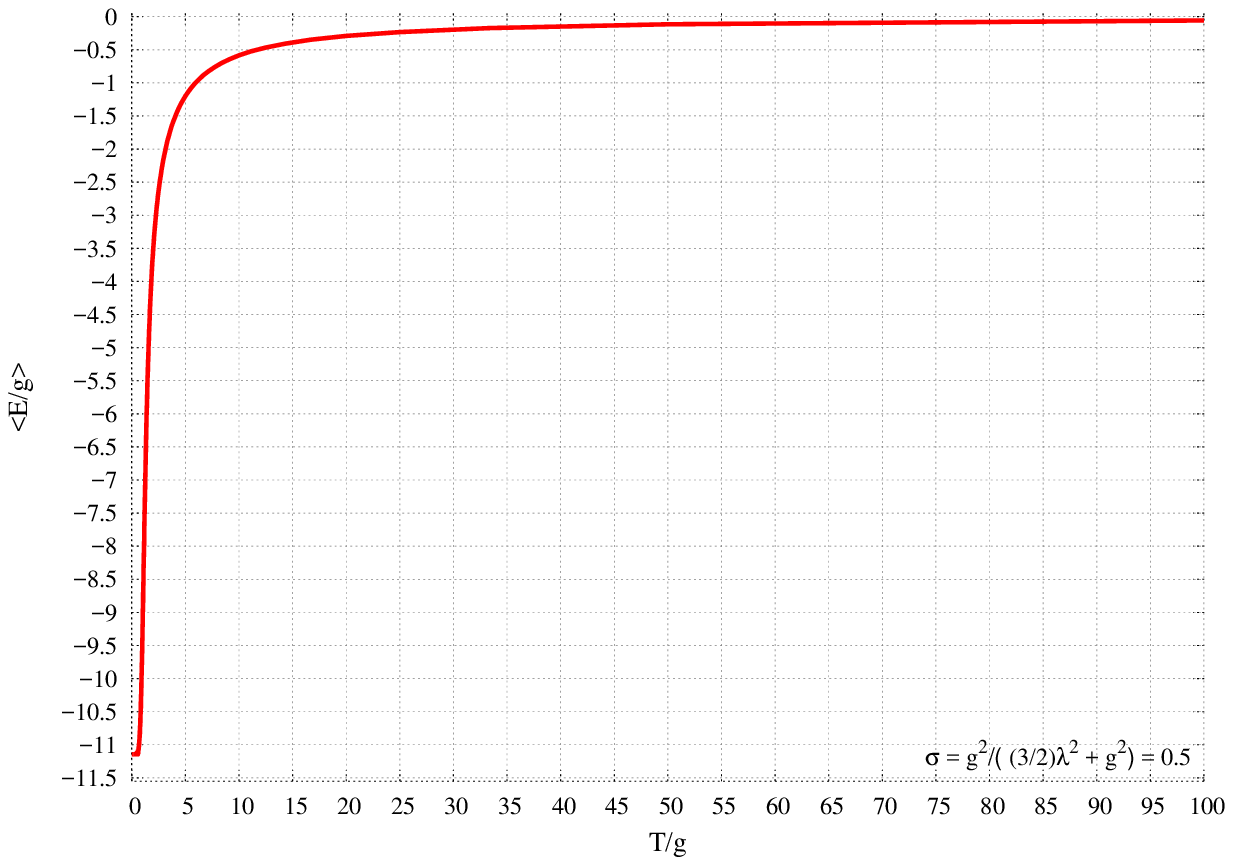}
}
\subfloat[Entropy, 4-site symmetric]{
\includegraphics[width=3.1in]{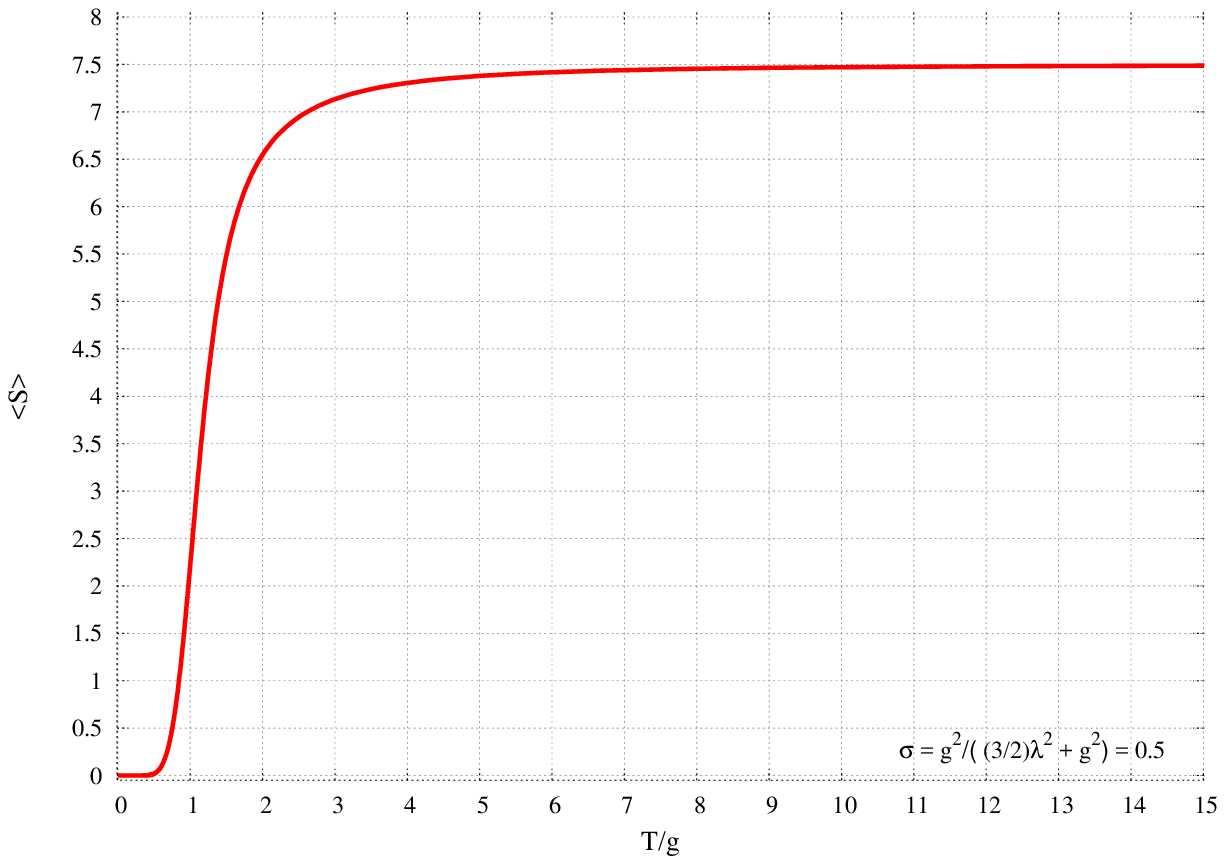}
}

\caption{The plots above are for the four-site Abelian KT chain model at fixed coupling ratio $\sigma = 0.5$ at symmetric hopping coupling. The ratio of the coupling $\sigma$ is defined as $\sigma \equiv \frac{g^2}{\frac{3}{2}\lambda^2 + g^2}$. This corresponds to $\lambda/g = 0.8165$. (a) The mean energy $\langle E/g \rangle$ against temperature $T/g$. (b) The mean entropy $\langle S \rangle$ against temperature $T/g$.}
\label{fig:4-site-KT-E-S-sym}

\end{figure}

\begin{figure}[h!]

\centering
\includegraphics[width=4in]{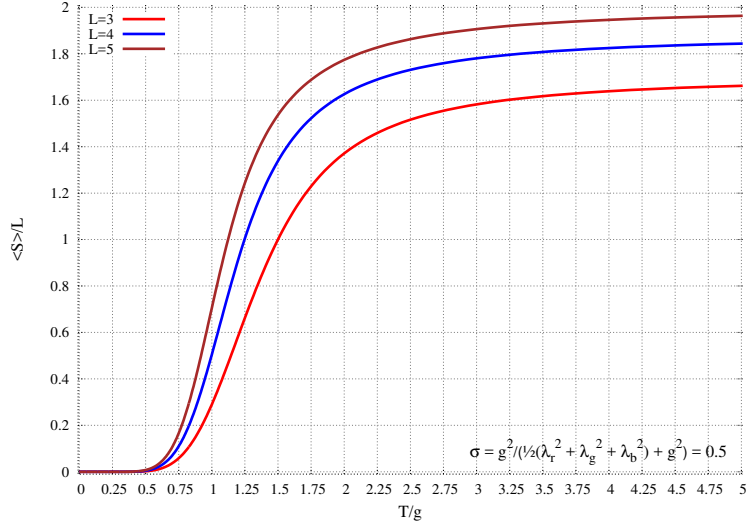}
\caption{Comparing the entropy per site of three-, four- and five-site KT chain models against temperature $T/g$. The plot is for $\sigma=0.5$, asymmetric case. The hopping couplings are $\lambda_r/g = 0.8255, \lambda_g/g = 0.3005$ and $\lambda_b/g = 1.1083$. Note that at large $T/g$ the entropy per site approaches $\log(D_L)/L$, where $D_L$ is the dimension of the Hilbert space of $L$-site Abelian KT chain model.}
\label{fig:3-4-5-site-S-T-asym}

\end{figure}

We also compute the thermodynamic quantities for the Abelian KT chain model: the mean energy, the mean entropy and the specific heat. In Figs.~\ref{fig:3-site-KT-E-S-asym},~\ref{fig:3-site-KT-E-S-sym},~\ref{fig:4-site-KT-E-S-asym} and \ref{fig:4-site-KT-E-S-sym}, we show the mean energy and mean entropy of the 3-site and 4-site KT chain models against temperature $T/g$ for asymmetric and symmetric cases, respectively. In Fig. \ref{fig:3-4-5-site-S-T-asym} we compare the entropy per site of three-, four- and five-site KT chain models against temperature $T/g$.

\begin{figure}[htp]

\centering
\includegraphics[width=4in]{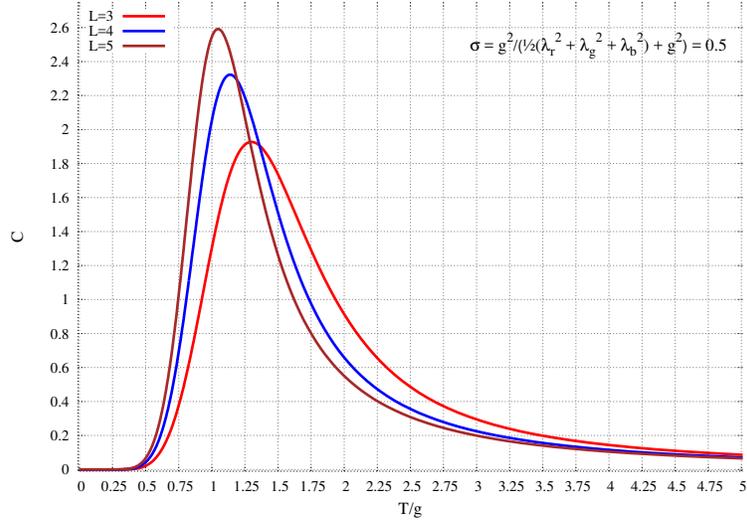}
\caption{Comparing the specific heat of the three-, four- and five-site KT chain models against temperature $T/g$. The plot is for $\sigma=0.5$, asymmetric case. The hopping couplings are $\lambda_r/g = 0.8255, \lambda_g/g = 0.3005$ and $\lambda_b/g = 1.1083$. Note that the critical temperature $T_c/g$ systematically shifts towards left as the lattice size is increased. The peak value of the specific heat also increases as the lattice size is increased, suggesting a phase transition in the infinite volume limit.}
\label{fig:3-4-5-site-C-T-asym}

\end{figure}

In Fig.~\ref{fig:3-4-5-site-C-T-asym} we provide the specific heat of the three, four and five-site KT chain models against temperature $T/g$. We see that the specific heat falls off to zero exponentially quickly as $T/g \to 0$. This fall off is expected since it indicates that the system possesses an energy gap, which we have already seen earlier. As $T/g \to \infty$ the specific heat falls off at a slower (power law) rate indicating that the states are being occupied as temperature increases. There is a critical temperature $T_c$ at which the specific heat attains its maximum.
Note that the critical temperature $T_c/g$ systematically shifts towards the low temperature region as the lattice size is increased. The peak value of the specific heat also increases as the lattice volume is increased, suggesting a possible phase transition in the infinite volume limit.

We attempt to fit the specific heat data to the following functional form \cite{Guttmann:1975}
\beq
C \sim \begin{cases}
\;\; A_+ \left| \frac{T}{T_c} - 1 \right|^{-\alpha_+} + B_+ &\quad\mbox{for} \;\; T > T_c\ , \\
\;\; A_- \left| 1 - \frac{T}{T_c} \right|^{-\alpha_-} + B_- &\quad\mbox{for} \;\; T < T_c\ . \\
\end{cases}
\label{eq:functional-form}
\eeq

Note that these expressions hold only in the vicinity of $T_c$. There are four fit parameters on each side and performing a reliable fit to all four parameters is a highly nontrivial issue. The critical temperature region is more readily attainable on the high-temperature side and so we proceed to perform the fit to the high-temperature side, $T > T_c$, of the specific heat data. 

\begin{figure}[htp]

\subfloat[$L=3$ asymmetric case.]{
\includegraphics[width=3.5in]{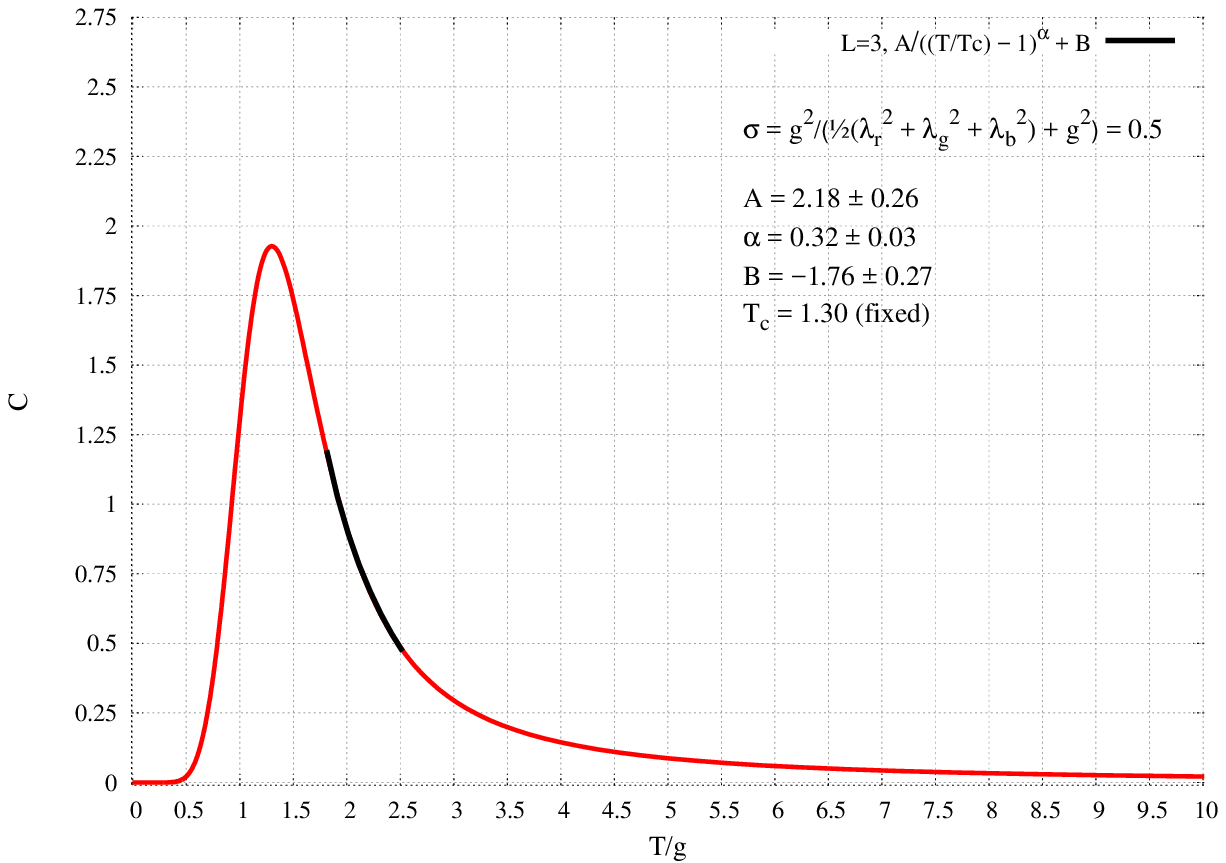}
}

\subfloat[$L=4$ asymmetric case.]{
\includegraphics[width=3.5in]{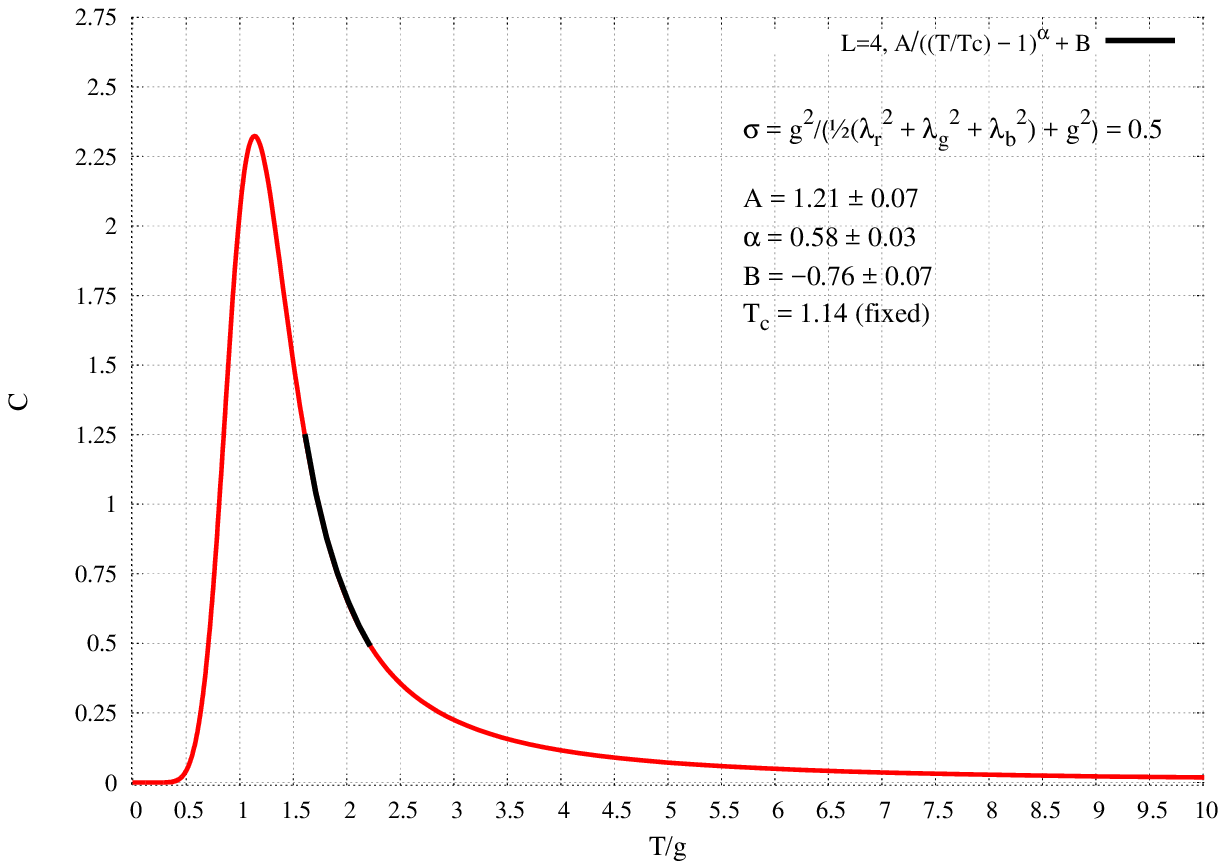}
}

\subfloat[$L=5$ asymmetric case.]{
\includegraphics[width=3.5in]{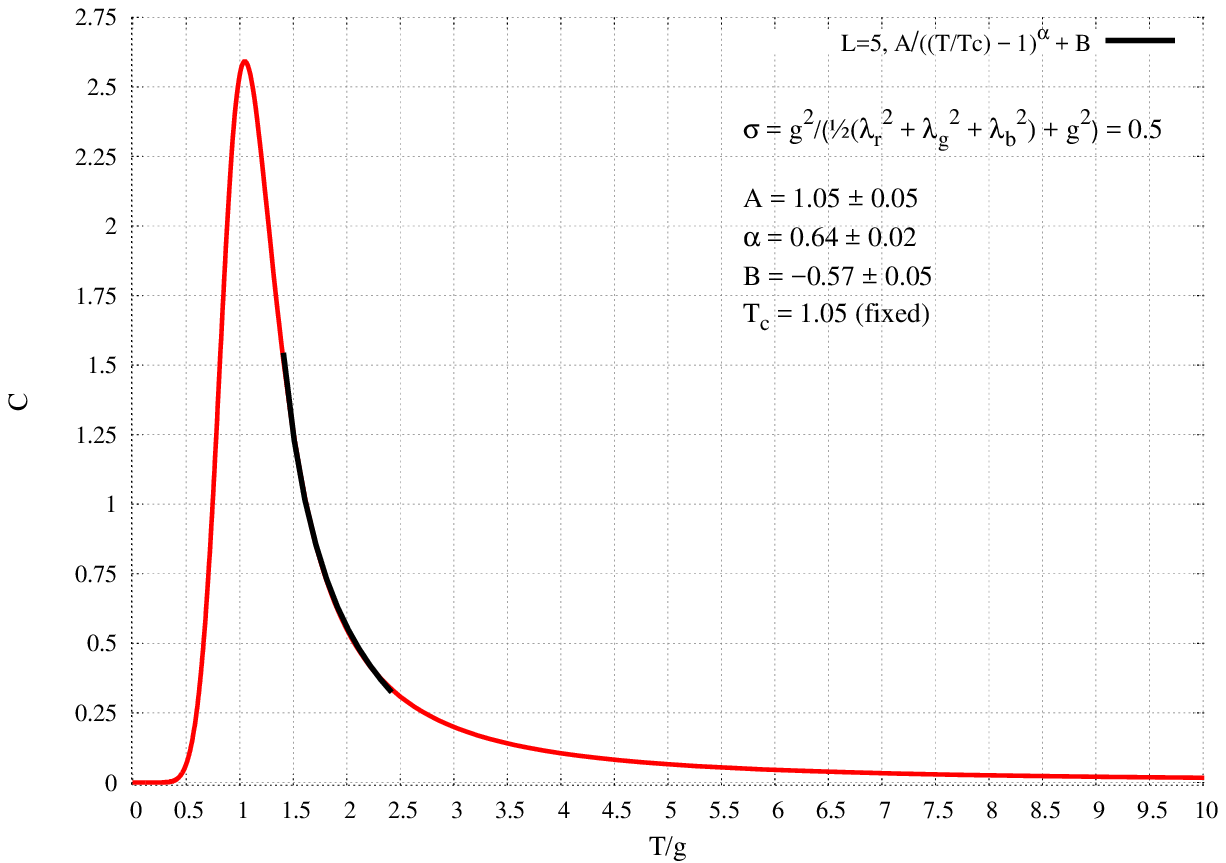}
}

\caption{The specific heat $C$ of $L = 3, 4, 5$ Abelian KT chain models against temperature $T/g$. The plot is for $\sigma=0.5$ and asymmetric case for the hopping couplings with $\lambda_r/g = 0.8255, \lambda_g/g = 0.3005$ and $\lambda_b/g = 1.1083$. We fit the data in the region $T > T_c$ to the functional form described in the text.}
\label{fig:L-site-KT-C-asym-FIT}

\end{figure}

In Fig.~\ref{fig:L-site-KT-C-asym-FIT}, we fit the $T > T_c$ region of the specific heat data of the three-, four- and five-site Abelian KT chain model with asymmetric hopping couplings to the functional form given in~\ref{eq:functional-form}. The critical parameters $A_+, \alpha_+$ and $B_+$, extracted in each case are provided in Table \ref{tab:C-data}.

\begin{table}[h!]
\centering
\begin{tabular}{|c|c|c|c|c|}
\hline
Lattice size $L$ & $T_c/g$ & $A_+$ & $\alpha_+$ & $B_+$ \\
\hline
\hline
3 & $1.30$ & $2.18 \pm 0.26$ & $0.32 \pm 0.03$ & $-1.76 \pm 0.27$ \\
4 & $1.14$ & $1.21 \pm 0.07$ & $0.58 \pm 0.03$ & $-0.76 \pm 0.07$ \\
5 & $1.05$ & $1.05 \pm 0.05$ & $0.64 \pm 0.02$ & $-0.57 \pm 0.05$ \\
\hline
\end{tabular}
\caption{The critical parameters extracted from the specific heat data for $L = 3, 4, 5$ cases of the Abelian KT chain model. The data are for $\sigma=0.5$ and asymmetric case for the hopping couplings with $\lambda_r/g = 0.8255, \lambda_g/g = 0.3005$ and $\lambda_b/g = 1.1083$. The fit is performed for a fixed value of the critical temperature $T_c$ for each $L$ case and in the region $T > T_c$.}
\label{tab:C-data}
\end{table}

\section{Conclusions and Discussions}
\label{sec:conclusions}

In this work, we have studied the spectrum of Abelian KT chains made of $L$ copies of Abelian KT tensor models, connected by Gu-Qi-Stanford type  hopping terms. Unlike their large $N$ cousins~\cite{Narayan:2017qtw}, they do not exhibit fast scrambling or maximality of chaos. In contrast, they seem to fall into the class of quasi-many-body localized (qMBL) system as evinced by the lack of level repulsion in the spectrum. We give a detailed characterization of the energy eigenstates, which we hope will lead to a more deeper understanding of tensor models. As we have discussed in the body of the paper, the spectral statistics of Abelian KT chains seem to show evidence of quasi-many-body localization. It would be good to confirm this by using other diagnostics of MBL phase available in the literature, and check how much of this behavior can be attributed to a finite size effect as has been discussed in \cite{2014arXiv1409.8054D, 2016PhRvL.117x0601Y, 2015AnPhy.362..714P}. Some of the proposed diagnostics are based on entanglement. Consider a system living in one spatial dimension which exhibits an ergodic phase, i.e., a phase where ETH holds. In this ergodic phase, if we follow the evolution of the isolated system from an initial product state, one often sees a ballistic spread in the entanglement, i.e., a linear growth of entanglement entropy with time. In contrast, MBL systems are expected to exhibit a slower growth of entanglement, with the entanglement growing logarithmically in time \cite{2006JSMTE..03..001D, 2008PhRvB..77f4426Z, 2012PhRvL.109a7202B, 2013PhRvL.111l7201S}. Another diagnostic is the area law for entanglement entropy instead of volume law as is usual for excited states in an ergodic system \cite{2013PhRvL.110z0601S}. It would be interesting to check whether the qMBL behavior of Abelian KT chains also extend to their entanglement entropy.

The large $N$ limit of these tensor models(either on the lattice or not) have shown signs of chaos,  therefore  it is  clear that the behavior of the  KT model we have studied is a feature of small $N$,  which is   analogous to the fact that there is no thermalization for small $N$ in the SYK models~\cite{You:2016ldz,Li:2017hdt}.  One can wonder how much of this behavior comes as a result of placing the tensor model on lattice with hopping terms. In the large $N$ limit, this set-up did not lead to MBL behavior either for the KT chain or for SYK\cite{Gu:2016oyy,Narayan:2017qtw}. Therefore,  we expect that this is not the reason behind  the signs of qMBL here. But as we commented before, one has to see if it is a finite size effect in the lattice. Our spectral analysis suggests that is not the case. We are taking values of the couplings  $\lambda$ that are big enough for the bands to overlap. But the lack of level repulsion in the level spacing distribution indicates that there is not much mixing between the bands. However, a  more detailed study with a  larger lattice is needed to say anything conclusively on this matter. 
Another interesting point is that MBL is generally a result of the emergence of conserved quantities which grow extensively with the lattice size e.g. \cite{PhysRevLett.111.127201}.In our model we already know there are such charges which arise from the $(Z_2)^L$ symmetry. But these are present in the large N lattice models as well and are insufficient to cause localization in such models. So, we do not expect these charges to be a significant reason behind the spectral statistics that we see in our model. It would be interesting to see whether there are some other charges in this model which grow extensively with lattice size and play a more dominant role in the spectral statistics.

We expect that the methods we describe in this work can be extended straightforwardly to the Abelian Gurau-Witten model~\cite{Gurau:2016lzk, Witten:2016iux} and more general tensor models on the lattice~\cite{Narayan:2017qtw}. 

In this regard, it is important to  mention that a  numerical study of the Abelian Gurau-Witten model has already appeared in the literature \cite{Krishnan:2016bvg}.

Unlike our work, in \cite{Krishnan:2016bvg} the authors study the spectral statistics of the Abelian GW model without any extension to the lattice and they look at the spectrum of the entire Hilbert space instead of the sector invariant under the $O(2)^6$ symmetry in that model. The main difference of their results with ours is that their model seems to exhibit level repulsion and their study of spectral form factor seems to indicate a random matrixlike behavior in the model.These differences may be due to some fundamental difference between the two models or because they are looking at the full Hilbert space instead of the $O(2)^6$-invariant sector. We think a more careful study of both the Abelian models  is necessary to clarify the reason behind the apparent differences between them.  However if such a difference indeed exists, then it opens up the possibility that by interpolating between the Abelian Gurau-Witten model and the Abelian Klebanov-Tarnopolsky\footnote{In large $N$, such a generalization for (non-Abelian) tensor model is analyzed in the context of maximal chaos in the adjoining paper~\cite{Narayan:2017qtw}.} model one can set up a tensor model with a quasi-many body localization transition. Here we would like to point out that since the non-Abelian tensor chains are known to exhibit random matrixlike  behavior, interspersing them with Abelian tensor models may be another way to set up a system with qMBL transition,  One might be able to explore this transition using some of the available criteria (see eg.  \cite{PhysRevX.5.041047}). 


\vskip 2cm\noindent
{\it Acknowledgements}
We thank Prithvi Narayan for collaboration during the initial stages of this project. We thank  Sumilan Banerjee, Subhro Bhattacharjee, Chethan Krishnan and K. V. Pavan Kumar for extensive discussions about this work. We also thank the anonymous referee for the comments on the draft. J.Y. thanks the Galileo Galilei Institute for Theoretical Physics (GGI) for the hospitality and INFN for partial support during the completion of this work, within the program ``New Developments in AdS3/CFT2 Holography''. J.Y. and V.G. also thank the International Centre for Theoretical Physics (ICTP) for the hospitality and Asia Pacific Center for Theoretical Physics (APCTP) for partial support during the completion of this work, within the program ``Spring School on Superstring Theory and Related Topics''. A.J. thanks the Harish-Chandra Research Institute, where part of this work was completed, for its hospitality. A.J. also thanks Indo-French Centre for the Promotion of Advanced Research (IFCPAR/CEFIPRA) for partial support. We thank the Simons Foundation for partial support. We gratefully acknowledge support from the International Centre for Theoretical Sciences (ICTS), Tata Institute of Fundamental Research, Bangalore.

\appendix

\section{Energy Eigenstates in the Singlet Sector of the Four-site KT Model}
\label{app:4site}

In this appendix we describe with more detail the energy spectrum of the 4-site KT chain model, first in the case where we have a generic asymmetric hopping couplings, the classification is based on the symmetries, or more concretely based on charges associated to those symmetries. For the generic  asymmetric coupling case we give a description of the middle states (states with zero energy). When all hopping couplings are the same we have additional symmetries which enlarge the subspace of middle states. We give a description of this sector.

In the last part of the appendix we describe some special states which  we dubbed protected, they are independent of the hopping coupling constants, and they are energy eigenstates  for both the generic asymmetric hopping coupling case and the symmetric hopping coupling case.

\subsection{Middle states for asymmetric couplings of the three hopping terms}
\label{app:4site middle asym}

There is a $\mathbb{Z}_2^4$ symmetry in the model corresponding to $a_i\rightarrow -a_i$ and $a_i^\dag\rightarrow -a_i^\dag$ for all $i$'s at a particular site. Hence, we can group the states with particular charges under this symmetry and the action of the Hamiltonian will be closed within each such sector.
 
The singlet sector of the Hamiltonian has overlap with 8 such subsectors, i.e., those with charges $(+, +, +, +)$, $(+, +, -, -)$, $(-, +, +, -)$, $(-, -, +, +)$, $(+, -, -, +)$, $(+, -, +, -)$, $(-, +, -, +)$ and $(-, -, -, -)$. The middle states in the singlet sector belong to only 4 of these subsectors, i.e., those with charges $(+, +, +, +)$, $(-, +, -, +)$, $(+, -, +, -)$ and $(-, -, -, -)$. These middle states are enumerated below.

\subsubsection{The $(+, +, +, +)$ subsector}
 
There are two middle states of the form $|A_\pm B_\mp C_\pm D_\mp \rangle$. These 2 states are related in the following way.
\begin{equation}
\begin{split}
|A_- B_+ C_- D_+\rangle = \hat{T} |A_+ B_- C_+ D_- \rangle.
\end{split}
\end{equation} 
There are 24 other middle states of the form $|\ad_{ij}\bd_{jk}\cd_{kl}\dd_{li}\rangle$ where $(i, j, k, l)$ is some permutation of $(1, 2, 3, 4)$. Thus, in total there are $26$ middle states in the $(+, +, +, +)$ sector.

\subsubsection{The $(-, +, -, +)$ subsector}

There is a state of the form 
\begin{align}
|B_+D_-\sum_{i=1}^4 \Big(\at_{\hat{i}} \cs_i+\as_i \ct_{\hat{i}} \Big)\rangle & \equiv \sum_{i=1}^4 \Big(B^{\dag}_+D^{\dag}_-(a^{\dag^3})_{\hat{i}} c^{\dag}_i|\ \rangle+B^\dag_+D^\dag_-a^{\dag}_i (c^{\dag^3})_{\hat{i}}|\ \rangle \Big)\ ,\\
|D_+B_-\sum_{i=1}^4 \Big(\ct_{\hat{i}} \as_i+\cs_i \at_{\hat{i}} \Big)\rangle & \equiv \sum_{i=1}^4 \Big(D^{\dag}_+B^{\dag}_-(c^{\dag^3})_{\hat{i}} a^{\dag}_i|\ \rangle+D^\dag_+B^\dag_-c^{\dag}_i (a^{\dag^3})_{\hat{i}}|\ \rangle \Big)\ , 
\end{align}
These two states are related to each other by 
\begin{equation}
|D_+B_-\sum_{i=1}^4 \Big(\ct_{\hat{i}} \as_i+\cs_i \at_{\hat{i}} \Big)\rangle =\hat{T}^2|B_+D_-\sum_{i=1}^4 \Big(\at_{\hat{i}} \cs_i+\as_i \ct_{\hat{i}} \Big)\rangle\ .
\end{equation}
Then there are 24 states as given below:
\begin{align}
|\at_{\hat{j}} \cs_j \bd_{\widehat{ij}}\dd_{ij}\rangle-|\ct_{\hat{i}}\as_i \bd_{\widehat{ij}}\dd_{ij}\rangle\ ,\\
 \hat{T}^2 \Big(|\at_{\hat{j}} \cs_j \bd_{\widehat{ij}}\dd_{ij}\rangle-|\ct_{\hat{i}}\as_i \bd_{\widehat{ij}}\dd_{ij}\rangle \Big) \ ,
\end{align}
where $(i,j)$ is an ordered pair chosen from the set $\{ 1, 2, 3, 4 \}$  with $i \neq j$. In total, we have $26$ middle states in the $(-, +, -, +)$ sector.

\subsubsection{The $(+, -, +, -)$ subsector}

There are again 26 states in this sector. These states are obtained by translating the states in the previous sector by 1 step.

\subsubsection{The $(-, -, -, -)$ subsector}

Using translation operator, it is convenient to define a projection operator onto $\mathbb{Z}_4$ charge eigenspace:
\begin{equation}
\projt_p\equiv {1\over 4} \sum_{n=0}^2 e^{i{2\pi p\over 4} n}\trans^n\ ,\hspace{1cm} (p=0,1,2,3)\ .
\end{equation}
There are 29 middle states in this sector. Two of them are obtained from linear combinations of the following vectors.
\begin{align}
v_1 =& \sum_{(ijkl)} \left(|a_i b_jc_k d_l  \rangle + | \at_{\hat{i}} \bt_{\hat{j}} \ct_{\hat{k}} \dt_{\hat{l}}\rangle\right) \cr
&+  2\projt_{2}\left[\sum _{i,j} \left(-|\at_{\hat{i}} \bt_{\hat{j}}\cs_j\ds_i\rangle-|\at_{\hat{j}} \bt_{\hat{i}}\cs_i\ds_j\rangle + |\at_{\hat{i}} \bs_{j}\ct_{\hat{j}}\ds_i\rangle+|\at_{\hat{i}} \bs_{i}\ct_{\hat{j}}\ds_j \right) \right]\ .
\end{align}
Note that here in the first term, the sum runs over all permutations $(i, j, k, l)$ of $(1, 2, 3, 4)$ and $\hat{i}$ is defined in \eqref{def:notations for oscillators}. In the second term, the sum is over all ordered pairs $(i,j)$ chosen from the set $\{1, 2, 3, 4\}$. Also, we define two states by
\begin{equation}
|v_2 \rangle= \sum_{i=1}^4|\at_{\hat{i}} \bs_i \ct_{\hat{i}} \ds_i\rangle\;,\qquad |v_3\rangle = \sum_{i=1}^4 | \bt_{\hat{i}} \cs_i \dt_{\hat{i}} \as_i \rangle=\trans |v_2\rangle\ .
\end{equation}
The 2 middle states that can be constructed out of linear combinations of these 3 states are found to be
\begin{equation}
|v_2\rangle +|v_3\rangle\;,\qquad  |v_1\rangle  +|v_2\rangle- |v_3\rangle\ .
\end{equation}
The first one is a Bloch state with Bloch momentum $0$ and the second one has Bloch momentum $\pi$. To enumerate the other 27 middle states, it would be convenient to define
\begin{equation}
|mn\rangle_p \equiv \projt_{p}|(\at \bt\cs\ds)_{\hat{m}\hat{n}mn} \rangle \hspace{1cm} (p=0,1,2,3\;,\;\; m,n=1,2,3,4)\ ,
\end{equation}
where $\frac{\pi}{2}p$ is the Bloch momentum of the state. The remaining 27 middle states are given in Table \ref{tab:0-modes-4-site-KT}.
\begin{table}[h!]
\centering
{
\renewcommand{\arraystretch}{1.2}
	\begin{tabular}{|>{\centering\arraybackslash}m{0.4cm}|>{\centering\arraybackslash}m{12cm}|>{\centering\arraybackslash}m{1.5cm}|}
\hline
$p$ & Energy eigenstate & {\tiny $\#$ of states} \\
\hline
\multirow{7}{*}{$2$} & $\frac{1}{2}|43\rangle_2+\frac{1}{2}|42\rangle_2-\frac{1}{2}|41\rangle_2-\frac{1}{2}|34\rangle_2+|31\rangle_2-\frac{1}{2}|24\rangle_2+|21\rangle_2+\frac{3}{2}|14\rangle_2+|11\rangle_2$ &  \\ \cline{2-2}
 & $\frac{1}{2}|43\rangle_2+\frac{1}{2}|42\rangle_2-\frac{1}{2}|41\rangle_2-\frac{1}{2}|34\rangle_2+|32\rangle_2+\frac{1}{2}|24\rangle_2+|21\rangle_2+\frac{1}{2}|14\rangle_2+|22\rangle_2$ &  \\ \cline{2-2}
 & $\frac{3}{2}|43\rangle_2-\frac{1}{2}|42\rangle_2-\frac{1}{2}|41\rangle_2-\frac{1}{2}|34\rangle_2+|32\rangle_2+|31\rangle_2+\frac{1}{2}|24\rangle_2+\frac{1}{2}|14\rangle_2+|33\rangle_2$ &  \\ \cline{2-2}
  & $\frac{1}{2}|43\rangle_2+\frac{1}{2}|42\rangle_2+\frac{1}{2}|41\rangle_2+\frac{1}{2}|34\rangle_2+\frac{1}{2}|24\rangle_2+\frac{1}{2}|14\rangle_2+|44\rangle_2$ &  7\\ \cline{2-2}
  & $|42\rangle_2-|41\rangle_2-|24\rangle_2+|21\rangle_2+|14\rangle_2-|12\rangle_2$ &  \\ \cline{2-2}
  & $-|43\rangle_2+|41\rangle_2+|34\rangle_2-|31\rangle_2-|14\rangle_2+|13\rangle_2$ &  \\ \cline{2-2}
  & $|43\rangle_2-|42\rangle_2-|34\rangle_2+|32\rangle_2+|24\rangle_2-|23\rangle_2$ &  \\ 
\hline 
\multirow{6}{*}{$0$} & $|43\rangle_0+|42\rangle_0-|31\rangle_0-|21\rangle_0+|11\rangle_0-|44\rangle_0$ &  \\ \cline{2-2}
 & $|43\rangle_0+|42\rangle_0+|41\rangle_0-|24\rangle_0-|23\rangle_0-|21\rangle_0+|22\rangle_0-|44\rangle_0$ &  \\ \cline{2-2}
 & $|43\rangle_0+|42\rangle_0+|41\rangle_0-|34\rangle_0-|32\rangle_0-|31\rangle_0+|33\rangle_0-|44\rangle_0$ & 6 \\ \cline{2-2}
 & $|24\rangle_0-|42\rangle_0-|32\rangle_0+|23\rangle_0-|12\rangle_0+|21\rangle_0$ &  \\ \cline{2-2}
 & $-|32\rangle_0+|23\rangle_0+|13\rangle_0-|31\rangle_0-|34\rangle_0+|43\rangle_0$ &  \\ \cline{2-2} 
 & $-|34\rangle_0+|43\rangle_0+|42\rangle_0-|24\rangle_0-|14\rangle_0+|41\rangle_0$ &  \\ \hline
\multirow{12}{*}{$\pm 1$} & $-\frac{1}{2}(1 \pm i)|43\rangle_{\pm 1}-\frac{1}{2}(1 \pm i)|42\rangle_{\pm 1} \mp i|41\rangle_{\pm 1}-\frac{1}{2}(1 \pm i)|34\rangle_{\pm 1}-$ &\\ & $\frac{1}{2}(1 \pm i)|32\rangle_{\pm 1} \mp i|31\rangle_{\pm 1}-\frac{1}{2}(1 \pm i)|24\rangle_{\pm 1}-\frac{1}{2}(1 \pm i)|23\rangle_{\pm 1} \mp i|21\rangle_{\pm 1}+|11\rangle_{\pm 1}$ & \\ \cline{2-2}
 & $-\frac{1}{2}(1 \mp i)|43\rangle_{ \pm 1}+\frac{1}{2}(1 \mp i)|42\rangle_{ \pm 1}-\frac{1}{2}(1 \mp i)|34\rangle_{ \pm 1}+\frac{1}{2}(1 \mp i)|32\rangle_{ \pm 1}+$ & \\ 
& $\frac{1}{2}(1 \pm i)|24\rangle_{ \pm 1}+\frac{1}{2}(1 \pm i)|23\rangle_{ \pm 1} \pm i|21\rangle_{ \pm 1}+|22\rangle_{\pm \frac{\pi}{2}}$ & \\ \cline{2-2}
 & $\frac{1}{2}(1 \mp i)|43\rangle_{ \pm 1}-\frac{1}{2}(1 \mp i)|42\rangle_{ \pm 1}+\frac{1}{2}(1 \pm i)|34\rangle_{ \pm 1}+\frac{1}{2}(1 \pm i)|32\rangle_{ \pm 1}-$ &\\ & $ \frac{1}{2}(1 \mp i)|24\rangle_{ \pm 1}+\frac{1}{2}(1 \mp i)|23\rangle_{ \pm 1} \pm i|31\rangle_{ \pm 1}+|33\rangle_{ \pm 1}$ & $7 \times 2$ \\ \cline{2-2}
 & $\frac{1}{2}(1 \pm i)|43\rangle_{\pm 1}+\frac{1}{2}(1 \pm i)|42\rangle_{\pm 1}+\frac{1}{2}(1 \mp i)|34\rangle_{\pm 1}-\frac{1}{2}(1 \mp i)|32\rangle_{\pm 1}+ $ & $= 14$ \\ 
& $ \frac{1}{2}(1 \mp i)|24\rangle_{\pm 1}-\frac{1}{2}(1 \mp i)|23\rangle_{\pm 1} \pm i|41\rangle_{\pm 1}+|44\rangle_{\pm 1}$ &  \\ \cline{2-2} 
  & $|43\rangle_{\pm 1}+|34\rangle_{\pm 1}+|21\rangle_{\pm 1}+|12\rangle_{\pm 1}$ &  \\ \cline{2-2} 
  & $|42\rangle_{\pm 1}+|24\rangle_{\pm 1}+|31\rangle_{\pm 1}+|13\rangle_{\pm 1}$ &  \\ \cline{2-2} 
 & $|41\rangle_{\pm 1}+|14\rangle_{\pm 1}+|32\rangle_{\pm 1}+|23\rangle_{\pm 1}$ &  \\ \hline 
\end{tabular}}
\caption{The middle states in the $(-, -, -, -)$ sector of the four-site KT chain model. For a state with label $p$, the Bloch momentum is $\frac{\pi}{2}p$.}
\label{tab:0-modes-4-site-KT}
\end{table}

\subsection{Middle states for symmetric couplings of the three hopping terms}
\label{app:4site middle sym}

In this section we will look at the middle states for the case when $\lambda_r=\lambda_g=\lambda_b=\lambda$. We will try in most cases to write the states using  the $\mathbb{Z}_3$ charges and $\mathbb{Z}_4$ charges (when translation symmetry is also present in the respective sector).

\subsubsection{The $(+,+,+,+)$ subsector}

There are $132$ states in this sector with zero energy. Generically these states can depend on the coupling constants, namely, they will be linear combinations of the basis with coefficients which are dimensionless functions of $g$ and $\lambda$. However what happens is that almost all of them are independent of the coupling, the subsector of these $132$, which is independent of the coupling, has $104$ members.

For the case of symmetric hopping, it is useful to utilize $\mathbb{Z}_4\times \mathbb{Z}_3$ symmetry where $\mathbb{Z}_4$ and $\mathbb{Z}_3$ is generated by $\trans$ and $\om$ defined in \eqref{def: omega} and \eqref{def: translation}, respectively. Hence, we define a projection operator onto $\mathbb{Z}_4\times \mathbb{Z}_3$ eigenstates:
\begin{equation}
\projec_{p,q}\equiv{1\over 12} \sum_{n=0}^2\sum_{m=0}^3 e^{{2\pi i\over 4}p m} e^{{2\pi i\over 3}q n }  \hat{T}^m \hat{\Omega}^n\ , \hspace{1cm} (p=0,1,2,3\;,\; q=0,1,2)\ .
\end{equation}
In addition, it is also convenient to introduce a transposition operator
\begin{equation}
\reflection_{(jk)} a_{ i}\reflection_{(jk)}^{-1}\equiv a_{(jk) i}\ , \hspace{5mm}\mbox{and, similar for $b, c, d$}\;\; (\; (jk)\in S_3 \;)\ .\label{def:transposition operator}
\end{equation}
We now highlight some of them. We have $26$ of the form
\begin{align}
&\projec_{0,0}|A_+B_-C_+D_-\rangle\;,\qquad  \projec_{2,0}|A_+B_-C_+D_-\rangle\\
&\projec_{n,q}|\ad_{12} \bd_{23} \cd_{34} \dd_{41}\rangle\;,\quad \projec_{n,q}\reflection_{(34)}|\ad_{12} \bd_{23} \cd_{34} \dd_{41}\rangle\ , \hspace{1cm} (n=0,1,2,3\;\mbox{and}\; q=0,1,2)\ .
\end{align}
%
In total we have 26. These 26 states are the middle states of the theory with generic asymmetric couplings. We have an additional 6 of the form:
\begin{align}
e^{0}_{0,q}=&\projec_{0,q} (1-\reflection_{34}) |A_+ \bd_{12}C_- \dd_{34}\rangle\ ,\\
e^{0}_{2,q}=&\projec_{2,q}  (1+\reflection_{34}) |A_+ \bd_{12}C_- \dd_{34}\rangle\ ,
\end{align}
%
where $q=0,1,2$. We now describe other states that appear. They are linear combinations of the following states:
\bea
|A_\pm B_\mp\cd_{ij} \dd_{kl}\rangle\quad\text{and}\quad T_3|ijkl\rangle_{\mp}\ ,
\eea
where $T_3$ is the translation by 3 sites on the lattice. The states are
\begin{align}
&(1+\trans^3\reflection_{(12)} )(|A_- B_+\cd_{12} \dd_{34}\rangle+|A_+ B_-\cd_{12} \dd_{34}\rangle  -  |A_+ B_-\cd_{34} \dd_{12}\rangle   -|A_- B_+\cd_{34} \dd_{12}\rangle )\ ,\cr
&(1-\trans^3\reflection_{(14)} )(-|A_- B_+ \cd_{12} \dd_{34}\rangle+|A_+ B_- \cd_{13} \dd_{24}\rangle-|A_- B_+ \cd_{24} \dd_{13}\rangle+|A_+ B_- \cd_{34} \dd_{12}\rangle )\ ,\cr
&(1-\trans^3\reflection_{(13)})(|A_- B_+ \cd_{12} \dd_{34}\rangle+|A_+ B_- \cd_{14} \dd_{23}\rangle-|A_- B_+ \cd_{23} \dd_{14}\rangle - |A_+ B_- \cd_{34} \dd_{12}\rangle)\ , \cr
&(1-\trans^3\reflection_{(24)})(|A_- B_+ \cd_{12} \dd_{34}\rangle -|A_- B_+ \cd_{14} \dd_{23}\rangle +|A_+ B_- \cd_{23} \dd_{14}\rangle -|A_+ B_- \cd_{34} \dd_{12}\rangle )\ ,
\end{align}
where the transposition operator $\reflection_{(jk)}$ is defined in \eqref{def:transposition operator}. We then have 26 states which have zero energy for any value of the couplings. The other 88 out of the 104 are middle states (states at zero energy) just in the symmetric coupling. Some of them are middle states for the partial symmetric points like $\lambda_r = \lambda_{g,b}$, which means that in the general asymmetric coupling case they have eigenvalue $\lambda_r - \lambda_{g,b}$.

The count leaves 28 states out of the 134 which do depend on the coupling. On dimensional grounds the coefficients of the linear combinations are dimensionless functions of $g$ and $\lambda$. They are given by at most quadratic functions in $\frac{g}{\lambda}$ and $\frac{\lambda}{g}$.

All other states we have not described explicitly are given by linear combinations of the following states:
\bea
&&|\ad_{ij} \bd_{pq} \cd_{mn} \dd_{kl} \rangle,\quad  |\ad_{ij} \bd_{kl} \cd_{pq} \dd_{mn} \rangle,\nonumber\\
&& |A_{\sigma_1}A_{\sigma_2}\cd_{ij}\dd_{kl} \rangle \quad \text{and translations},\nonumber\\
&& |A_{\sigma_1}B_{\sigma_2}C_{\sigma_3}D_{\sigma_4}\rangle\ ,
\eea
where $(kl)=(\widehat{ij})$, $(pq)=(\widehat{mn})$ and $\sigma_1, \sigma_2, \sigma_3, \sigma_4 = \pm$.

\subsubsection{The $(-, -, -, -,)$ subsector}

There are 134  middle states. All of them  are linear combinations whose coefficients do not depend on the couplings. We show 72 of them which are very easily written down. All  these states are of the  bi-cubic form, namely, $a^3 b_i c^3 d_j$ or one of the other five possibilities:
\bea
\label{eq:mmmmsymm0}
&&|\at_{\hat{i}} \bs_i \ct_{\hat{j}} \ds_{j} \rangle\quad \text{and}\quad |\as_i \bt_{\hat{i}} \cs_j \dt_{\hat{j}} \rangle \quad\text{there are 28 states in each,}\nonumber\\
&&|\at_{\hat{i}} \bt_{\hat{i}} \cs_i \ds_i\rangle\;\;\text{and 3 translations, for  each translation there are 4 states.}\nonumber
\eea
The first line corresponds to all choices for three different labels of $a$'s and $c$'s that is, $4^2$. The choice of label for $b$ and $d$ is uniquely determined by the previous choice, except that we can interchange $b$ and $d$ labels. This gives $4^2 \times 2$. In multiplying by 2 we are overcounting the choices with labels ${(1, 1), (2, 2), (3, 3), (4, 4)}$. This gives $4^2 \times 2 - 4 = 28$. The same  reorganization of the fields is true for the combination $a_i b^3 c_j d^3$, changing the role of $a, c$ with $b, d$.

The other states that come in the pack of 4, correspond to picking the same choice of labels for the bicubics, e.g. $a_1a_2a_3b_1b_2b_3c_4d_4$ the cubic terms in $a$'s are the same as for $b$'s and this defines uniquely the $c, d$ labels. This gives exactly the count 4.

The states  above are already quite simple, but the writing  does not refer at all about the $\mathbb{Z}_4\times\mathbb{Z}_3$ charges. Let us summarize how the states in \eqref{eq:mmmmsymm0} decomposes under these charges, the first line can be replaced by
\bea
&& \projec_{p,0}|\at_{\widehat{1}}\bs_1\ct_{\widehat{1}}\ds_1\rangle,\quad p=0,1,\nonumber\\
&& \projec_{p,q}|\at_{\widehat{i}}\bs_i\ct_{\widehat{i}}\ds_i\rangle,\quad p=0,1,\,q=0,1,2, \quad\text{one}\,\, i\neq 1,\nonumber\\
&& \projec_{p,q}|\at_{\widehat{1}}\bs_1\ct_{\widehat{i}}\ds_i\rangle,\quad p=0,1,2,3,\,q=0,1,2, \quad\text{one}\,\, i\neq 1,\nonumber\\
&& \projec_{p,q}|\at_{\widehat{i}}\bs_i\ct_{\widehat{j}}\ds_j\rangle,\quad p=0,1,2,3,\,q=0,1,2,\, (i,j)=(2,3),(2,4),(3,4)\ .
\eea
So the first line decomposes as $(2+6+12+36=56)$, and the second line in \eqref{eq:mmmmsymm0} decomposes as,
\bea
&& \projec_{p,0}|\at_{\widehat{1}}\bt_{\widehat{1}}\cs_1\ds_1\rangle,\quad p=0,1,2,3\nonumber\\
&&\projec_{p,q}|\at_{\widehat{i}}\bt_{\widehat{i}}\cs_i\ds_i\rangle,\quad p=0,1,2,3,\,\,q=0,1,2, \quad\text{one}\,\, i\neq 1\ .
\eea
So we have the count $(4+12=16)$. In total, we have 72 states. All these states appear in the linear combinations for the middle states states of the $(-, -, -, -)$ sector in the asymmetric case.

Now, the other states will be given by linear combinations of the following states:
\bea
&&|\at_{\hat{i}} \at_{\hat{j}} \cs_i \ds_j\rangle,\,|\at_{\hat{i}} \bt_{\hat{j}} \cs_j \ds_i\rangle\quad\text{and the four translations},\nonumber\\ 
&&|\at_{\hat{i}} \bs_i \ct_{\hat{j}} \ds_j\rangle, \,|\at_{\hat{i}} \bs_j \ct_{\hat{j}} \ds_i\rangle\quad\text{and translation by one site},\nonumber\\
&&| \at_{\hat{i}} \bt_{\hat{j}}\ct_{\hat{k}}\dt_{\hat{l}}\rangle,\,\, \text{and}\,\, |a_i b_j c_k d_l\rangle\quad\text{with} \,\, (i,j,k,l)=(P(1),P(2),P(3),P(4))\ .
\eea
It is worth mentioning that the bicubic states do not mix with the other two states that appear in the linear combinations, while the other  two, do mix.

\subsubsection{The $(+,-,+,-,)$ subsector}

There are 112 states in this subsector. Out of them, 62 are independent of the coupling. We highlight some of them below. We use the  similar projector operator as introduced in the previous section, with the only catch that now, we do not have the $\mathbb{Z}_4$ symmetry(in this case instead  there is a $\mathbb{Z}_2$, but we choose not to use it for the following discussion).
\begin{equation}
\projz_{q}\equiv{1\over 3} \sum_{n=0}^2  e^{{2\pi i\over 3}q n } \hat{\Omega}^n\hspace{1cm} (q=0,1,2)
\end{equation}
We have 24 states given by,
\bea
\projz_{q}| -\ad_{12}\bt_{\widehat{m}}\cd_{\widehat{12}}\ds_m+\ad_{12}\bs_{n}\cd_{\widehat{12}}\dt_{\widehat{n}}\rangle,\quad(m,n)=(1,2),(2,1),(3,4),(4,3)\nonumber,\\
\projz_{q}| -\ad_{23}\bt_{\widehat{m}}\cd_{\widehat{23}}\ds_m+\ad_{23}\bs_{n}\cd_{\widehat{23}}\dt_{\widehat{n}}\rangle,\quad(m,n)=(2,3),(3,2),(1,4),(4,1)\nonumber,\\
\eea 
With $q=0,1,2$ so in total we have 24 states. All the remaining states are essentially linear combinations of states that we used above (with many more terms)
\bea
&&|\ad_{ij}\bt_{\hat{p}} \cd_{kl}d_p\rangle, \, |\ad_{ij}b_q \cd_{kl}\dt_{\hat{q}}\rangle, \, |\ad_{ij}\bt_{\hat{p}} \cd_{pk}d_l\rangle, \,
|\ad_{ij}b_{k} \cd_{ql}\dt_{\hat{q}} \rangle, \nn \\
&&|\ad_{kl}\bt_{\hat{m}} \cd_{jm}d_i\rangle, |\ad_{kl}b_{j} \cd_{in}\dt_{\hat{n}}\rangle \, |A_{\pm}\bt_{\hat{p}}C_{\mp}d_p\rangle.
\eea
The states which do depend on the coupling constant are given by linear combinations of the following states (in addition to the states already listed):
\bea
&&|A_{\pm}\bt_{\hat{i}}\cd_{ij}\dt_{\hat{j}}\rangle, \, |A_{\pm}\bt_{\hat{i}}\cd_{ji}\dt_{\hat{j}}\rangle, \, |\ad_{ij}\bt_{\hat{i}}C_{\pm}\dt_{\hat{j}}\rangle, \nn \\
&&|\ad_{ji}\bt_{\hat{i}}C_{\pm}\dt_{\hat{j}}\rangle, \, |\ad_{ij}b_{k}C_{\pm}d_l\rangle, \, |A_{\pm}b_{i}\cd_{jk}d_{l}\rangle, \, |A_{\pm}b_pC_{\mp}\dt_{\hat{p}}\rangle. \nn
\eea
The middle states in the $(-,+,-,+)$ subsector are obtained by single translations from those in the $(+, -, +, -)$ subsector.

\subsubsection{The $(+,+,-,-)$ subsector}

There are 112 middle states in this subsector. It looks like all states in this sector depend on the coupling, we will display some of the states and the structure for the values $g = 1, \lambda = 2$. 
\bea
&&\projz_{q}|\ad_{12}\bd_{\widehat{12}}\ct_{\widehat{1}}\ds_{1}-\ad_{14}\bd_{\widehat{14}}\ct_{\widehat{1}}\ds_{1}\rangle,\nonumber\\
&&\projz_{q}|\ad_{24}\bd_{\widehat{24}}\cs_1\dt_{\widehat{1}}-\ad_{23}\bd_{\widehat{23}}\cs_1\dt_{\widehat{1}}\rangle,\nonumber\\
&&\projz_{q}|-\ad_{24}\bd_{\widehat{24}}\ct_{\widehat{2}}\ds_{2}+\ad_{23}\bd_{\widehat{23}}\ct_{\widehat{2}}\ds_{2}\rangle,\nonumber\\
&&\projz_{q}|\ad_{12}\bd_{\widehat{12}}\ct_{\widehat{2}}\ds_{2}-\ad_{24}\bd_{\widehat{24}}\ct_{\widehat{2}}\ds_{2}\rangle,\nonumber\\
&&\projz_{q}|-\ad_{12}\bd_{\widehat{12}}\cs_4\dt_{\widehat{4}}+\ad_{23}\bd_{\widehat{23}}\cs_4\dt_{\widehat{4}}\rangle,\nonumber\\
&&\projz_{q}|-\ad_{12}\bd_{\widehat{12}}\cs_3\dt_{\widehat{3}}+\ad_{24}\bd_{\widehat{24}}\cs_3\dt_{\widehat{3}}\rangle,
\eea
although, $q=0,1,2$, there are just 16 states, because in the first two lines above the states with zero charge vanish, reducing the count from 18 to 16. The rest of the states in this subsector are given as similar linear combinations of the following states
\bea
&&|\ad_{ij}\bd_{kl}c_p\dt_{\hat{p}}\rangle,|\ad_{ij}\bd_{kl}\ct_{\hat{p}}d_p\rangle,|\ad_{ip}\bd_{jk}c_l\dt_{\hat{p}}\rangle,|\ad_{ip}\bd_{jk}\ct_{\hat{p}}d_l\rangle, \nn \\
&&|\ad_{ij}\bd_{kp}c_l\dt_{\hat{p}}\rangle,\,\,|\ad_{ij}\bd_{pk}\ct_{\hat{p}}d_l\rangle,\,\,|A_{\pm}B_{\mp}\ct_{\hat{p}}d_p\rangle, \nn \\
&&|A_{\pm}B_{\pm}\ct_{\hat{p}}d_p\rangle,\,\,|A_{\pm}B_{\mp}c_p\dt_{\hat{p}}\rangle,\,\,|A_{\pm}B_{\pm}c_p\dt_{\hat{p}}\rangle, \nn\\
&&|A_{\pm}\bd_{ij}c_k d_l\rangle,\,\,|\ad_{ij}B_{\pm}c_k d_l\rangle,\,\,|A_{\pm}\bd_{ij}\ct_{\hat{i}}\dt_{\hat{j}}\rangle,\,\,|\ad_{ij}B_{\pm}\ct_{\hat{i}}\dt_{\hat{j}}\rangle \nn.
\eea
By translational invariance $(+, +, -, -)$ is related to $(-, +, +, -)$, $(-, -, +, +)$, $(+, -, -, +)$ through translations of one, two and three sites, respectively.

The energy spectrum will be equal for these four different sectors, and the states, in particular, the middle states will be given by translations of the one discussed for the case of $(+, +, -, -)$ sector. 

\subsection{Protected states with energies $\pm 4g$}
\label{app:4gprotected}

As defined in the comments upon the 2-site model, the protected states are the states whose energies are independent of the couplings of the hopping terms.

In the 4-site model, apart from the middle states that we enumerated before, there are five protected states with energy $4g$ and five other protected states with energy $-4g$. These are distributed in the $(+,+,+,+)$ , $(+,-,+,-)$ and $(-,+,-,+)$ subsectors.

\subsubsection{Protected states in $(+,+,+,+)$ subsector}

In the $(+, +, +, +)$ subsector we find three protected states with energy $E = 4g$ and three protected states with energy $E =- 4g$ . They are characterized as follows:
\begin{align}
&\projec_{2,q} |A_+ \bd_{12}C_+ \dd_{34}\rangle\ ,\\
&\reflection_{23}\projec_{2,q} |A_+ \bd_{12}C_+ \dd_{34}\rangle\ ,
\end{align}
where, the first line correspond to states with $E=4g$,the second to energy $E=-4g$ and $q=0,1,2$.

%
%
%
\subsubsection{Protected states in $(+,-,+,-)$ subsector}

In the $(+, -, +, -)$ subsector, we find one protected state with energy $E = 4g$ and $1$ protected state with energy $E =- 4g$. These states are
\begin{equation}
\begin{split}
\sum_{i=1}^4\Big(|A_\pm \bs_i C_\pm \dt_{\hat{i}}\rangle+|A_\pm \bt_{\hat{i}} C_\pm \ds_i\rangle\Big)\ ,
\end{split}
\end{equation}
where the state with the $(+)$ sign has energy $4g$ and the one with the $(-)$ sign has energy $-4g$. This couple of states  are both charge zero  states of the $\mathbb{Z}_3$, symmetry, actually they are sum of two zero charge states,
\bea
\projz_{q=0}\left((1+\widehat{T}^2)|A_\pm \bs_1 C_\pm \dt_{\hat{1}}\rangle\right)+\projz_{q=0}\left((1+\widehat{T}^2)|A_\pm \bs_2 C_\pm \dt_{\hat{2}}\rangle \right)\ .
\eea
In the first state in the sum, we  can actually remove the projection operator since, the state $(1+\widehat{T}^2)|A_\pm \bs_1 C_\pm \dt_{\hat{1}}\rangle$ it is by itself an invariant.

\subsubsection{Protected states in $(-,+,-,+)$ subsector}

In the $(-, +, -, +)$ subsector, we find one protected state with energy $E = 4g$ and $1$ protected state with energy $E =- 4g$. These states are obtained by translating the protected states in the $(+,-,+,-)$ subsector by 1 step. Therefore these states are
\begin{equation}
\begin{split}
\sum_{i=1}^4\Big(|B_\pm \cs_i D_\pm \at_{\hat{i}}\rangle+|B_\pm \ct_{\hat{i}} D_\pm\as_i\rangle\Big)=\hat{T}\Big[\sum_{i=1}^4\Big(|A_\pm \bs_i C_\pm \dt_{\hat{i}}\rangle+|A_\pm \bt_{\hat{i}} C_\pm \ds_i\rangle\Big)\Big]
\end{split}\ ,
\end{equation}
where, as before, the state with the $(+)$ sign has energy $4g$ and the one with the $(-)$ sign has energy $-4g$. We can also used the projector operator as in the previous section,
\bea
&&\widehat{T}\left(\projz_{q=0}\left((1+\widehat{T}^2)|A_\pm \bs_1 C_\pm \dt_{\hat{1}}\rangle\right)+\projz_{q=0}\left((1+\widehat{T}^2)|A_\pm \bs_2 C_\pm \dt_{\hat{2}}\rangle \right)\right)\ ,\nonumber\\
&&\projz_{q=0}\left((\widehat{T}+\widehat{T}^3)|A_\pm \bs_1 C_\pm \dt_{\hat{1}}\rangle\right)+\projz_{q=0}\left((\widehat{T}+\widehat{T}^3)|A_\pm \bs_2 C_\pm \dt_{\hat{2}}\rangle \right)\ ,
\eea
where in the second line we used the fact that translations and the $\mathbb{Z}_3$  transformations commute.


\bibliography{AbelianTens_prd}

\begin{thebibliography}{10}

\bibitem{1958PhRv..109.1492A}
P.~W. {Anderson}.
\newblock {Absence of Diffusion in Certain Random Lattices}.
\newblock {\em Physical Review}, 109:1492--1505, March 1958.

\bibitem{Balasubramanian:2016ids}
Vijay Balasubramanian, Ben Craps, Bart{\l}omiej Czech, and G{\'a}bor
  S{\'a}rosi.
\newblock {Echoes of chaos from string theory black holes}.
\newblock {\em JHEP}, 03:154, 2017.

\bibitem{Banerjee:2016ncu}
Sumilan Banerjee and Ehud Altman.
\newblock {Solvable model for a dynamical quantum phase transition from fast to
  slow scrambling}.
\newblock {\em Phys. Rev.}, B95(13):134302, 2017.

\bibitem{2012PhRvL.109a7202B}
J.~H. {Bardarson}, F.~{Pollmann}, and J.~E. {Moore}.
\newblock {Unbounded Growth of Entanglement in Models of Many-Body
  Localization}.
\newblock {\em Physical Review Letters}, 109(1):017202, July 2012.

\bibitem{2006AnPhy.321.1126B}
D.~M. {Basko}, I.~L. {Aleiner}, and B.~L. {Altshuler}.
\newblock {Metal insulator transition in a weakly interacting many-electron
  system with localized single-particle states}.
\newblock {\em Annals of Physics}, 321:1126--1205, May 2006.

\bibitem{Berkooz:2016cvq}
Micha Berkooz, Prithvi Narayan, Moshe Rozali, and Joan Sim{\'o}n.
\newblock {Higher Dimensional Generalizations of the SYK Model}.
\newblock {\em JHEP}, 01:138, 2017.

\bibitem{Bohigas:1983er}
O.~Bohigas, M.~J. Giannoni, and C.~Schmit.
\newblock {Characterization of chaotic quantum spectra and universality of
  level fluctuation laws}.
\newblock {\em Phys. Rev. Lett.}, 52:1--4, 1984.

\bibitem{Bonzom:2012hw}
Valentin Bonzom, Razvan Gurau, and Vincent Rivasseau.
\newblock {Random tensor models in the large N limit: Uncoloring the colored
  tensor models}.
\newblock {\em Phys. Rev.}, D85:084037, 2012.

\bibitem{Carrozza:2015adg}
Sylvain Carrozza and Adrian Tanasa.
\newblock {$O(N)$ Random Tensor Models}.
\newblock {\em Lett. Math. Phys.}, 106(11):1531--1559, 2016.

\bibitem{Cotler:2016fpe}
Jordan~S. Cotler, Guy Gur-Ari, Masanori Hanada, Joseph Polchinski, Phil Saad,
  Stephen~H. Shenker, Douglas Stanford, Alexandre Streicher, and Masaki Tezuka.
\newblock {Black Holes and Random Matrices}.
\newblock {\em JHEP}, 05:118, 2017.

\bibitem{Das:2017pif}
Sumit~R. Das, Antal Jevicki, and Kenta Suzuki.
\newblock {Three Dimensional View of the SYK/AdS Duality}.
\newblock {\em JHEP}, 09:017, 2017.

\bibitem{Davison:2016ngz}
Richard~A. Davison, Wenbo Fu, Antoine Georges, Yingfei Gu, Kristan Jensen, and
  Subir Sachdev.
\newblock {Thermoelectric transport in disordered metals without
  quasiparticles: the SYK models and holography}.
\newblock {\em Phys. Rev.}, B95(15):155131, 2017.

\bibitem{2013arXiv1305.5127D}
W.~{De Roeck} and F.~{Huveneers}.
\newblock {Asymptotic localization of energy in non-disordered oscillator
  chains}.
\newblock {\em ArXiv e-prints}, May 2013.

\bibitem{2014CMaPh.332.1017D}
W.~{De Roeck} and F.~{Huveneers}.
\newblock {Asymptotic Quantum Many-Body Localization from Thermal Disorder}.
\newblock {\em Communications in Mathematical Physics}, 332:1017--1082,
  December 2014.

\bibitem{2014arXiv1409.8054D}
W.~{De Roeck} and F.~{Huveneers}.
\newblock {Can translation invariant systems exhibit a Many-Body Localized
  phase?}
\newblock {\em ArXiv e-prints}, September 2014.

\bibitem{2014PhRvB..90p5137D}
W.~{De Roeck} and F.~{Huveneers}.
\newblock {Scenario for delocalization in translation-invariant systems}.
\newblock {\em Physical Review}, B90(16):165137, October 2014.

\bibitem{2006JSMTE..03..001D}
G.~{DeChiara}, S.~{Montangero}, P.~{Calabrese}, and R.~{Fazio}.
\newblock {Entanglement entropy dynamics of Heisenberg chains}.
\newblock {\em Journal of Statistical Mechanics: Theory and Experiment},
  3:03001, March 2006.

\bibitem{1991PhRvA..43.2046D}
J.~M. {Deutsch}.
\newblock {Quantum statistical mechanics in a closed system}.
\newblock {\em Phys. Rev.}, A43:2046--2049, February 1991.

\bibitem{Dyer:2016pou}
Ethan Dyer and Guy Gur-Ari.
\newblock {2D CFT Partition Functions at Late Times}.
\newblock {\em JHEP}, 08:075, 2017.

\bibitem{1962JMP.....3..140D}
F.~J. {Dyson}.
\newblock {Statistical Theory of the Energy Levels of Complex Systems. I}.
\newblock {\em Journal of Mathematical Physics}, 3:140--156, January 1962.

\bibitem{1962JMP.....3..157D}
F.~J. {Dyson}.
\newblock {Statistical Theory of the Energy Levels of Complex Systems. II}.
\newblock {\em Journal of Mathematical Physics}, 3:157--165, January 1962.

\bibitem{Garcia-Garcia:2017pzl}
Antonio~M. García-García and Jacobus J.~M. Verbaarschot.
\newblock {Analytical Spectral Density of the Sachdev-Ye-Kitaev Model at finite
  N}.
\newblock {\em Phys. Rev.}, D96(6):066012, 2017.

\bibitem{2014JSMTE..10..010G}
T.~{Grover} and M.~P.~A. {Fisher}.
\newblock {Quantum disentangled liquids}.
\newblock {\em Journal of Statistical Mechanics: Theory and Experiment},
  10:10010, October 2014.

\bibitem{Gu:2017ohj}
Yingfei Gu, Andrew Lucas, and Xiao-Liang Qi.
\newblock {Energy diffusion and the butterfly effect in inhomogeneous
  Sachdev-Ye-Kitaev chains}.
\newblock {\em SciPost Phys.}, 2:018, 2017.

\bibitem{Gu:2016oyy}
Yingfei Gu, Xiao-Liang Qi, and Douglas Stanford.
\newblock {Local criticality, diffusion and chaos in generalized
  Sachdev-Ye-Kitaev models}.
\newblock {\em JHEP}, 05:125, 2017.

\bibitem{Guhr:1997ve}
Thomas Guhr, Axel Muller-Groeling, and Hans~A. Weidenmuller.
\newblock {Random matrix theories in quantum physics: Common concepts}.
\newblock {\em Phys. Rept.}, 299:189--425, 1998.

\bibitem{Gurau:2011xq}
Razvan Gurau.
\newblock {The complete 1/N expansion of colored tensor models in arbitrary
  dimension}.
\newblock {\em Annales Henri Poincare}, 13:399--423, 2012.

\bibitem{Gurau:2016cjo}
Razvan Gurau.
\newblock {Invitation to Random Tensors}.
\newblock {\em SIGMA}, 12:094, 2016.

\bibitem{Gurau:2016lzk}
Razvan Gurau.
\newblock {The complete $1/N$ expansion of a SYK--like tensor model}.
\newblock {\em Nucl. Phys.}, B916:386--401, 2017.

\bibitem{Guttmann:1975}
A.~J. Guttmann.
\newblock {Analysis of experimental specific heat data near the critical
  temperature. I. Theory}.
\newblock {\em Journal of Physics C: Solid State Physics}, 8:4037, 1975.

\bibitem{2016JSP...163..998I}
J.~Z. {Imbrie}.
\newblock {On Many-Body Localization for Quantum Spin Chains}.
\newblock {\em Journal of Statistical Physics}, 163:998--1048, June 2016.

\bibitem{Jevicki:2016ito}
Antal Jevicki and Kenta Suzuki.
\newblock {Bi-Local Holography in the SYK Model: Perturbations}.
\newblock {\em JHEP}, 11:046, 2016.

\bibitem{Jevicki:2016bwu}
Antal Jevicki, Kenta Suzuki, and Junggi Yoon.
\newblock {Bi-Local Holography in the SYK Model}.
\newblock {\em JHEP}, 07:007, 2016.

\bibitem{Jian:2017jfl}
Chao-Ming Jian, Zhen Bi, and Cenke Xu.
\newblock {A model for continuous thermal Metal to Insulator Transition}.
\newblock {\em Phys. Rev.}, B96(11):115122, 2017.

\bibitem{Jian:2017unn}
Shao-Kai Jian and Hong Yao.
\newblock {Solvable Sachdev-Ye-Kitaev models in higher dimensions: from
  diffusion to many-body localization}.
\newblock {\em Phys. Rev. Lett.}, 119(20):206602, 2017.

\bibitem{kagan1984localization}
Yu~Kagan and LA~Maksimov.
\newblock Localization in a system of interacting particles diffusing in a
  regular crystal.
\newblock {\em Zhurnal Eksperimental'noi i Teoreticheskoi Fiziki}, 87:348--365,
  1984.

\bibitem{KitaevTalks}
A.~Kitaev.
\newblock {A simple model of quantum holography}.
\newblock {\em {}}, pages
  \url{http://online.kitp.ucsb.edu/online/entangled15/kitaev/},
  \url{http://online.kitp.ucsb.edu/online/entangled15/kitaev2/}, Talks at KITP,
  April 7, 2015 and May 27, (2015), 2015.

\bibitem{kitaevfirsttalk}
A.~Kitaev.
\newblock {Hidden correlations in the Hawking radiation and thermal noise}.
\newblock {\em {}}, pages
  \url{http://online.kitp.ucsb.edu/online/joint98/kitaev/}, KITP seminar, Feb.
  12, (2015), 2015.

\bibitem{Klebanov:2016xxf}
Igor~R. Klebanov and Grigory Tarnopolsky.
\newblock {Uncolored random tensors, melon diagrams, and the Sachdev-Ye-Kitaev
  models}.
\newblock {\em Phys. Rev.}, D95(4):046004, 2017.

\bibitem{Krishnan:2017ztz}
Chethan Krishnan, K.~V.~Pavan Kumar, and Sambuddha Sanyal.
\newblock {Random Matrices and Holographic Tensor Models}.
\newblock {\em JHEP}, 06:036, 2017.

\bibitem{Krishnan:2016bvg}
Chethan Krishnan, Sambuddha Sanyal, and P.~N. Bala~Subramanian.
\newblock {Quantum Chaos and Holographic Tensor Models}.
\newblock {\em JHEP}, 03:056, 2017.

\bibitem{Li:2017hdt}
Tianlin Li, Junyu Liu, Yuan Xin, and Yehao Zhou.
\newblock {Supersymmetric SYK model and random matrix theory}.
\newblock {\em JHEP}, 06:111, 2017.

\bibitem{Liu:2016rdi}
Yizhuang Liu, Maciej~A. Nowak, and Ismail Zahed.
\newblock {Disorder in the Sachdev-Yee-Kitaev Model}.
\newblock {\em Phys. Lett.}, B773:647--653, 2017.

\bibitem{Maldacena:2015waa}
Juan Maldacena, Stephen~H. Shenker, and Douglas Stanford.
\newblock {A bound on chaos}.
\newblock {\em JHEP}, 08:106, 2016.

\bibitem{Maldacena:2016hyu}
Juan Maldacena and Douglas Stanford.
\newblock {Remarks on the Sachdev-Ye-Kitaev model}.
\newblock {\em Phys. Rev.}, D94(10):106002, 2016.

\bibitem{Maldacena:2016upp}
Juan Maldacena, Douglas Stanford, and Zhenbin Yang.
\newblock {Conformal symmetry and its breaking in two dimensional Nearly
  Anti-de-Sitter space}.
\newblock {\em PTEP}, 2016(12):12C104, 2016.

\bibitem{Mandal:2017thl}
Gautam Mandal, Pranjal Nayak, and Spenta~R. Wadia.
\newblock {Coadjoint orbit action of Virasoro group and two-dimensional quantum
  gravity dual to SYK/tensor models}.
\newblock {\em JHEP}, 11:046, 2017.

\bibitem{2015ARCMP...6...15N}
R.~{Nandkishore} and D.~A. {Huse}.
\newblock {Many-Body Localization and Thermalization in Quantum Statistical
  Mechanics}.
\newblock {\em Annual Review of Condensed Matter Physics}, 6:15--38, March
  2015.

\bibitem{Narayan:2017qtw}
Prithvi Narayan and Junggi Yoon.
\newblock {SYK-like Tensor Models on the Lattice}.
\newblock {\em JHEP}, 08:083, 2017.

\bibitem{Oganesyan:2007aa}
V.~Oganesyan and D.~A. Huse.
\newblock {Localization of interacting fermions at high temperature}.
\newblock {\em Phys. Rev.}, B75:155111, 2007.

\bibitem{Papadodimas:2015xma}
Kyriakos Papadodimas and Suvrat Raju.
\newblock {Local Operators in the Eternal Black Hole}.
\newblock {\em Phys. Rev. Lett.}, 115(21):211601, 2015.

\bibitem{2015AnPhy.362..714P}
Z.~{Papi{\'c}}, E.~M. {Stoudenmire}, and D.~A. {Abanin}.
\newblock {Many-body localization in disorder-free systems: The importance of
  finite-size constraints}.
\newblock {\em Annals of Physics}, 362:714--725, November 2015.

\bibitem{Peng:2016mxj}
Cheng Peng, Marcus Spradlin, and Anastasia Volovich.
\newblock {A Supersymmetric SYK-like Tensor Model}.
\newblock {\em JHEP}, 05:062, 2017.

\bibitem{Polchinski:2016xgd}
Joseph Polchinski and Vladimir Rosenhaus.
\newblock {The Spectrum in the Sachdev-Ye-Kitaev Model}.
\newblock {\em JHEP}, 04:001, 2016.

\bibitem{2008Natur.452..854R}
M.~{Rigol}, V.~{Dunjko}, and M.~{Olshanii}.
\newblock {Thermalization and its mechanism for generic isolated quantum
  systems}.
\newblock {\em Nature}, 452:854--858, April 2008.

\bibitem{Sachdev:2015efa}
Subir Sachdev.
\newblock {Bekenstein-Hawking Entropy and Strange Metals}.
\newblock {\em Phys. Rev.}, X5(4):041025, 2015.

\bibitem{Sachdev:1992fk}
Subir Sachdev and Jinwu Ye.
\newblock {Gapless spin fluid ground state in a random, quantum Heisenberg
  magnet}.
\newblock {\em Phys. Rev. Lett.}, 70:3339, 1993.

\bibitem{2014AIPC.1610...11S}
M.~{Schiulaz} and M.~{M{\"u}ller}.
\newblock {Ideal quantum glass transitions: Many-body localization without
  quenched disorder}.
\newblock In {\em American Institute of Physics Conference Series}, volume 1610
  of {\em American Institute of Physics Conference Series}, pages 11--23,
  August 2014.

\bibitem{2013PhRvL.111l7201S}
M.~{Serbyn}, Z.~{Papi{\'c}}, and D.~A. {Abanin}.
\newblock {Local Conservation Laws and the Structure of the Many-Body Localized
  States}.
\newblock {\em Physical Review Letters}, 111(12):127201, September 2013.

\bibitem{2013PhRvL.110z0601S}
M.~{Serbyn}, Z.~{Papi{\'c}}, and D.~A. {Abanin}.
\newblock {Universal Slow Growth of Entanglement in Interacting Strongly
  Disordered Systems}.
\newblock {\em Physical Review Letters}, 110(26):260601, June 2013.

\bibitem{PhysRevLett.111.127201}
Maksym Serbyn, Z.~Papi\ifmmode~\acute{c}\else \'{c}\fi{}, and Dmitry~A. Abanin.
\newblock Local conservation laws and the structure of the many-body localized
  states.
\newblock {\em Phys. Rev. Lett.}, 111:127201, Sep 2013.

\bibitem{PhysRevX.5.041047}
Maksym Serbyn, Z.~Papi\ifmmode~\acute{c}\else \'{c}\fi{}, and Dmitry~A. Abanin.
\newblock Criterion for many-body localization-delocalization phase transition.
\newblock {\em Phys. Rev. X}, 5:041047, Dec 2015.

\bibitem{Shenker:2013pqa}
Stephen~H. Shenker and Douglas Stanford.
\newblock {Black holes and the butterfly effect}.
\newblock {\em JHEP}, 03:067, 2014.

\bibitem{1994PhRvE..50..888S}
M.~{Srednicki}.
\newblock {Chaos and quantum thermalization}.
\newblock {\em Phys. Rev.}, E50:888--901, August 1994.

\bibitem{2008PhRvB..77f4426Z}
M.~{{\v Z}nidari{\v c}}, T.~{Prosen}, and P.~{Prelov{\v s}ek}.
\newblock {Many-body localization in the Heisenberg XXZ magnet in a random
  field}.
\newblock {\em Phys. Rev.}, B77(6):064426, February 2008.

\bibitem{1967SIAMR...9....1W}
E.~P. {Wigner}.
\newblock {Random Matrices in Physics}.
\newblock {\em SIAM Review}, 9:1--23, January 1967.

\bibitem{Witten:2016iux}
Edward Witten.
\newblock {An SYK-Like Model Without Disorder}.
\newblock 2016.

\bibitem{2016PhRvL.117x0601Y}
N.~Y. {Yao}, C.~R. {Laumann}, J.~I. {Cirac}, M.~D. {Lukin}, and J.~E. {Moore}.
\newblock {Quasi-Many-Body Localization in Translation-Invariant Systems}.
\newblock {\em Physical Review Letters}, 117(24):240601, December 2016.

\bibitem{You:2016ldz}
Yi-Zhuang You, Andreas W.~W. Ludwig, and Cenke Xu.
\newblock {Sachdev-Ye-Kitaev Model and Thermalization on the Boundary of
  Many-Body Localized Fermionic Symmetry Protected Topological States}.
\newblock {\em Phys. Rev.}, B95(11):115150, 2017.

\end{thebibliography}
\bibliographystyle{plain}

\end{document}